\documentclass[aps,floats]{revtex4-2}
\usepackage{amsmath,amssymb}
\usepackage{graphicx,epsfig}
\usepackage[greek,english]{babel}
\usepackage{bbold}

\numberwithin{equation}{section}
 
\begin{document}
\bibliographystyle {plain}

\pdfoutput=1
\def\oppropto{\mathop{\propto}} 
\def\opsimeq{\mathop{\simeq}}
\def\opoverderline{\mathop{\overline}}
\def\operarrow{\mathop{\longrightarrow}}
\def\opsim{\mathop{\sim}}

\def\opmin{\mathop{\min}} 
\def\opmax{\mathop{\max}} 
\def\oplim{\mathop{\lim}} 

\newcommand{\asc}{\mathrel{\rotatebox[origin=c]{45}{$\leftrightarrow$}}}
\newcommand{\des}{\mathrel{\rotatebox[origin=c]{-45}{$\leftrightarrow$}}}

%%%%%%%%%%%%%%%%%%%%%%%%%%%%%%%%%%%%%%%%%%%%%%%%%%%%%%%%%%%%%%%%%%%%%%%%%%%%
\title{ Real-Space renormalization for the drawups and drawdowns of stochastic trajectories: \\
universality classes for diffusion processes and random walks via the scale function    
} 

%%%%%%%%%%%%%%%%%%%%%%%%%%%%%%%%%%%%%%%%%%%%%%%%%%%%%%%%%%%%%%%%%%%%%%%%%%%%

\author{C\'ecile Monthus}
\affiliation{Universit\'e Paris-Saclay, CNRS, CEA, Institut de Physique Th\'eorique, 91191 Gif-sur-Yvette, France}

%%%%%%%%%%%%%%%%%%%%%%%%%%%%%%%%%%%%%%%%%%%%%%%%%%%%%%%%%%%%%%%%%%%%%%%%%%%%

\begin{abstract}
For one-dimensional diffusion processes with arbitrary diffusion coefficient $D(x)$ and Ito force $F_I(x)$, we analyze  the statistical properties of the renormalized trajectories at scale $\Gamma$, where one keeps only the alternating sequence of local minima and local maxima that are separated by drawups and drawdowns of amplitude bigger than $\Gamma$. The elementary RG-step between $\Gamma$ and $\Gamma+d \Gamma$ thus amounts to eliminate the drawups and drawdowns of amplitude $\Gamma$, a real-space-RG-procedure that is well-known for Brownian trajectories in the context of the Strong-Disorder-RG approach applied to the disorder landscape of random models. The main result is that the universality classes are determined by the scale function $s(x)$ satisfying $D(x) s''(x)+F_I(x) s'(x)=0$ that governs the exit probabilities from an interval : all the diffusions sharing the same scale function $s(x)$ have the same statistics of renormalized trajectories, even if their initial microscopic trajectories are completely different.  The simplest universality classes for diffusions on the whole line $x \in ]-\infty,+\infty[$ and on the half-line $x \in ]0,+\infty[$ are illustrated with several examples of processes, with or without steady states: Cauchy, Student, Bessel, Fisher-Snedecor, Geometric Brownian. This framework concerning diffusion processes is then adapted to discrete-time and to continuous-time random walks on the one-dimensional lattice with arbitrary nearest-neighbor jump probabilities or rates.

\end{abstract}

\maketitle

% \tableofcontents

\section{ Introduction }

The Strong-Disorder-RG approach (see the reviews \cite{igloi05review,igloiMiniReview} and references therein)
has been first formulated  
for quantum antiferromagnetic spin chains \cite{ma79,danielantiferro}
and for the Random Transverse Field Ising Chain \cite{fisher92,rgfinitesize,danielreview}.
 The last case can be reformulated 
as a real-space-RG concerning the disorder landscape behaving as a random walk 
\cite{igloi98,dharyoung,microcano}.
As a consequence,  the Strong-Disorder-RG approach has been applied to random walks in Brownian landscapes
\cite{igloi98c,rgsinaishort,rgsinailong,rgrfim,rgreadiff,locgolosov,energysinai},
but can be reformulated as an exact procedure to analyze the statistics of local extrema of a Brownian motion
as a function of the scale $\Gamma$.
It is then natural to apply the same idea to analyze the statistics of local extrema
of other stochastic processes \cite{rgtoy,schehr10}, with explicit results for the toy model \cite{rgtoy},
and for the Brownian motion with various constraints, for Bessel processes and for some continuous time random walks \cite{schehr10}.

The goal of the present paper is to generalize this idea \cite{rgtoy,schehr10} to any one-dimensional diffusion process
involving an arbitrary diffusion coefficient $D(x)$ and an arbitrary Ito force $F_I(x)$,
and to discrete-time and to continuous-time random walks on the one-dimensional lattice with arbitrary nearest-neighbor jump probabilities or rates.
We will describe in great detail the cases of diffusion processes on the whole line $x \in ]-\infty,+\infty[$
and on the half-line $x \in ]0,+\infty[$ and identify the simplest universality classes with explicit examples.
In the renormalized trajectory at scale $\Gamma$, one keeps only
the alternating sequence of local minima and local maxima,
that are separated by drawups and drawdowns of amplitude bigger than $\Gamma$,
so that the elementary RG-step amounts to decimate the drawups and drawdowns of 
amplitude $\Gamma$. Besides this RG-flow analysis, one can also compute  
the statistics of successive drawups and drawdowns via direct probabilistic methods,
and it will be useful to consider both perspectives.
Note that the statistics of the maximal drawdown
has been also much studied in the mathematical literature (see \cite{Taylor75,Lehoczky77,2009,Z2012,Z2015,proof,Salminen,BookZhang} and references therein),
in particular in relation with financial applications,
while the notion of $\Gamma$-extrema has been also analyzed 
in mathematics under the name h-extrema \cite{Neveu,Foggionato}.
More generally, the present analysis of the renormalized trajectories based on the drawups and drawdowns
of amplitude bigger than $\Gamma$ belongs to the field of extreme value statistics
that has attracted a lot of interest in statistical physics recently
(see the book \cite{BookExtremes} and the reviews \cite{SatyaReview,schehr_EVS} and references therein).

%%%%%%%%%%%%%%%%%%%%%%%%%%%%%%%%%%%

The paper is organized as follows.
In Section \ref{sec_nota}, we introduce the notations for the properties of diffusion processes
that will be useful in the other sections of the main text.
In section \ref{sec_WholeLine}, we  analyze the renormalized trajectories for diffusions on 
the whole line $x \in ]-\infty,+\infty[$
before describing examples in section \ref{sec_WholeLineExamples}.
In section \ref{sec_HalfLine}, we  analyze the renormalized trajectories for diffusions on 
the half line $x \in ]0,+\infty[$, before describing examples with various possible boundary behaviors at the origin
in section \ref{sec_HalfLineExamples}.
Our conclusions are summarized in section \ref{sec_conclusion}.

In the Appendices, we describe the following complementary results : 

$\bullet$ In the Appendices from A to F, we explain how the results of the main text
 concerning only the values of extrema can be generalized to take into account
the durations of the drawups and drawdowns of the renormalized trajectories, in order to
obtain explicit results for the time-Laplace transforms 
as a function of the Laplace variable $p>0$ (instead of $p=0$ of the main text).

$\bullet$ In the two last Appendices, we explain how the framework concerning diffusion processes
can be adapted 
for random walks on the one-dimensional lattice defined either in discrete-time in Appendix \ref{sec_DiscreteTimeRW}
or in continuous-time in Appendix \ref{sec_ContinuousTimeRW}.

%%%%%%%%%%%%%%%%%%%%%%%%%%%%%%%%%%%%%%%%%%%%%%%%

\section{ Notations for the basic properties of the diffusion process}

\label{sec_nota}

The diffusive trajectory $X(t)$ is generated by the
Ito Stochastic Differential Equations involving 
the Ito force $F_I(x)$
and the diffusion coefficient $D(x)$ 
\begin{eqnarray}
dX(t) =  F_I( X (t) ) \ dt + \sqrt{ 2 D ( X (t) ) } \ dB(t)
 \label{Ito}
\end{eqnarray}
while the Brownian increments $dB(t) =B(t+dt)-B(t)$ satisfies
 \begin{eqnarray}
{\mathbb E} [dB(t) ] && =0 
\nonumber \\
{\mathbb E} [dB^2(t)] && = dt 
\label{Brownian}
\end{eqnarray}

\subsection{ Ito formula for the dynamics of an arbitrary observable $O(X(t))$ }

The Ito SDE of Eq. \ref{Ito} 
and the Taylor expansion of $O(x+\epsilon) $ up to second order in $\epsilon$
\begin{eqnarray}
 O(x+\epsilon )  &&  =O(x) 
 + \epsilon O'(x  ) 
  +\frac{\epsilon^2}{2} O'' (x  )+...
 \label{TaylorO}
\end{eqnarray}
leads to the Ito SDE for the observable $O(X(t) )$ involving all the terms up to order $dt$
\begin{eqnarray}
d O(X(t) )  &&  \equiv  
O(X(t) + d X(t)) - O(X(t) )
%  \nonumber \\  && 
  =   dX(t)  O'(X(t)  )   +\frac{[dX(t)]^2}{2} O'' (X(t)  ) +...
  \nonumber \\
  && =  
  \left[ F_I( X (t) ) \ dt + \sqrt{ 2 D ( X (t) ) } \ dB(t)\right]   O'(X(t)  )   
  +\frac{\left[ F_I( X (t) ) \ dt + \sqrt{ 2 D ( X (t) ) } \ dB(t)\right] ^2}{2} O'' (X(t)  ) 
 +...
    \nonumber \\
  && = \left[ F_I( X (t) )   O'(X(t)  ) +   D ( X (t) ) O'' (X(t)  )  \right] dt 
  +    \sqrt{ 2 D ( X (t) ) } O'(X(t)  ) \ dB(t)      
\nonumber \\
  &&  \equiv  F^{[O]}_I (X(t)) dt +       \sqrt{ 2 D ( X (t) ) } O'(X(t)  ) \ dB(t)   
  \label{ItoFormula}
\end{eqnarray}
where the Ito force $F^{[O]}_I (X(t)) $ is obtained from the observable $O(x)$ 
\begin{eqnarray}
F^{[O]}_I(x) = F_I( x )   O'(x  ) +   D ( x ) O'' (x  )    \equiv {\cal F}^{\dagger} O (x  )
\label{ItoFOx}
\end{eqnarray}
via the application of the second-order differential generator
\begin{eqnarray}
  {\cal F}^{\dagger} =    F_I(x) \frac{\partial}{\partial x}  + D(x)  \frac{\partial^2}{\partial x^2}
\label{adjointIto}
\end{eqnarray}
that governs the dynamics of the averaged values ${\mathbb E} [ O(X(t) ) ] $ of observables 
\begin{eqnarray}
\partial_t {\mathbb E} [ O(X(t) ) ]= {\mathbb E} [ F^{[O]}_I (X(t)) ]  =  {\mathbb E} [ ({\cal F}^{\dagger} O ) (X(t)) ] 
=  {\mathbb E} [ F_I( X (t) )   O'(X(t)  ) +   D ( X (t) ) O''(X(t)  ) ]  
  \label{ItoFormulaAv}
\end{eqnarray}
and that is the adjoint of the Fokker-Planck generator ${\cal F}$ as recalled in the next subsection.

%%%%%%%%%%%%%%%%%%%%%%%%%%%%%

\subsection{ Fokker-Planck generator ${\cal F}$ governing the dynamics of the probability $P_t(x)$ to be at position $x$ at time $t$ }

The probability $P_t(x)$ to be at position $x$ at time $t$
\begin{eqnarray}
P_t(x)= {\mathbb E} [ \delta(X(t) -x) ]
\label{Ptxdef}
\end{eqnarray}
follows the dynamics of Eq. \ref{ItoFormulaAv}
for the special case $O(X(t) )= \delta(X(t) -x)$
\begin{eqnarray}
\partial_t  P_t(x) && = \partial_t {\mathbb E} [ \delta(X(t) -x) ]
=  {\mathbb E} [ F_I( X (t) )   \delta'(X(t) -x ) +   D ( X (t) ) \delta''(X(t) -x ) ]   
\nonumber \\
&& = - \frac{\partial}{\partial x}   {\mathbb E} [ F_I( X (t) )   \delta(X(t) -x ) ]
+\frac{\partial^2}{\partial x^2}   {\mathbb E} [    D ( X (t) ) \delta(X(t) -x ) ] 
\nonumber \\
&& = - \frac{\partial}{\partial x}   \left[ F_I( x ) P_t(x) \right]
+\frac{\partial^2}{\partial x^2}   \left[ D( x ) P_t(x) \right] \equiv   {\cal F}P_t(x)
  \label{ItoFormulaAvdelta}
\end{eqnarray}
This Fokker-Planck dynamics for $P_t(x)$ governed by the
Fokker-Planck generator : 
\begin{eqnarray}
  {\cal F} =  - \frac{\partial}{\partial x} \left( F_I(x)   - \frac{\partial}{\partial x} D(x)  \right)
\label{FPgenerator}
\end{eqnarray}
can be interpreted as the continuity equation
\begin{eqnarray}
 \partial_t P_t( x )    =  - \frac{\partial}{\partial x} J_t(x)
\label{continuity}
\end{eqnarray}
involving the divergence of the current 
\begin{eqnarray}
 J_t( x )  \equiv F_I( x ) P_t(x) -\frac{\partial}{\partial x}   \left[ D( x ) P_t(x) \right]
\label{fokkerplanckcurrent}
\end{eqnarray}

When it exists, the equilibrium steady state $P_{eq}(x)$ associated to the vanishing steady current
\begin{eqnarray}
0= J_{eq}( x )  \equiv F_I( x ) P_{eq}(x) -\frac{\partial}{\partial x}   \left[ D( x ) P_{eq}(x) \right]
\label{eqcurrent}
\end{eqnarray}
can be written in the Boltzmann form involving some normalization Z and some potential $U(x)$
\begin{eqnarray}
 P_{eq}(x) = \frac{ e^{- U(x)} }{Z}  = \frac{ e^{ -U_I(x)} }{D(x) Z}
\label{Peq}
\end{eqnarray}
while it is convenient to introduce also the other potential $U_I(x)$ even if it is directly related to $U(x)$ via
\begin{eqnarray}
U_I(x)  \equiv U(x)- \ln D(x)
\label{UUI}
\end{eqnarray}
since its derivative is simply given by the ratio of the Ito force $F_I(x)$ and the diffusion coefficient $D(x)$
\begin{eqnarray}
U_I'(x) && \equiv - \frac{F_I(x)}{D(x) }
\nonumber \\
 U'(x) && \equiv \frac{D'(x)-F_I(x)}{D(x) }
\label{U'}
\end{eqnarray}
It is actually useful to introduce these potentials $U(x)$ and $U_I(x)$ even when $P_{eq}(x) $ is not normalizable
in order to rewrite the Fokker-Planck generator $  {\cal F} $ and its adjoint $ {\cal F}^{\dagger}   $
in fully-factorized forms as recalled in the next subsection.

%%%%%%%%%%%%%%%%%%%%%%%%%%%%%%%%%%%%%%%

\subsection{ Fully-factorized forms of the Fokker-Planck generator $  {\cal F} $ and its adjoint $ {\cal F}^{\dagger}   $
in terms of the potential $U(x)$ and the diffusion coefficient $D(x)$ }

The two potentials $U(x)$and $U_I(x)$ introduced above
 are useful to rewrite the Fokker-Planck generator $ {\cal F} $ of Eq. \ref{FPgenerator}
in the fully factorized forms
\begin{eqnarray}
 {\cal F}    && =\frac{\partial}{\partial x}  D(x) e^{- U(x) }  \frac{\partial}{\partial x}  e^{ U(x) }  
 \nonumber \\
 && =\frac{\partial}{\partial x}   e^{- U_I(x) }  \frac{\partial}{\partial x}  e^{ U(x) }  
\label{fokkerplanckgeneratorfactorizedU}
\end{eqnarray}
that make obvious that the two independent solutions annihilated by
Fokker-Planck generator ${\cal F}$ are
\begin{eqnarray}
0 && =  {\cal F}P_{eq}(x)   \text{ with $P_{eq}(x)= e^{ - U(x) }  $ associated to $J_{eq}=0$ }
 \nonumber \\
0&&= {\cal F} P_{neq}(x)
\text{ with} \ P_{neq}(x)= e^{- U(x)} \int_0^x dy \frac{ e^{U(y) }  } { D(y) }  =  e^{- U(x)} \int_0^x dy \ e^{U_I(y) }
\text{ associated to } J_{neq}(x)= -1
\label{fokkerplanckannihil}
\end{eqnarray}

The corresponding fully factorized form of the adjoint
\begin{eqnarray}
  {\cal F}^{\dagger}  && = e^{ U(x) }   \frac{\partial}{\partial x} D(x) e^{- U(x) }   \frac{\partial}{\partial x}
  \nonumber \\
 && = e^{ U(x) }   \frac{\partial}{\partial x}  e^{- U_I(x) }   \frac{\partial}{\partial x}
  \label{FPgeneratorprodadjointfactorized}
\end{eqnarray}
 can be used to analyze the observables $O(x)$
whose averaged-values ${\mathbb E} [ O(X(t) ) ] $ are conserved by the dynamics 
of Eq. \ref{ItoFormulaAv} : they are annihilated
by the adjoint ${\cal F}^{\dagger} $ of Eq. \ref{FPgeneratorprodadjointfactorized}
\begin{eqnarray}
0= {\cal F}^{\dagger} O (x) 
=e^{ U(x) }   \frac{d}{d x} \left[  e^{- U_I(x) }  \frac{d O(x)}{d x} \right]
\label{adjointzero}
\end{eqnarray}
so that the two independent solutions are the constant unity $O_1(x)=1$ 
associated to the conservation of the total probability, and the so-called scale-function $O_2(x)=s(x)$
 in the mathematical literature \cite{BookKarlin,BookYor,HandBook,BookBorodin}.
\begin{eqnarray}
s(x)  \equiv  \int^x dy e^{U_I(y) } = \int^x dy \frac{ e^{U(y) }  } { D(y) } 
\ \ \ \text{ with its positive derivative} \ \ s'(x)= e^{U_I(x) }=\frac{ e^{U(x) }  } { D(x) }
\label{Sx}
\end{eqnarray}

To make the full link with the mathematical literature \cite{BookKarlin,BookYor,HandBook,BookBorodin}, 
let us mention that the function $e^{-U(x)}$ introduced above
is called the 'speed measure density' $m'(x)$ 
\begin{eqnarray}
 e^{- U(x)} = m'(x) 
\label{rhoeqmx}
\end{eqnarray}
so that the adjoint operator of Eq. \ref{FPgeneratorprodadjointfactorized} reads in terms of $m'(x)$ of Eq. \ref{rhoeqmx}
and $s'(x)$ of Eq. \ref{Sx}
\begin{eqnarray}
  {\cal F}^{\dagger} &&     = e^{ U(x) }   \frac{\partial}{\partial x} e^{- U_I(x) } \frac{\partial}{\partial x}
   \nonumber \\
   && = \left( \frac{ 1}{m'(x) }  \frac{\partial}{\partial x} \right) \left( \frac{ 1}{s'(x) }  \frac{\partial}{\partial x} \right)
   =  \frac{d}{d m(x) } \frac{d}{d s(x) } 
  \label{FPgeneratorprodadjointmath}
\end{eqnarray}

Let us now recall how the scale function $s(x)$
determines the exit probabilities out of an interval $[x_L,x_R]$, that will play a major role in all the further sections.

%%%%%%%%%%%%%%%%%%%%%%%%%%%%%%%%%%%%%%%%%

\subsection{ Exit probabilities $E^{\pm}_{[x_L,x_R]}(x)$
 out of the interval $[x_L,x_R] $ as a function of the initial condition $x$}

When the diffusion process starts at $x$, the probability $E^+_{[x_L,x_R]}(x)$ to reach the right boundary $x_R$
before the left boundary $x_L$, 
and the complementary
probability $E^-_{[x_L,x_R]}(x)= 1- E^+_{[x_L,x_R]}(x)$ to reach the left boundary $x_L$ before the right boundary $x_R$,
 can be written in terms of the scale function $s(x)$
\begin{eqnarray}
E^+_{[x_L,x_R]}(x) && = \frac{ s(x)-s(x_L) }{ s(x_R)-s(x_L) } 
\nonumber \\
E^-_{[x_L,x_R]}(x)&& = \frac{ s(x_R)-s(x) }{ s(x_R)-s(x_L) } 
  \label{ExitRx}
\end{eqnarray}
Indeed, as explained in textbooks \cite{gardiner,vankampen,risken},
the probability $E^+_{[x_L,x_R]}(x)$ satisfies Eq. \ref{adjointzero} concerning conserved observables $O(x)$
 \begin{eqnarray}
0= {\cal F}^{\dagger} E^+_{[x_L,x_R]}(x)
\label{adjointzeroExit}
\end{eqnarray}
and can be thus written as a linear combination of the two independent solutions $O_1(x)=1$ and $O_2(x)=s(x)$ of Eq. \ref{adjointzero}
where the two constants are determined by the two boundary conditions at $x=x_R$ and $x=x_L$
 \begin{eqnarray}
 E^+_{[x_L,x_R]}(x=x_R) && =1
 \nonumber \\
E^+_{[x_L,x_R]}(x=x_L) && =0
\label{BCExit}
\end{eqnarray}
so that one obtains Eq. \ref{ExitRx}.

%%%%%%%%%%%%%%%%%%%%%%%%%%%%%%%%%%%%%%%%%%

\subsection{ Classifications of the boundary behaviors of the scale function $s(x \to 0^+)$ for diffusions on $x \in ]0,+\infty[$}

For a diffusion defined on the half-line $x \in ]0,+\infty[$,
the probability to reach the boundary $x_L=0$ before $x_R$ when starting at $x \in ]0,x_R[$ given by Eq. \ref{ExitRx}
depends on the behavior of the scale function $s(x_L \to 0^+)$ near the origin as follows
\begin{eqnarray}
E^-_{[0,x_R]}(x) = \frac{ s(x_R)-s(x) }{ s(x_R)-s(0^+) } 
= \begin{cases} 
0 \ \ & \text{if } \ s(0^+)=-\infty \ \ \ \text{ The boundary $x=0$ is never reached (Never Reached) } \\
>0 & \text{if } \ s(0^+) > -\infty \ \ \ \text{Cases (I) and (R) depending on Eq. \ref{MFTPCriterion}}
\end{cases}
  \label{ExitRx0}
\end{eqnarray}
 
When $s(0^+) $ is finite, the remaining question is whether the boundary $x_L=0$ can be reached in finite-time or not,
using for criterion the Mean-First-Passage-Time $\tau^-_{[0,x]}(x) $ at $x_L=0$ when starting at $x$ in the presence of a reflecting boundary at $x$
\begin{eqnarray}
\tau^-_{[0,x]}(x) 
&& \equiv \int_0^{x}d z \frac{e^{U(z)}}{D(z)}\int_z^{x}d y e^{-U(y)}
= \int_0^{x}d y e^{-U(y)}\int_0^{y}d z \frac{e^{U(z)}}{D(z)}
\nonumber \\
&& =\int_0^{x}d y m'(y) \left[ s(y)-s(0) \right] = \int_0^{x}d y \frac{ s(y)-s(0) }{ D(y) s'(y) }
  \label{MFTP}
\end{eqnarray}
also known as 'Feller integral' in the mathematical literature. Its convergence or divergence
does not depend on the precise value of $x$ 
and one obtains the following discussion
\begin{eqnarray}
\tau^-_{[0,x]}(x)  
= \begin{cases} 
+ \infty \ \ & \text{the time needed to reach the boundary $0$ is infinite time (Infinite-time) } \\
<+ \infty  & \text{ the particle hits the boundary $0$ in finite time and is reflected (Reflected)} 
\end{cases}
  \label{MFTPCriterion}
\end{eqnarray}
Simple examples of the three possible behaviors (N),(I),(R)
will be discussed in section \ref{sec_HalfLineExamples}.

%%%%%%%%%%%%%%%%%%%%%%%%%%%%%%%%%%%%%%%%%%

\section{ Analysis of renormalized trajectories for diffusions on $x \in ]-\infty,+\infty[$}

\label{sec_WholeLine}

In this section, we consider only the positions of maxima and minima 
but not the durations between them that are discussed in Appendices.

\subsection{ Direct computation via probability analysis }

\subsubsection{ Probability $P_{\Gamma}^+(x \vert x_0) $ that an ascending bond 
at scale $\Gamma$ starting at $x_0$ ends at the position $x$ with $x \in [x_0+\Gamma,+\infty[$} 

The probability to reach $x$ before hitting $(x-\Gamma)$ when starting at $(x-dx)$
given by the exit formula of Eq. \ref{ExitRx}
\begin{eqnarray}
E^+_{[x_L=x-\Gamma,x_R=x]}(x-dx) && = \frac{ s(x-dx)-s(x-\Gamma) }{ s(x)-s(x-\Gamma) } 
= 1 - dx  \frac{ s'(x) }{ s(x)-s(x-\Gamma) } 
\equiv 1 - dx R^+_{\Gamma}(x)
  \label{ExitRxplus}
\end{eqnarray}
involves the function
\begin{eqnarray}
R^+_{\Gamma}(x) \equiv  \frac{ s'(x) }{ s(x)-s(x-\Gamma) } 
=  \frac{  e^{U_I(x) }   }{ \int_{x-\Gamma}^x dy e^{U_I(y) } } 
= \frac{ \frac{ e^{U(x) }  } { D(x) } }{ \int_{x-\Gamma}^x dy \frac{ e^{U(y) }  } { D(y) } } 
  \label{Rxplus}
\end{eqnarray}
that can be interpreted as the rate at which the running maximum at $x$ is stopped by a drawdown $\Gamma$.

As a consequence, the probability that an ascending bond
 reaches at least $x$ without any drawdown $\Gamma$
has the form of a survival probability involving the rate $R^+_{\Gamma}(z) $
\begin{eqnarray}
C_{\Gamma}^+(x \vert x_0) \equiv \int_x^{+\infty} dy P_{\Gamma}^+(y \vert x_0) =
 e^{ - \int_{x_0+\Gamma}^x dz R_{\Gamma}^+(z)  } \ \ \ \ \text{ for } \ \ x \in [x_0+\Gamma,+\infty[
  \label{Cplus}
\end{eqnarray}
while the probability distribution $P_{\Gamma}^+(x \vert x_0) $ itself is obtained via derivation 
\begin{eqnarray}
P_{\Gamma}^+(x \vert x_0) = - \partial_x C_{\Gamma}^+(x \vert x_0) 
= R_{\Gamma}^+(x) e^{ - \int_{x_0+\Gamma}^x dz R_{\Gamma}^+(z)  } \ \ \ \ \text{ for } \ \ x \in [x_0+\Gamma,+\infty[
  \label{Pplus}
\end{eqnarray}

In particular, the probability of an ascending bond that is about to be decimated at scale $\Gamma$ reduces to
\begin{eqnarray}
P_{\Gamma}^+(x=x_0+\Gamma \vert x_0) 
= R_{\Gamma}^+(x_0+\Gamma) = \frac{ s'(x_0+\Gamma) }{ s(x_0+\Gamma)-s(x_0) } 
= \partial_{\Gamma} \ln \left[ s(x_0+\Gamma)-s(x_0) \right]
  \label{Pplusdecim}
\end{eqnarray}

%%%%%%%%%%%%%%%%%%%%%%%%%%%%

\subsubsection{ Probability $P_{\Gamma}^-(x \vert x_0) $ that a descending bond 
at scale $\Gamma$ starting at $x_0$ ends at the position $x$ with $x \in ]-\infty,x_0-\Gamma]$} 

The probability to reach $x$ before hitting $(x+\Gamma)$ when starting at $(x+dx)$
given by the exit formula of Eq. \ref{ExitRx}
\begin{eqnarray}
E^-_{[x_L=x,x_R=x+\Gamma]}(x+dx) && = \frac{ s(x+\Gamma)-s(x+dx) }{ s(x+\Gamma)-s(x) } 
= 1 - dx  \frac{ s'(x) }{ s(x+\Gamma)-s(x) } 
\equiv 1 - dx R^-_{\Gamma}(x)
  \label{ExitRxmoins}
\end{eqnarray}
involves the function
\begin{eqnarray}
R^-_{\Gamma}(x) \equiv  \frac{ s'(x) }{ s(x+\Gamma)-s(x) } 
=  \frac{ e^{U_I(x) }   }{ \int_x^{x+\Gamma} dy e^{U_I(y) } } 
=  \frac{ \frac{ e^{U(x) }  } { D(x) } }{ \int_x^{x+\Gamma} dy \frac{ e^{U(y) }  } { D(y) } } 
  \label{Rxmoins}
\end{eqnarray}
that can be interpreted as the rate at which the running minimum at $x$ is stopped by a drawup $\Gamma$.

As a consequence, the probability that a descending bond
reaches at least $x$ without any drawup $\Gamma$
has the form of a survival probability involving the rate $R^-_{\Gamma}(z) $
\begin{eqnarray}
C_{\Gamma}^-(x \vert x_0) \equiv \int_{-\infty}^x dy P_{\Gamma}^-(y \vert x_0) =
 e^{ - \int_x^{x_0-\Gamma} dz R_{\Gamma}^-(z)  } \ \ \ \ \text{ for } \ \ x \in ]-\infty,x_0-\Gamma]
  \label{Cmoins}
\end{eqnarray}
while the probability distribution $P_{\Gamma}^-(x \vert x_0) $ itself is obtained via derivation 
\begin{eqnarray}
P_{\Gamma}^-(x \vert x_0) =  \partial_x C_{\Gamma}^-(x\vert x_0) 
= R_{\Gamma}^-(x) e^{ - \int_x^{x_0-\Gamma} dz R_{\Gamma}^-(z)  } \ \ \ \ \text{ for } \ \  x \in ]-\infty,x_0-\Gamma]
  \label{Pmoins}
\end{eqnarray}

In particular, the probability of a descending bond that is about to be decimated at scale $\Gamma$ reduces to
\begin{eqnarray}
P_{\Gamma}^-(x=x_0-\Gamma \vert x_0) 
= R_{\Gamma}^-(x_0-\Gamma) =  \frac{ s'(x_0-\Gamma) }{ s(x_0)-s(x_0-\Gamma) }
=  \partial_{\Gamma} \ln \left[ s(x_0)-s(x_0-\Gamma) \right]
  \label{Pmoinsdecim}
\end{eqnarray}

%%%%%%%%%%%%%%%%%%%%%%%%%%%%%%

\subsubsection{ Densities $n_{\Gamma}^{\pm}(x)$ of maxima and minima 
should be stable by the Markov chain generated by the conditional probabilities $P_{\Gamma}^{\pm}(x \vert x_0) $ }

At scale $\Gamma$, the density $n_{\Gamma}^{+}(x)$ of maxima and the density $n_{\Gamma}^{-}(x)$ of minima 
should be stationary under the Markov chain generated by the conditional probabilities $P_{\Gamma}^{\pm}(x \vert x_0) $ of Eqs \ref{Pplus}
and \ref{Pmoins}
that play the role of transition matrices
\begin{eqnarray}
n_{\Gamma}^{+}(x) && = \int_{-\infty}^{x-\Gamma} dx_0 P_{\Gamma}^+(x \vert x_0) n_{\Gamma}^-(x_0)  
= R_{\Gamma}^+(x)  \int_{-\infty}^{x-\Gamma} dx_0 e^{ - \int_{x_0+\Gamma}^x dz R_{\Gamma}^+(z)  } n_{\Gamma}^-(x_0)  
\equiv R_{\Gamma}^+(x) \Phi_{\Gamma}^+(x)
\nonumber \\
n_{\Gamma}^{-}(x)&& = \int_{x+\Gamma}^{+\infty} dx_0 P_{\Gamma}^-(x \vert x_0) n_{\Gamma}^+(x_0)
= R_{\Gamma}^-(x) \int_{x+\Gamma}^{+\infty} dx_0 e^{ - \int_x^{x_0-\Gamma} dz R_{\Gamma}^-(z)  }n_{\Gamma}^+(x_0)
\equiv R_{\Gamma}^-(x) \Phi_{\Gamma}^-(x)
  \label{MarkovChain}
\end{eqnarray}
where we have introduced the two auxiliary functions
\begin{eqnarray}
\Phi_{\Gamma}^{+}(x) && \equiv  \int_{-\infty}^{x-\Gamma} dx_0 e^{ - \int_{x_0+\Gamma}^x dz R_{\Gamma}^+(z)  } n_{\Gamma}^-(x_0)  
\nonumber \\
\Phi_{\Gamma}^{-}(x) && \equiv \int_{x+\Gamma}^{+\infty} dx_0 e^{ - \int_x^{x_0-\Gamma} dz R_{\Gamma}^-(z)  }n_{\Gamma}^+(x_0)
  \label{Phi}
\end{eqnarray}
Their partial derivatives with respect to $x$ 
\begin{eqnarray}
\partial_x \Phi_{\Gamma}^{+}(x) && =  n_{\Gamma}^-(x-\Gamma) - R_{\Gamma}^+(x) \Phi_{\Gamma}^{+}(x)  
=n_{\Gamma}^-(x-\Gamma) - n_{\Gamma}^+(x)
\nonumber \\
\partial_x\Phi_{\Gamma}^{-}(x) && = - n_{\Gamma}^+(x+\Gamma)+ R_{\Gamma}^-(x) \Phi_{\Gamma}^{-}(x)
= - n_{\Gamma}^+(x+\Gamma)+ n_{\Gamma}^-(x)
  \label{Phideri}
\end{eqnarray}
shows that the difference $ \left[ \Phi_{\Gamma}^{-}(x) - \Phi_{\Gamma}^{+}(x+\Gamma)\right] $ does not depend on $x$
\begin{eqnarray}
\partial_x \left[ \Phi_{\Gamma}^{-}(x) - \Phi_{\Gamma}^{+}(x+\Gamma)\right] 
= 0
  \label{vanish}
\end{eqnarray}
and reduces to a function $ \kappa_{\Gamma} $ of $\Gamma$
\begin{eqnarray}
\Phi_{\Gamma}^{+}(x+\Gamma) - \Phi_{\Gamma}^{-}(x)   = \kappa_{\Gamma}
  \label{vanishinteg}
\end{eqnarray}

Putting everything together, one obtains that the first equation of Eq. \ref{Phideri}
becomes a closed differential equation for $\Phi_{\Gamma}^{+}(x) $
\begin{eqnarray}
\partial_x \Phi_{\Gamma}^{+}(x) && =  R_{\Gamma}^-(x-\Gamma) \Phi_{\Gamma}^{-}(x-\Gamma)
 - R_{\Gamma}^+(x) \Phi_{\Gamma}^{+}(x)  
\nonumber \\
 && =R_{\Gamma}^-(x-\Gamma) \left[\Phi_{\Gamma}^{+}(x) - \kappa_{\Gamma} \right]
 - R_{\Gamma}^+(x) \Phi_{\Gamma}^{+}(x) 
  \label{Phideriplus}
\end{eqnarray}
where one can plug the explicit expressions of $R_{\Gamma}^{\pm}(x) $ of Eqs \ref{Rxplus}
and \ref{Rxmoins}
\begin{eqnarray}
\partial_x \Phi_{\Gamma}^{+}(x)    
 = \frac{ s'(x-\Gamma) }{ s(x)-s(x-\Gamma) }  \left[\Phi_{\Gamma}^{+}(x) - \kappa_{\Gamma} \right]
 - \frac{ s'(x) }{ s(x)-s(x-\Gamma) }  \Phi_{\Gamma}^{+}(x) 
  \label{Phideriplusexpli}
\end{eqnarray}
to obtain
\begin{eqnarray}
\partial_x \left( \Phi_{\Gamma}^{+}(x)  \left[ s(x)-s(x-\Gamma)\right]  \right)
&& =  \left( \partial_x\Phi_{\Gamma}^{+}(x) \right) \left[ s(x)-s(x-\Gamma)\right]  
+ \Phi_{\Gamma}^{+}(x)  \left[ s'(x)-s'(x-\Gamma)\right]
\nonumber \\
 && =  s'(x-\Gamma)   \left[\Phi_{\Gamma}^{+}(x) - \kappa_{\Gamma} \right]
 -  s'(x)   \Phi_{\Gamma}^{+}(x) + \Phi_{\Gamma}^{+}(x)  \left[ s'(x)-s'(x-\Gamma)\right] 
 \nonumber \\
 &&=  -\kappa_{\Gamma}s'(x-\Gamma) 
  \label{Phideriplusexplicombi}
\end{eqnarray}
The integration introduces another function $c_{\Gamma} $ of $\Gamma$ 
needed to write $\Phi_{\Gamma}^{+}(x)$ 
\begin{eqnarray}
 \Phi_{\Gamma}^{+}(x)   &&= \frac{c_{\Gamma} - \kappa_{\Gamma}s(x-\Gamma) }{s(x)-s(x-\Gamma)}
  \label{PhiplusSol}
\end{eqnarray}
while $ \Phi_{\Gamma}^{-}(x)$ of Eq. \ref{vanishinteg}
becomes
\begin{eqnarray}
 \Phi_{\Gamma}^{-}(x) = \Phi_{\Gamma}^{+}(x+\Gamma) - \kappa_{\Gamma}
 = \frac{c_{\Gamma} - \kappa_{\Gamma}s(x) - \kappa_{\Gamma} \left[s(x+\Gamma)-s(x) \right] }{s(x+\Gamma)-s(x)}
  =  \frac{c_{\Gamma} - \kappa_{\Gamma}s(x+\Gamma) }{s(x+\Gamma)-s(x)}
  \label{PhimoinsSol}
\end{eqnarray}

So using $R_{\Gamma}^{\pm}(x) $ of Eqs \ref{Rxplus}
and \ref{Rxmoins}, one obtains that the densities $n_{\Gamma}^{\pm}(x)$ of maxima and minima
of Eqs \ref{MarkovChain}
read in terms of the scale function $s(.)$ 
\begin{eqnarray}
n_{\Gamma}^{+}(x) && =  R_{\Gamma}^+(x) \Phi_{\Gamma}^+(x) 
=  \frac{ s'(x) \left[ c_{\Gamma} - \kappa_{\Gamma}s(x-\Gamma)\right] }{[s(x)-s(x-\Gamma)]^2}
\nonumber \\
n_{\Gamma}^{-}(x)&& = R_{\Gamma}^-(x) \Phi_{\Gamma}^-(x)
=   \frac{ s'(x) [c_{\Gamma} - \kappa_{\Gamma}s(x+\Gamma)] }{[s(x+\Gamma)-s(x)]^2}
  \label{ngpmsol}
\end{eqnarray}

The fact that the family of possible steady densities $ n_{\Gamma}^{\pm}(x)$ of the Markov chain of Eq \ref{MarkovChain}
involves two parameters $(c_{\Gamma},\kappa_{\Gamma})$
is reminiscent of the two independent steady solutions $P_{eq}$ and $P_{neq}$ of Eq. \ref{fokkerplanckannihil}
of the initial model, and the parameter $\kappa_{\Gamma}$ has indeed the interpretation of a steady current,
as explained in the next subsection.

\subsubsection{ Physical meaning of the parameter $\kappa_{\Gamma}$ as a steady current 
for the Markov chain at scale $\Gamma$}

\label{sub_kappacurrent}

To have an interpretation of the parameter $\kappa_{\Gamma}$ introduced in Eq. \ref{vanishinteg},
it is useful to see that $\Phi_{\Gamma}^+(x) $ of Eq. \ref{Phi}
represents the probability that 
$x$ belongs to an ascending bond starting at the minimum $x_0 \in ]-\infty,x-\Gamma[$
drawn with the steady density $n_{\Gamma}^-(x_0)$ of minima, so it is convenient to add the contribution 
$x_0 \in ]x-\Gamma,x[$ to obtain the steady flow upward through $x$
\begin{eqnarray}
J^+_{\Gamma}(x) = \Phi_{\Gamma}^{+}(x) +  \int_{x-\Gamma}^x dx_0 n_{\Gamma}^-(x_0) 
  \label{JPhip}
\end{eqnarray}
Similarly, $\Phi_{\Gamma}^-(x) $ of Eq. \ref{Phi} represents the probability that 
$x$ belongs to a descending bond starting at the maximum $x_0 \in ]x+\Gamma,+\infty[$
drawn with the steady density $n_{\Gamma}^+(x_0)$, so it is convenient to add the contribution 
$x_0 \in ]x,x+\Gamma[$ to obtain the steady flow downward through $x$
\begin{eqnarray}
J^-_{\Gamma}(x) =\Phi_{\Gamma}^{-}(x) +\int_x^{x+\Gamma} dx_0 n_{\Gamma}^+(x_0)
  \label{JPhim}
\end{eqnarray}
The two spatial derivatives satisfy local equations using Eq. \ref{Phideri}
\begin{eqnarray}
\partial_x J^+_{\Gamma}(x) && = \partial_x \Phi_{\Gamma}^{+}(x) +   n_{\Gamma}^-(x) -n_{\Gamma}^-(x-\Gamma)
= n_{\Gamma}^-(x) - n_{\Gamma}^+(x)
\nonumber \\
\partial_xJ^-_{\Gamma}(x) && =\partial_x\Phi_{\Gamma}^{-}(x) + n_{\Gamma}^+(x+\Gamma)- n_{\Gamma}^+(x)
=   n_{\Gamma}^-(x)- n_{\Gamma}^+(x)
  \label{JPhideri}
\end{eqnarray}
and their difference vanishes
\begin{eqnarray}
\partial_x  J^+_{\Gamma}(x) - \partial_x J^-_{\Gamma}(x)  =0 \ \ \ \ \ \to J^+_{\Gamma}(x) -J^-_{\Gamma}(x)=J_{\Gamma} 
  \label{JPhidiffx}
\end{eqnarray}

To obtain the corresponding uniform value $J_{\Gamma} $, one can use
that the integrals appearing in Eqs \ref{JPhip}
and \ref{JPhim}
also appear in the integration of the first Eq. \ref{Phideri}
\begin{eqnarray}
\Phi_{\Gamma}^{+}(x+\Gamma)-\Phi_{\Gamma}^{+}(x)
&& =\int_x^{x+\Gamma} dy \partial_y \Phi_{\Gamma}^{+}(y)   
=\int_x^{x+\Gamma} dy \left[ n_{\Gamma}^-(y-\Gamma) - n_{\Gamma}^+(y) \right]
\nonumber \\
&&= \int_{x-\Gamma}^x dx_0  n_{\Gamma}^-(x_0) - \int_x^{x+\Gamma} dx_0 n_{\Gamma}^+(x_0) 
  \label{Phiderip}
\end{eqnarray}
so that one obtains that the difference of the two flows of Eqs \ref{JPhip}
and \ref{JPhim}
\begin{eqnarray}
J^+_{\Gamma}(x) - J^-_{\Gamma}(x) && = \Phi_{\Gamma}^{+}(x) - \Phi_{\Gamma}^{-}(x)
+  \int_{x-\Gamma}^x dx_0 n_{\Gamma}^-(x_0) -\int_x^{x+\Gamma} dx_0 n_{\Gamma}^+(x_0)
=  \Phi_{\Gamma}^{+}(x) - \Phi_{\Gamma}^{-}(x)
+\Phi_{\Gamma}^{+}(x+\Gamma)-\Phi_{\Gamma}^{+}(x)
\nonumber \\
&&= \Phi_{\Gamma}^{+}(x+\Gamma) -\Phi_{\Gamma}^{-}(x) =  \kappa_{\Gamma} \equiv J_{\Gamma}
  \label{Jtot}
\end{eqnarray}
So the parameter $\kappa_{\Gamma} =J_{\Gamma}$ is the uniform current 
$J_{\Gamma}=J^+_{\Gamma}(x) -J^-_{\Gamma}(x)$.

%%%%%%%%%%%%%%%%%%%%%%%%%%%%%%%%%%%%%%%%%%%

\subsection{ Analysis via the RG-flows as a function of the renormalization scale $\Gamma$ }

\label{sec_RG}

\subsubsection{ Observables of the renormalized trajectory at scale $\Gamma$ }

In this subsection, we consider three types of observables at scale $\Gamma$:

$\bullet$ The bond density $N_{\Gamma}(x_0\asc x)$ of ascending bonds involving $x \geq x_0+\Gamma$,
and bond density $N_{\Gamma}(x_0\des x)$ of descending bonds involving $x \leq x_0-\Gamma$

$\bullet$ The densities $n_{\Gamma}^-(y)$ and $n_{\Gamma}^+(y)$ of minima and maxima
at scale $\Gamma$, that can be obtained as marginals of the two bond densities $N_{\Gamma}(x_0\asc x) $ and $N_{\Gamma}(x_0\des x) $
\begin{eqnarray}
n_{\Gamma}^-(y) && =\int_{y+\Gamma}^{+\infty} dx N_{\Gamma}(y\asc x) = \int_{y+\Gamma}^{+\infty} dx_0 N_{\Gamma}(x_0\des y)
\nonumber \\
n_{\Gamma}^+(y) && =\int_{-\infty}^{y-\Gamma} dx_0 N_{\Gamma}(x_0\asc y) = \int_{-\infty}^{y-\Gamma} dx N_{\Gamma}(y\des x)
  \label{ngamma}
\end{eqnarray}

$\bullet$ The conditional probabilities of ascending bonds and descending bonds
computed as the ratios
\begin{eqnarray}
P_{\Gamma}(x_0\nearrow x) && = \frac{N_{\Gamma}(x_0\asc x)}{n^-_{\Gamma}(x_0)} = P_{\Gamma}^+(x \vert x_0)
\nonumber \\
P_{\Gamma}(x_0\searrow x) && = \frac{ N_{\Gamma}(x_0\des x)}{n^+_{\Gamma}(x_0)} =P_{\Gamma}^-(x \vert x_0)
  \label{Pgamma}
\end{eqnarray}
that should coincide with the conditional probabilities $P_{\Gamma}^{\pm}(x \vert x_0) $ of Eqs \ref{Pplus}
and \ref{Pmoins}.
To write the RG-flows, it is useful to introduce also the backward conditional probabilities
\begin{eqnarray}
 P_{\Gamma}(x\swarrow x_0)\equiv \frac{N_{\Gamma}(x\asc x_0)}{n^+_{\Gamma}(x_0)}
 =\frac{n^-_{\Gamma}(x)\,P_{\Gamma}(x\nearrow x_0)}{n^+_{\Gamma}(x_0)}
\nonumber \\
P_{\Gamma}(x\nwarrow x_0)\equiv\frac{N_{\Gamma}(x\des x_0)}{n^-_{\Gamma}(x_0)}
=\frac{n^+_{\Gamma}(x)\,P_{\Gamma}(x\searrow x_0)}{n^-_{\Gamma}(x_0)}
  \label{PgammaBack}
\end{eqnarray}

%%%%%%%%%%%%%%%%%%%%%%%%%%%%%%%%%%%%

\subsubsection{ RG-flows as a function of the renormalization scale $\Gamma$ }

$\bullet$ The RG-flow for bond density $N_{\Gamma}(x_0\asc x)$ of ascending bonds 
\begin{eqnarray}
 \partial_{\Gamma} N_{\Gamma}(x_0\asc x) &&=- P_{\Gamma}(x_0+\Gamma\nwarrow x_0)N_{\Gamma}(x_0\asc x)
 -  N_{\Gamma}(x_0\asc x) P_{\Gamma}(x\searrow x-\Gamma)
\nonumber \\
&&+\int_{x_0+\Gamma}^{x}d zN_{\Gamma}(x_0\asc z) P_{\Gamma}(z\searrow z-\Gamma) P_{\Gamma}(z-\Gamma\nearrow x)
  \label{RGNasc}
\end{eqnarray}
contains two loss-terms associated to the decimation of its left-neighbor and of its right-neighbor,
and one gain term corresponding to the merging of three bonds via the decimation of the middle bond.

Similarly, the RG-flow for the bond density $N_{\Gamma}(x_0\des x)$ of descending bonds reads
\begin{eqnarray}
\partial_{\Gamma} N_{\Gamma}(x_0\des x) && =-P_{\Gamma}(x_0-\Gamma\swarrow x_0)N_{\Gamma}(x_0\des x)
- N_{\Gamma}(x_0\des x) P_{\Gamma}(x\nearrow x+\Gamma)
\nonumber \\
&& +\int_{x}^{x_0-\Gamma}d z N_{\Gamma}(x_0\des z) P_{\Gamma}(z\nearrow z+\Gamma)
P_{\Gamma}(z+\Gamma\searrow x)
  \label{RGNdes}
\end{eqnarray}

$\bullet$ The RG-flow for the density $n_{\Gamma}^-(y)$ of minima
can be obtained from its definition of Eq. \ref{ngamma} and the RG-flow of Eq. \ref{RGNasc}
\begin{eqnarray}
\partial_{\Gamma} n_{\Gamma}^-(y) && = - N_{\Gamma}(y\asc y+\Gamma) 
+ \int_{y+\Gamma}^{+\infty} dx \bigg[ \partial_{\Gamma} N_{\Gamma}(y\asc x) \bigg]
\nonumber \\
&& =  - n_{\Gamma}^-(y)  P_{\Gamma}(y\nearrow y+\Gamma)
 - P_{\Gamma}(y+\Gamma\nwarrow y)n_{\Gamma}^-(y) 
  \label{RGnmin}
\end{eqnarray}
and reduces as it should to the two loss-terms associated to the decimation of its right-bond and of its left-bond.
Similarly, the RG-flow for the density $n_{\Gamma}^+(y)$ of maxima reads
\begin{eqnarray}
\partial_{\Gamma} n_{\Gamma}^+(y) &&  =  - n_{\Gamma}^+(y)  P_{\Gamma}(y\searrow y-\Gamma) 
 - P_{\Gamma}(y-\Gamma\swarrow y)n_{\Gamma}^+(y) 
  \label{RGnmax}
\end{eqnarray}

$\bullet$ The RG-flow for the conditional probability of ascending bonds 
can be obtained from its definition of Eq. \ref{Pgamma} and the RG-flow of Eqs \ref{RGNasc} and \ref{RGnmin}
\begin{eqnarray}
\partial_{\Gamma} P_{\Gamma}(x_0\nearrow x) && = \frac{ \partial_{\Gamma} N_{\Gamma}(x_0\asc x)}{n^-_{\Gamma}(x_0)}
- N_{\Gamma}(x_0\asc x)\frac{ \partial_{\Gamma} n^-_{\Gamma}(x_0)}{ [ n^-_{\Gamma}(x_0) ]^2}
\nonumber \\
&& =- P_{\Gamma}(x_0+\Gamma\nwarrow x_0) P_{\Gamma}(x_0\nearrow x)
 -  P_{\Gamma}(x_0\nearrow x) P_{\Gamma}(x\searrow x-\Gamma)
\nonumber \\
&&+\int_{x_0+\Gamma}^{x}d z
P_{\Gamma}(x_0\nearrow z)
P_{\Gamma}(z\searrow z-\Gamma) P_{\Gamma}(z-\Gamma\nearrow x)
\nonumber \\
&& + P_{\Gamma}(x_0\nearrow x)  \bigg[ P_{\Gamma}(x_0\nearrow x_0+\Gamma)
 + P_{\Gamma}(x_0+\Gamma\nwarrow x_0)\bigg]
 \nonumber \\
&& =
   P_{\Gamma}(x_0\nearrow x) \left[ - P_{\Gamma}(x\searrow x-\Gamma)
 +   P_{\Gamma}(x_0\nearrow x_0+\Gamma) \right]
\nonumber \\
&&+\int_{x_0+\Gamma}^{x}d z
P_{\Gamma}(x_0\nearrow z)
P_{\Gamma}(z\searrow z-\Gamma) P_{\Gamma}(z-\Gamma\nearrow x)
  \label{RGPasc}
\end{eqnarray}
Similarly, the RG-flow for the conditional probability of descending bonds reads
\begin{eqnarray}
\partial_{\Gamma} P_{\Gamma}(x_0\searrow x)
&& = P_{\Gamma}(x_0\searrow x) \left[-P_{\Gamma}(x\nearrow x+\Gamma)+ P_{\Gamma}(x_0\searrow x_0-\Gamma)\right]
\nonumber \\
&&+\int_{x}^{x_0-\Gamma} d z P_{\Gamma}(x_0\searrow z) P_{\Gamma}(z\nearrow z+\Gamma) P_{\Gamma}(z+\Gamma\searrow x)
  \label{RGPdes}
\end{eqnarray}

%%%%%%%%%%%%%%%%%%%%%%%%%%%%%%%%%%%%

\subsubsection{ Solutions of the RG-flows for the conditional probabilities $P_{\Gamma}(x_0\nearrow x) $ and $P_{\Gamma}(x_0\searrow x) $ }

Since the conditional probabilities $P_{\Gamma}(x_0\nearrow x) $ and $P_{\Gamma}(x_0\searrow x) $
should coincide with
$P_{\Gamma}^{\pm}(x \vert x_0) $ of Eqs \ref{Pplus}
and \ref{Pmoins}
\begin{eqnarray}
P_{\Gamma}(x_0\nearrow x) &&  = P_{\Gamma}^+(x \vert x_0) 
= R_{\Gamma}^+(x) e^{ - \int_{x_0+\Gamma}^x dz R_{\Gamma}^+(z)  }
\nonumber \\
P_{\Gamma}(x_0\searrow x) &&  =P_{\Gamma}^-(x \vert x_0) 
= R_{\Gamma}^-(x) e^{ - \int_x^{x_0-\Gamma} dz R_{\Gamma}^-(z)  }
  \label{PgammaAnsatz}
\end{eqnarray}
it is interesting to plug these Ansatz involving two functions $R_{\Gamma}^{\pm}(x) $ of a single variable
into the RG-flows of Eqs \ref{RGPasc} and \ref{RGPdes} to see what are the important properties of 
two functions $R_{\Gamma}^{\pm}(x) $ from the RG perspective.
The two derivatives of Eqs \ref{PgammaAnsatz} with respect to $\Gamma$ 
\begin{eqnarray}
\partial_{\Gamma} P_{\Gamma}(x_0\nearrow x) &&   = 
\left[ \partial_{\Gamma} \ln R_{\Gamma}^+(x) 
+   R_{\Gamma}^+(x_0+\Gamma) 
- \int_{x_0+\Gamma}^x dz \partial_{\Gamma} R_{\Gamma}^+(z) 
\right] R_{\Gamma}^+(x)
e^{ - \int_{x_0+\Gamma}^x dz R_{\Gamma}^+(z)  }
\nonumber \\
\partial_{\Gamma} P_{\Gamma}(x_0\searrow x) &&   = 
\left[ \partial_{\Gamma} \ln R_{\Gamma}^-(x) 
+ R_{\Gamma}^-(x_0-\Gamma)
- \int_x^{x_0-\Gamma} dz \partial_{\Gamma} R_{\Gamma}^-(z)
 \right] R_{\Gamma}^-(x)  e^{ - \int_x^{x_0-\Gamma} dz R_{\Gamma}^-(z)  }
  \label{PgammaAnsatzderi}
\end{eqnarray}
and the two convolutions appearing in Eq \ref{RGPasc}
\begin{eqnarray}
&& \int_{x_0+\Gamma}^{x}d z
P_{\Gamma}(x_0\nearrow z)
P_{\Gamma}(z\searrow z-\Gamma) 
P_{\Gamma}(z-\Gamma\nearrow x)
%\nonumber \\&&
= \int_{x_0+\Gamma}^{x}d z
R_{\Gamma}^+(z) e^{ - \int_{x_0+\Gamma}^z dy R_{\Gamma}^+(y)  }
R_{\Gamma}^-( z-\Gamma) 
R_{\Gamma}^+(x) e^{ - \int_{z}^x dy R_{\Gamma}^+(y)  }
\nonumber \\
&&
=R_{\Gamma}^+(x) e^{ - \int_{x_0+\Gamma}^x dy R_{\Gamma}^+(y)  }
\int_{x_0+\Gamma}^{x}d z
R_{\Gamma}^+(z) 
R_{\Gamma}^-( z-\Gamma) = 
P_{\Gamma}(x_0\nearrow x)
\int_{x_0+\Gamma}^{x}d z
R_{\Gamma}^+(z) 
R_{\Gamma}^-( z-\Gamma)
  \label{RGPascconvol}
\end{eqnarray}
and Eq \ref{RGPdes}
\begin{eqnarray}
&& \int_{x}^{x_0-\Gamma} d z P_{\Gamma}(x_0\searrow z) P_{\Gamma}(z\nearrow z+\Gamma) P_{\Gamma}(z+\Gamma\searrow x)
% \nonumber \\ &&
= \int_{x}^{x_0-\Gamma} d z 
R_{\Gamma}^-(z) e^{ - \int_z^{x_0-\Gamma} dy R_{\Gamma}^-(y)  }
R_{\Gamma}^+(z+\Gamma) 
R_{\Gamma}^-(x) e^{ - \int_x^{z} dy R_{\Gamma}^-(y)  }
\nonumber \\ &&
=R_{\Gamma}^-(x) e^{ - \int_x^{x_0-\Gamma} dy R_{\Gamma}^-(y)  }
 \int_{x}^{x_0-\Gamma} d z 
R_{\Gamma}^-(z) 
R_{\Gamma}^+(z+\Gamma) 
=
P_{\Gamma}(x_0\searrow x)
 \int_{x}^{x_0-\Gamma} d z 
R_{\Gamma}^-(z) 
R_{\Gamma}^+(z+\Gamma) 
  \label{RGPdesconvol}
\end{eqnarray}
can be plugged into the RG-flow of Eq. \ref{RGPasc} divided by $P_{\Gamma}(x_0\nearrow x) $
\begin{eqnarray}
0 && = - \partial_{\Gamma} \ln P_{\Gamma}(x_0\nearrow x) 
    - P_{\Gamma}(x\searrow x-\Gamma)
 +   P_{\Gamma}(x_0\nearrow x_0+\Gamma)
\nonumber \\
&&+\frac{1}{P_{\Gamma}(x_0\nearrow x)} \int_{x_0+\Gamma}^{x}d z
P_{\Gamma}(x_0\nearrow z)
P_{\Gamma}(z\searrow z-\Gamma) P_{\Gamma}(z-\Gamma\nearrow x)
\nonumber \\
&& = -  \partial_{\Gamma} \ln R_{\Gamma}^+(x) 
-   R_{\Gamma}^+(x_0+\Gamma) 
+ \int_{x_0+\Gamma}^x dz \partial_{\Gamma} R_{\Gamma}^+(z) 
    - R_{\Gamma}^-( x-\Gamma)
 +   R_{\Gamma}^+( x_0+\Gamma)
\nonumber \\
&&
+\int_{x_0+\Gamma}^{x}d z
R_{\Gamma}^+(z) 
R_{\Gamma}^-( z-\Gamma)
\nonumber \\
&& = -  \partial_{\Gamma} \ln R_{\Gamma}^+(x) 
    - R_{\Gamma}^-( x-\Gamma)
+ \int_{x_0+\Gamma}^x dz \left[ \partial_{\Gamma} R_{\Gamma}^+(z) 
+R_{\Gamma}^+(z) R_{\Gamma}^-( z-\Gamma) \right]
  \label{RGPascSolv}
\end{eqnarray}

and into the RG-flow of Eq. \ref{RGPdes} divided by $P_{\Gamma}(x_0\searrow x) $

\begin{eqnarray}
0 && = - \partial_{\Gamma} \ln P_{\Gamma}(x_0\searrow x)
-P_{\Gamma}(x\nearrow x+\Gamma)+ P_{\Gamma}(x_0\searrow x_0-\Gamma)
\nonumber \\
&&+\frac{1}{P_{\Gamma}(x_0\searrow x)} \int_{x}^{x_0-\Gamma} d z P_{\Gamma}(x_0\searrow z) P_{\Gamma}(z\nearrow z+\Gamma) P_{\Gamma}(z+\Gamma\searrow x)
\nonumber \\
&& =- \partial_{\Gamma} \ln R_{\Gamma}^-(x) 
- R_{\Gamma}^-(x_0-\Gamma)
+ \int_x^{x_0-\Gamma} dz \partial_{\Gamma} R_{\Gamma}^-(z)
-R_{\Gamma}^+( x+\Gamma)
+ R_{\Gamma}^-( x_0-\Gamma)
\nonumber \\
&&+
 \int_{x}^{x_0-\Gamma} d z 
R_{\Gamma}^-(z) 
R_{\Gamma}^+(z+\Gamma) 
\nonumber \\
&& =- \partial_{\Gamma} \ln R_{\Gamma}^-(x) -R_{\Gamma}^+( x+\Gamma)
+ \int_x^{x_0-\Gamma} dz \left[ \partial_{\Gamma} R_{\Gamma}^-(z)
+  R_{\Gamma}^-(z) R_{\Gamma}^+(z+\Gamma) \right]
  \label{RGPdesSolv}
\end{eqnarray}
to obtain that the two equations 
\begin{eqnarray}
  \partial_{\Gamma} \ln R_{\Gamma}^+(x) 
    + R_{\Gamma}^-( x-\Gamma)
&& = \int_{x_0+\Gamma}^x dz R_{\Gamma}^+(z)  \left[  \partial_{\Gamma} \ln R_{\Gamma}^+(z) 
+ R_{\Gamma}^-( z-\Gamma) \right] 
\nonumber \\
\partial_{\Gamma} \ln R_{\Gamma}^-(x) +R_{\Gamma}^+( x+\Gamma)
&& = \int_x^{x_0-\Gamma} dz R_{\Gamma}^-(z) \left[ \partial_{\Gamma} \ln R_{\Gamma}^-(z)
+   R_{\Gamma}^+(z+\Gamma) \right] 
  \label{RGPascdesSolv}
\end{eqnarray}
Since the left handsides do not depend on $x_0$ that appear in the boundaries on the integrals on the right handsides,
these equations can be satisfied only if the two integrands vanish, and then the left handsides automatically vanish,
so that one obtains the two following RG-flows for the two functions $R_{\Gamma}^{\pm}(.)$
\begin{eqnarray}
  \partial_{\Gamma}   \ln R_{\Gamma}^+(x)    && =-  R_{\Gamma}^-( x-\Gamma) 
    \nonumber \\
\partial_{\Gamma}  \ln R_{\Gamma}^-(x) && = - R_{\Gamma}^+( x+\Gamma)
  \label{RGforRpm}
\end{eqnarray}

These RG-flows are satisfied by the explicit forms of Eqs \ref{Rxplus}
and \ref{Rxmoins} that correspond to the log-derivatives
\begin{eqnarray}
 \partial_{\Gamma}   \ln R^+_{\Gamma}(x) = \partial_{\Gamma}   \ln \frac{ s'(x) }{ s(x)-s(x-\Gamma) } 
= - \frac{ s'(x-\Gamma) }{ s(x)-s(x-\Gamma) }  =-  R_{\Gamma}^-( x-\Gamma) 
  \label{Rxplusln}
\end{eqnarray}

\begin{eqnarray}
 \partial_{\Gamma}   \ln R^-_{\Gamma}(x) =  \partial_{\Gamma}   \ln  \frac{ s'(x) }{ s(x+\Gamma)-s(x) } 
= -  \frac{ s'(x+\Gamma) }{ s(x+\Gamma)-s(x) } =- R_{\Gamma}^+( x+\Gamma)
  \label{Rxmoinsln}
\end{eqnarray}

%%%%%%%%%%%%%%%%%%%%%%%%%%%%%%%%%%%%

\subsubsection{ Solutions of the RG-flows for the densities $n_{\Gamma}^{\pm}(x) $ of maxima and minima  }

The RG-flows of Eqs \ref{RGnmin} and \ref{RGnmax} for the densities $n_{\Gamma}^{\pm}(x) $ of maxima and minima
can be rewritten in terms of the forward conditional probabilities alone
via the replacement of the backward conditional probabilities of Eq. \ref{PgammaBack}
\begin{eqnarray}
\partial_{\Gamma} n_{\Gamma}^-(x) &&   =  - n_{\Gamma}^-(x)  P_{\Gamma}(x\nearrow x+\Gamma)
 - n_{\Gamma}^+(x+\Gamma) P_{\Gamma}(x+\Gamma\searrow x)
 \nonumber \\
 \partial_{\Gamma} n_{\Gamma}^+(x) &&    =  - n_{\Gamma}^+(x)  P_{\Gamma}(x\searrow x-\Gamma) 
 - n_{\Gamma}^-(x-\Gamma) P_{\Gamma}(x-\Gamma\nearrow x)
  \label{RGnmaxforward}
\end{eqnarray}
where one can plug the probabilities of about-to-be-decimated bonds of Eqs \ref{Pplusdecim}
and \ref{Pmoinsdecim}
\begin{eqnarray}
\partial_{\Gamma} n_{\Gamma}^-(x) &&   =  - n_{\Gamma}^-(x)   \frac{ s'(x+\Gamma) }{ s(x+\Gamma)-s(x) }
 - n_{\Gamma}^+(x+\Gamma)  \frac{ s'(x) }{ s(x+\Gamma)-s(x) }
 \nonumber \\
 \partial_{\Gamma} n_{\Gamma}^+(x) &&    =  - n_{\Gamma}^+(x)  \frac{ s'(x-\Gamma) }{ s(x)-s(x-\Gamma) }
 - n_{\Gamma}^-(x-\Gamma)\frac{ s'(x) }{ s(x)-s(x-\Gamma) }
  \label{RGnmaxexpli}
\end{eqnarray}

Let us plug the form of Eq. \ref{ngpmsol}
\begin{eqnarray}
n_{\Gamma}^{-}(x)&& 
=   \frac{ s'(x) [c_{\Gamma} - \kappa_{\Gamma}s(x+\Gamma)] }{[s(x+\Gamma)-s(x)]^2}
\nonumber \\
n_{\Gamma}^{+}(x) &&
=  \frac{ s'(x) \left[ c_{\Gamma} - \kappa_{\Gamma}s(x-\Gamma)\right] }{[s(x)-s(x-\Gamma)]^2}
  \label{ngpmsolbis}
\end{eqnarray}
with the corresponding derivatives with respect to $\Gamma$
\begin{eqnarray}
\partial_{\Gamma}n_{\Gamma}^{-}(x) &&  
= \frac{ s'(x) [\partial_{\Gamma}c_{\Gamma} -s(x+\Gamma) \partial_{\Gamma}\kappa_{\Gamma}] }{[s(x+\Gamma)-s(x)]^2}
+ \frac{ s'(x) [ -\kappa_{\Gamma}s'(x+\Gamma)] }{[s(x+\Gamma)-s(x)]^2}
-2 \frac{ s'(x)s'(x+\Gamma) [c_{\Gamma} - \kappa_{\Gamma}s(x+\Gamma)] }{[s(x+\Gamma)-s(x)]^3}
  \label{npmsolderi-}
\end{eqnarray}
and
\begin{eqnarray}
\partial_{\Gamma} n_{\Gamma}^{+}(x)&& 
=  \frac{ s'(x) \left[ \partial_{\Gamma}c_{\Gamma} -s(x-\Gamma)\partial_{\Gamma} \kappa_{\Gamma}\right] }{[s(x)-s(x-\Gamma)]^2}
+\frac{ s'(x) \left[  \kappa_{\Gamma}s'(x-\Gamma)\right] }{[s(x)-s(x-\Gamma)]^2}
-2 \frac{ s'(x)s'(x-\Gamma) \left[ c_{\Gamma} - \kappa_{\Gamma}s(x-\Gamma)\right] }{[s(x)-s(x-\Gamma)]^3}
  \label{npmsolderi+}
\end{eqnarray}
while the opposite of the right-handsides of Eqs \ref{RGnmaxexpli} reduce to
\begin{eqnarray}
 n_{\Gamma}^-(x)   \frac{ s'(x+\Gamma) }{ s(x+\Gamma)-s(x) }
+  n_{\Gamma}^+(x+\Gamma)  \frac{ s'(x) }{ s(x+\Gamma)-s(x) }
 && 
 =  \frac{ s'(x)s'(x+\Gamma) [c_{\Gamma} - \kappa_{\Gamma}s(x+\Gamma)] }{[s(x+\Gamma)-s(x)]^3} 
+  \frac{ s'(x) s'(x+\Gamma) \left[ c_{\Gamma} - \kappa_{\Gamma}s(x)\right] }{[s(x+\Gamma)-s(x)]^3} 
  \nonumber \\
    n_{\Gamma}^+(x)  \frac{ s'(x-\Gamma) }{ s(x)-s(x-\Gamma) }
+  n_{\Gamma}^-(x-\Gamma)\frac{ s'(x) }{ s(x)-s(x-\Gamma) }
&&  
= \frac{ s'(x)s'(x-\Gamma) \left[ c_{\Gamma} - \kappa_{\Gamma}s(x-\Gamma)\right] }{[s(x)-s(x-\Gamma)]^3} 
 +  \frac{ s'(x) s'(x-\Gamma) [c_{\Gamma} - \kappa_{\Gamma}s(x)] }{[s(x)-s(x-\Gamma)]^3}
 \nonumber \\
  \label{RGnmaxexpliright}
\end{eqnarray}
As a consequence, the RG-flows of Eq. \ref{RGnmaxexpli} reduce to
\begin{eqnarray}
0 &&  
=  \partial_{\Gamma}c_{\Gamma} -s(x+\Gamma) \partial_{\Gamma}\kappa_{\Gamma}
\nonumber \\
0 && 
=   \partial_{\Gamma}c_{\Gamma} -s(x-\Gamma)\partial_{\Gamma} \kappa_{\Gamma}
  \label{npmsolderi+flow}
\end{eqnarray}
and yield that $c_{\Gamma}=c$ and $\kappa_{\Gamma}=\kappa $ 
do not depend on the scale $\Gamma$,
so that the two densities of Eq. \ref{ngpmsol} become
\begin{eqnarray}
n_{\Gamma}^{-}(x)&& 
=   \frac{ s'(x) [c - \kappa s(x+\Gamma)] }{[s(x+\Gamma)-s(x)]^2} \geq 0
\nonumber \\
n_{\Gamma}^{+}(x) &&
=  \frac{ s'(x) \left[ c - \kappa s(x-\Gamma)\right] }{[s(x)-s(x-\Gamma)]^2} \geq 0
  \label{ngammackappa}
\end{eqnarray}
that should be positive $n_{\Gamma}^{\pm}(x) \geq 0$ for any $x$,
so that there are constraints on the possible values of the parameter $\kappa$ 
that depend on the possible values of the scale function $s(x)$.
For instance, if the scale function $s(x)$ for a diffusion on $x \in ]-\infty,+\infty[$ varies
between $s(-\infty)=-\infty$ and $s(+\infty)=+\infty$, then the positivity constraint for the two densities of Eq. \ref{ngammackappa}
for any $x$ yields that $\kappa$ should vanish $\kappa=0$.

When the parameter $\kappa$ that can be interpreted as a current (see \ref{sub_kappacurrent}) 
vanishes $\kappa=0$, then the corresponding densities 
\begin{eqnarray}
\text{ When  $\kappa=0 $ :  } \ \ \ n_{\Gamma}^{-}(x)&& 
=  c \frac{ s'(x)  }{[s(x+\Gamma)-s(x)]^2}
\nonumber \\
n_{\Gamma}^{+}(x) &&
= c \frac{ s'(x)  }{[s(x)-s(x-\Gamma)]^2}
  \label{ngammackappa0}
\end{eqnarray}
where the remaining constant $c$ is simply the global normalization,
have the following properties :

(i) the total density $n_{\Gamma}$ of maxima or minima
 \begin{eqnarray}
n_{\Gamma} = \int_{-\infty}^{+\infty} dx n^{\pm}_{\Gamma}(x) 
  \label{pinorma}
\end{eqnarray}
is finite when the following integral converges 
\begin{eqnarray}
  \int_{-\infty}^{+\infty} dx  \frac{s'(x) }{[s(x+\Gamma)-s(x)]^2} < +\infty
  \label{pinormafinite}
\end{eqnarray}

(ii) For large $ x  \to \pm \infty$, if $ \frac{s''(x)}{s'(x) } \to 0$,
then the denominator of Eq. \ref{ngpmsol}
can be approximated for finite $\Gamma$ by
$[s(x+\Gamma)-s(x) ]^2 \simeq [ \Gamma s'(x) ]^2 $, 
so that the asymptotic behaviors of the extrema densities of Eq. \ref{ngpmsol}
involving the derivative $s'(x)$ of the scale function of Eq. \ref{Sx}
\begin{eqnarray}
n_{\Gamma}^{\pm}(x) && \opsimeq_{x \to \pm \infty}
   \frac{ c }{\Gamma^2 s'(x) } =  \frac{ c }{\Gamma^2  } D(x) e^{-U(x) }
  \label{ngammackappa0symp}
\end{eqnarray}
i.e. the diffusion coefficient $D(x)$ appears as an additional factor with respect
to the asymptotic behavior of the steady state $P_{eq}(x) \propto e^{-U(x)}$ of Eq. \ref{Peq}.

(iii) they lead to detailed-balance properties for the renormalized trajectories at scale $\Gamma$,
as explained in detail in the next subsection.

%%%%%%%%%%%%%%%%%%%%%%%%%%%%%%%%%%%%%%%%%%%%

\subsection{ Detailed-Balance at scale $\Gamma$ when $\kappa=0$ (densities of Eq. \ref{ngammackappa0})
and similarity transformation towards symmetric blocks }

\label{subsec_similarity}

One of the most well-known properties of reversible Markov generators 
that is explained in textbooks \cite{gardiner,vankampen,risken} 
and that has been extensively used in many specific applications to various models (see for instance \cite{glauber,Felderhof,siggia,referee,kimball,peschel,jpb_antoine,locgolosov,texier,us_eigenvaluemethod,Castelnovo,c_lyapunov,us_gyrator,us_kemeny,c_pearson,c_boundarydriven,c_largedevDBsusy}), is that 
they can be transformed into symmetric operators  
via similarity transformations. 
For the initial Fokker-Planck generator ${\cal F}$ of Eq. \ref{FPgenerator} and \ref{fokkerplanckgeneratorfactorizedU},
this well-known property is that
\begin{eqnarray}
{\cal H} && \equiv - e^{ \frac{U(x)}{2} }  {\cal F}  e^{- \frac{U(x)}{2} } 
= -  e^{ \frac{U(x)}{2} }  \frac{\partial}{\partial x} e^{- \frac{U(x)}{2} } D(x) e^{ - \frac{U(x)}{2} }\frac{\partial}{\partial x}  e^{ \frac{U(x)}{2} }
\label{Hsusy}
\end{eqnarray}
is a hermitian quantum Hamiltonian.
 It is thus interesting to analyze how detailed-balance survives within the renormalized trajectories.

\subsubsection{ Detailed-Balance properties of the conditional probabilities $P_{\Gamma}^-(. \vert .) $ 
with respect to the steady densities $n_{\Gamma}^{\pm}$ of extrema }

The two functions $R^{\pm}_{\Gamma}(x)$ introduced in Eqs \ref{Rxplus}
and \ref{Rxmoins}
respectively are related by
\begin{eqnarray}
\frac{ R^-_{\Gamma}(x) }{ s'(x) } =  \frac{ 1 }{ s(x+\Gamma)-s(x) } = \frac{ R^+_{\Gamma}(x+\Gamma) }{ s'(x+\Gamma) } 
  \label{Rpm}
\end{eqnarray}

As a consequence, the integral of $R_{\Gamma}^-(z) $ that appears in the exponential of the conditional probability 
$P_{\Gamma}^-(x \vert x_0)$ of Eq. \ref{Pmoins} for descending bonds
\begin{eqnarray}
 \int_x^{x_0-\Gamma} dz R_{\Gamma}^-(z)
&& = \int_x^{x_0-\Gamma} dz \frac{ s'(z) }{ s(z+\Gamma)-s(z) }   
  =  \int_{x+\Gamma}^{x_0} dy \frac{ s'(y-\Gamma) }{ s(y)-s(y-\Gamma) }   
  \nonumber \\
&& =  
\int_{x+\Gamma}^{x_0} dy \left( \frac{ s'(y)}{ s(y)-s(y-\Gamma) } - \frac{ s'(y) - s'(y-\Gamma) }{ s(y)-s(y-\Gamma) }  \right) 
%  \nonumber \\ && 
= 
 \int_{x+\Gamma}^{x_0} dy R_{\Gamma}^+(y)   -  \left[ \ln \left( s(y)-s(y-\Gamma) \right) \right]_{x+\Gamma}^{x_0}
   \nonumber \\
&& = \int_{x+\Gamma}^{x_0} dy R_{\Gamma}^+(y)   +  \ln\left( \frac{  s(x+\Gamma)-s(x)  }{  s(x_0)-s(x_0-\Gamma)  } \right)
  \label{DBintegR}
\end{eqnarray}
   can be thus rewritten, up to boundary terms,
as an integral of the function $R_{\Gamma}^+(y) $ that appears in the 
exponential of the conditional probability 
$P_{\Gamma}^+(x_0 \vert x)$ of Eq. \ref{Pplus} for ascending bonds where the two end points $(x_0,x)$ are exchanged
\begin{eqnarray}
P_{\Gamma}^+(x_0 \vert x)  
= R_{\Gamma}^+(x_0) e^{ - \int_{x+\Gamma}^{x_0} dy R_{\Gamma}^+(y)  } 
= \frac{ s'(x_0) }{ s(x_0)-s(x_0-\Gamma) }  e^{ - \int_{x+\Gamma}^{x_0} dy R_{\Gamma}^+(y)  }
  \label{Pplusx0xexchanged}
\end{eqnarray}

Plugging Eq. \ref{DBintegR} into Eq. \ref{Pmoins} yields
that $P_{\Gamma}^-(x \vert x_0) $ can be rewritten in terms of $P_{\Gamma}^+(x_0 \vert x) $
 \begin{eqnarray}
P_{\Gamma}^-(x \vert x_0) 
&& = R_{\Gamma}^-(x) e^{ - \int_x^{x_0-\Gamma} dz R_{\Gamma}^-(z)  } 
=\frac{ s'(x) }{ s(x+\Gamma)-s(x) } e^{-  \int_x^{x_0-\Gamma} dz R_{\Gamma}^-(z) }
\nonumber \\
&&=\frac{ s'(x) }{ s(x+\Gamma)-s(x) } e^{- \int_{x+\Gamma}^{x_0} dy R_{\Gamma}^+(y)   -  \ln\left( \frac{  s(x+\Gamma)-s(x)  }{  s(x_0)-s(x_0-\Gamma)  } \right)}
% \nonumber \\ &&
=\frac{ s'(x) [ s(x_0)-s(x_0-\Gamma)] }{ [s(x+\Gamma)-s(x)]^2 } e^{- \int_{x+\Gamma}^{x_0} dy R_{\Gamma}^+(y)  }
\nonumber \\
&&=\frac{ s'(x) [ s(x_0)-s(x_0-\Gamma)]^2 }{ s'(x_0) [s(x+\Gamma)-s(x)]^2 } P_{\Gamma}^+(x_0 \vert x) 
\equiv \frac{ n_{\Gamma}^-(x) }{ n_{\Gamma}^+(x_0)} P_{\Gamma}^+(x_0 \vert x)
  \label{PmoinsDB}
\end{eqnarray}
where one recognizes the density $n_{\Gamma}^+(x_0) $ of maxima and the density $n_{\Gamma}^-(x) $ of minima  of Eq. \ref{ngammackappa0}. 
The physical meaning of this identity
 \begin{eqnarray}
P_{\Gamma}^-(x \vert x_0) n_{\Gamma}^+(x_0)
= P_{\Gamma}^+(x_0 \vert x)n_{\Gamma}^-(x)
  \label{DBRGwithn}
\end{eqnarray}
is thus the Detailed-Balance of the conditional probabilities $P_{\Gamma}^-(. \vert .) $ at scale $\Gamma$
with respect to the steady densities $n_{\Gamma}^{\pm}$ of extrema of the Markov chain of Eq. \ref{MarkovChain}. 
Let us stress the difference with the Detailed-Balance properties of the initial Fokker-Planck model
with respect to its steady state $P_{eq}(x) $ of Eq. \ref{Peq} 
that reads in terms of $s'(x)$ and the diffusion coefficient $D(x)$
 \begin{eqnarray}
P_{eq}(x) \propto e^{-U(x)}= m'(x) = \frac{e^{-U_I(x) } }{ D(x) s'(x)} = \frac{1}{ D(x) s'(x)}
  \label{Peqm's'}
\end{eqnarray}

The rewriting of the detailed balance property of Eq. \ref{DBRGwithn}
in terms of the notations of the renormalized perspective of Eq. \ref{Pgamma} reads
for the bond densities $N_{\Gamma}(..) $
\begin{eqnarray}
N_{\Gamma}(x_0\des x) \equiv n^+_{\Gamma}(x_0)P_{\Gamma}(x_0\searrow x) = 
n^-_{\Gamma}(x) P_{\Gamma}(x \nearrow x_0)  \equiv N_{\Gamma}(x \asc x_0)
  \label{DBRGwithNarrow}
\end{eqnarray}
and for the backward conditional probabilities of Eq. \ref{PgammaBack}
\begin{eqnarray}
 P_{\Gamma}(x\swarrow x_0)\equiv \frac{N_{\Gamma}(x\asc x_0)}{n^+_{\Gamma}(x_0)}
 =\frac{n^-_{\Gamma}(x)\,P_{\Gamma}(x\nearrow x_0)}{n^+_{\Gamma}(x_0)} = P_{\Gamma}(x_0\searrow x)
\nonumber \\
P_{\Gamma}(x\nwarrow x_0)\equiv\frac{N_{\Gamma}(x\des x_0)}{n^-_{\Gamma}(x_0)}
=\frac{n^+_{\Gamma}(x)\,P_{\Gamma}(x\searrow x_0)}{n^-_{\Gamma}(x_0)} =P_{\Gamma}(x_0 \nearrow x)
  \label{PgammaBackDB}
\end{eqnarray}

%%%%%%%%%%%%%%%%%%%%%%%%%%%%%%%%%%%%%%%%%%%%%%%

\subsubsection{ Similarity transformation of the conditional probabilities $P_{\Gamma}^{\pm}(. \vert .) $
towards symmetric blocks $ B_{\Gamma}(.,.) $}

The detailed-balance property of Eqs \ref{DBRGwithn} \ref{DBRGwithNarrow} yields that it can be useful to define the
similarity transformations for the conditional probabilities
 \begin{eqnarray}
P_{\Gamma}^-(x \vert x_0) && = \sqrt{ n_{\Gamma}^-(x)} B_{\Gamma}^-(x,x_0) \frac{1}{\sqrt{n_{\Gamma}^+(x_0) }  }
\nonumber \\
P_{\Gamma}^+(x \vert x_0) && = \sqrt{ n_{\Gamma}^+(x)} B_{\Gamma}^+(x,x_0) \frac{1}{\sqrt{n_{\Gamma}^-(x_0) }  }
  \label{SimilarityB}
\end{eqnarray}
so that the 
detailed-balance property of Eqs \ref{DBRGwithn} \ref{DBRGwithNarrow}
rewritten for the blocks $ B_{\Gamma}(.,.) $ 
 \begin{eqnarray}
B_{\Gamma}^-(x,x_0) \sqrt{n_{\Gamma}^+(x_0) n_{\Gamma}^-(x)} 
% =  N_{\Gamma}(x_0\des x) 
= P_{\Gamma}^-(x \vert x_0) n_{\Gamma}^+(x_0)
= P_{\Gamma}^+(x_0 \vert x)n_{\Gamma}^-(x)  
= \sqrt{ n_{\Gamma}^+(x_0) n_{\Gamma}^-(x)} B_{\Gamma}^+(x_0,x) 
  \label{SymBlockB}
\end{eqnarray}
yield that they are symmetric with respect to the exchange of their two end-points
 \begin{eqnarray}
  B_{\Gamma}^-(x,x_0) = B_{\Gamma}^+(x_0,x) 
  \label{SymBlockBex}
\end{eqnarray}
as can be also checked via
 their explicit forms in terms of the scale function $s(.)$ obtained from Eq. \ref{PmoinsDB} using Eq. \ref{ngammackappa0}
 \begin{eqnarray}
B_{\Gamma}^-(x,x_0)  && = \frac{1}{\sqrt{n_{\Gamma}^-(x) }  } P_{\Gamma}^-(x \vert x_0) \sqrt{n_{\Gamma}^+(x_0) }
=  \frac{s(x+\Gamma)-s(x)}{ \sqrt{ s'(x) }}  P_{\Gamma}^-(x \vert x_0)  \frac{ \sqrt{s'(x_0)} }{s(x_0)-s(x_0-\Gamma) }
 \nonumber \\
&& 
=    \frac{ \sqrt{ s'(x) s'(x_0)} }{s(x_0)-s(x_0-\Gamma) } e^{-  \int_x^{x_0-\Gamma} dz R_{\Gamma}^-(z) }
=\frac{ \sqrt{ s'(x) s'(x_0)}  }{ s(x+\Gamma)-s(x) } e^{- \int_{x+\Gamma}^{x_0} dy R_{\Gamma}^+(y)  }
\nonumber \\
&&=\frac{  [ s(x_0)-s(x_0-\Gamma)] }{ \sqrt{s'(x_0)}  } 
P_{\Gamma}^+(x_0 \vert x) 
\frac{ \sqrt{ s'(x) }  }{  [s(x+\Gamma)-s(x)] }
= \frac{1}{\sqrt{n_{\Gamma}^+(x_0) }  }P_{\Gamma}^+(x_0 \vert x) \sqrt{n_{\Gamma}^{-}(x) } = B_{\Gamma}^+(x_0, x)
  \label{ExpliciBlockB}
\end{eqnarray}
with the special case of blocks about to be decimated
 \begin{eqnarray}
B_{\Gamma}^-(x=x_0-\Gamma,x_0)  && =     \frac{ \sqrt{ s'(x_0-\Gamma) s'(x_0)} }{s(x_0)-s(x_0-\Gamma) } 
\nonumber \\
B_{\Gamma}^+(x_0=x+\Gamma, x) && =\frac{ \sqrt{ s'(x) s'(x+\Gamma)}  }{ s(x+\Gamma)-s(x) } 
  \label{ExpliciBlockBdecim}
\end{eqnarray}

Taking the logarithms of the explicit forms of Eq. \ref{ExpliciBlockB}
 \begin{eqnarray}
\ln B_{\Gamma}^-(x,x_0)  && =   \ln \sqrt{ s'(x) s'(x_0)} - \ln [s(x_0)-s(x_0-\Gamma) ] -  \int_x^{x_0-\Gamma} dz R_{\Gamma}^-(z)
\nonumber \\
\ln B_{\Gamma}^+(x_0, x) && =\ln \sqrt{ s'(x) s'(x_0)}  -\ln[ s(x+\Gamma)-s(x) ] - \int_{x+\Gamma}^{x_0} dy R_{\Gamma}^+(y)  
  \label{ExpliciBlockBln}
\end{eqnarray}
one obtains that their RG-flows reduce to
 \begin{eqnarray}
\partial_{\Gamma} \ln B_{\Gamma}^-(x,x_0)  && =    - \frac{s'(x_0-\Gamma)}{s(x_0)-s(x_0-\Gamma) } 
+ R_{\Gamma}^-(x_0-\Gamma)
-  \int_x^{x_0-\Gamma} dz \partial_{\Gamma} R_{\Gamma}^-(z)
= -  \int_x^{x_0-\Gamma} dz \partial_{\Gamma} R_{\Gamma}^-(z)
\nonumber \\
\partial_{\Gamma}\ln B_{\Gamma}^+(x_0, x) && =  - \frac{s'(x+\Gamma)}{ s(x+\Gamma)-s(x) }
 + R_{\Gamma}^+(x+\Gamma) 
 - \int_{x+\Gamma}^{x_0} dy \partial_{\Gamma}R_{\Gamma}^+(y)   
 = - \int_{x+\Gamma}^{x_0} dy \partial_{\Gamma}R_{\Gamma}^+(y) 
  \label{RGBlog}
\end{eqnarray}
where one can plug Eq. \ref{RGforRpm} to obtain
\begin{eqnarray}
\partial_{\Gamma} \ln B_{\Gamma}^-(x,x_0)  &&   
=   \int_x^{x_0-\Gamma} dz  R_{\Gamma}^-(z)R_\Gamma^+( z+\Gamma)
= \int_x^{x_0-\Gamma} dz  \frac{ s'(z)  s'(z+\Gamma) }{ [s(z+\Gamma)-s(z) ]^2} 
\nonumber \\
\partial_{\Gamma} \ln B_{\Gamma}^+(x_0, x) &&     
 =  \int_{x+\Gamma}^{x_0} dy R_\Gamma^-( y-\Gamma) R_{\Gamma}^+(y)  
 = \int_{x+\Gamma}^{x_0} dy  \frac{ s'(y-\Gamma)  s'(y) }{ [s(y)-s(y-\Gamma) ]^2} 
  \label{RGBln}
\end{eqnarray}
For the blocks $B_{\Gamma}^{\pm}$ themselves, these RG-flows 
are similar to Eqs \ref{RGPasc} and \ref{RGPdes}, except that only the 
merging term involving three blocks is present on the right handside,
i.e. 
\begin{eqnarray}
\partial_{\Gamma} B_{\Gamma}(x_0\searrow x)
&& = \int_{x}^{x_0-\Gamma} d z B_{\Gamma}(x_0\searrow z) B_{\Gamma}(z\nearrow z+\Gamma) B_{\Gamma}(z+\Gamma\searrow x)
\nonumber \\
\partial_{\Gamma} B_{\Gamma}(x_0\nearrow x) && = \int_{x_0+\Gamma}^{x}d z
B_{\Gamma}(x_0\nearrow z)
B_{\Gamma}(z\searrow z-\Gamma) B_{\Gamma}(z-\Gamma\nearrow x)
   \label{RGBascdes}
\end{eqnarray}
in agreement with \cite{rgtoy}.
Finally, let us mention that the product of conditional probabilities along a renormalized trajectory 
can be rewritten as a product of the symmetric blocks up to the boundary similarity factors 
 \begin{eqnarray}
&& P_{\Gamma}^+(x_{2n} \vert x_{2n-1}) P_{\Gamma}^-(x_{2n-1} \vert x_{2n-2})... P_{\Gamma}^+(x_2 \vert x_1) P_{\Gamma}^-(x_1 \vert x_0) 
\nonumber \\ && = 
\sqrt{ n_{\Gamma}^+(x_{2n})} B_{\Gamma}^+(x_{2n},x_{2n-1}) 
 B_{\Gamma}^-(x_{2n-1},x_{2n-2})
... B_{\Gamma}^+(x_2,x_1) 
 B_{\Gamma}^-(x_1,x_0) \frac{1}{\sqrt{n_{\Gamma}^+(x_0) }  }
  \label{SimilarityBtraj}
\end{eqnarray}
as in \cite{rgtoy}.

%%%%%%%%%%%%%%%%%%%%%%%%%%%%%%%%%%%%%%%%%%%%

\subsection{ Discussion }

In summary, for diffusions on the whole line $x\in ]-\infty,+\infty[$ involving an arbitrary diffusion coefficient $D(x)$
and an arbitrary Ito force $F_I(x)$,
all the properties of the renormalized trajectories can be computed in terms of the scale function $s(x)$ alone.
So all the diffusion processes that have the same scale function have exactly the same statistics of renormalized trajectories at scale $\Gamma$. As a consequence, the scale function $s(x)$ characterizes
the possible universality classes. In the next section, we describe the universality class $s(x)=x$ 
containing the Brownian motion and we give some other examples.

% %%%%%%%%%%%%%%%%%%%%%%%%%%%%%%%%%%%%%%%%%%%%%%%

% %%%%%%%%%%%%%%%%%%%%%%%%%%%%%%%%%%%%%%%%%%%%%%%

\section{ Examples of diffusions on the whole line $x \in ]-\infty,+\infty[$ }

\label{sec_WholeLineExamples}

In this section, we show how the present analysis based on the scale function $s(.)$ 
is useful to analyze specific examples and to identify universality classes.

\subsection{ Universality class $s(x)=x $ on $x \in ]-\infty,+\infty[$ : vanishing Ito force $F_I(x)=0$ and arbitrary $D(x)$ }

\label{subsec_universalityBrown}

In this section, we focus on diffusion processes whose scale function 
\begin{eqnarray}
s(x) && =x
\nonumber \\
s'(x) && =1 
  \label{spxlinear}
\end{eqnarray}
The corresponding Ito potential $U_I(x)$ and its derivative vanish
\begin{eqnarray}
 U_I(x) && = 0
 \nonumber \\
  U_I'(x) && \equiv - \frac{F_I(x)}{D(x) } =0
  \label{UIzero}
\end{eqnarray}
i.e. the Ito force vanishes
\begin{eqnarray}
 F_I(x) = 0
\label{FIzero}
\end{eqnarray}
but the diffusion coefficient $D(x)$ is arbitrary, with some well-known examples
\begin{eqnarray}
D(x)&& =1 \ \ \text{ Brownian }
\nonumber \\
D(x)&& = 1+x^2 \ \ \text{ Cauchy process, the special case $\mu=1$ of the student process of Eq. \ref{Dstudent}}
\label{DBC}
\end{eqnarray}
Remarkably, even if these processes have different speed-measures $m'(.)$ or potential $U(.)$
\begin{eqnarray}
m'(x)=e^{-U(x)} = \frac{1}{D(x)}
\label{steadyUniversalityBrown}
\end{eqnarray}
that can be normalizable (Cauchy process) or not (Brownian motion),
 they have exactly the same renormalized trajectories at scale $\Gamma$ described below,
 while their differences appear only when one considers the durations of the bonds, as discussed in the Appendices.

The scale function $s(x)=x$ of Eq. \ref{spxlinear}
yields that the two functions $
 $ coincide and reduce to
\begin{eqnarray}
R^{\pm}_{\Gamma}(x) 
= \frac{1}{\Gamma} 
  \label{RxBrown0}
\end{eqnarray}
and lead to the well-known distributions 
\begin{eqnarray}
P_{\Gamma}^+(x \vert x_0)  
&&= \frac{1}{\Gamma} e^{ - \frac{x-x_0-\Gamma}{\Gamma} } \ \ \ \ \text{ for } \ \ x \in [x_0+\Gamma,+\infty[
\nonumber \\
P_{\Gamma}^-(x \vert x_0)  
&&=\frac{1}{\Gamma} e^{ - \frac{x_0-\Gamma-x}{\Gamma} } \ \ \ \ \text{ for } \ \  x \in ]-\infty,x_0-\Gamma]
  \label{PplusmoinsBrown0}
\end{eqnarray}
with the uniform densities of extrema
\begin{eqnarray}
n_{\Gamma}^{\pm}(x) && =  \frac{ c }{\Gamma^2  }
  \label{npmsolfinalBrown}
\end{eqnarray}

%%%%%%%%%%%%%%%%%%%%%%%%%%%%%%%%%%%%%%

\subsection{ Student-process with the quadratic diffusion coefficient $D(x)=1+x^2$  and the linear Ito force $  F_I(x)  =  (1-\mu ) x $      }

\label{sec_student}

In this subsection, we focus on the Student process \cite{pearson_student_processes,pearson_student}
that belongs to the Pearson family (see  \cite{pearson1895,pearson_wong,diaconis,autocorrelation,pearson_class,pearson2012,PearsonHeavyTailed,pearson2018,c_pearson}
and references therein)
with its quadratic diffusion coefficient  
and its linear Ito force
\begin{eqnarray}
D(x)&& =1+x^2  
\nonumber \\
  F_I(x) && =  (1-\mu ) x
\label{Dstudent}
\end{eqnarray}
that converges towards the  steady state 
\begin{eqnarray}
P_{eq}(x) = \frac{1}
{Z \left( 1+x^2 \right)^{\frac{1+\mu}{2}}}
  \ \ \ {\rm } \ \ {\rm for } \ \ x \in ]-\infty,+\infty[
\ \ \ {\rm with } \ \ \mu>0
\label{student}
\end{eqnarray}
(where the normalization $Z=\frac{\sqrt{\pi}\Gamma_E(\frac{\mu}{2})}{\Gamma_E(\frac{\mu+1}{2}) }$ involves 
the Euler Gamma function $\Gamma_E(.)$)
displaying the power-law tails
\begin{eqnarray}
P_{eq}(x) \oppropto_{ x \to \pm \infty} \frac{1}{ \vert x \vert^{1+\mu} }
\label{studentpower}
\end{eqnarray}
so that the second moment $\int dx x^2 P_{eq}(x) $  is finite only for $\mu>2$.

The two potentials of Eq \ref{Peq} are both logarithmic
\begin{eqnarray}
U(x) && = \frac{1+\mu}{2} \ln \left( 1+x^2 \right)
\nonumber \\
U_I(x) && =\frac{\mu-1}{2} \ln \left( 1+x^2 \right)
\label{Ustudent}
\end{eqnarray}
and lead to
the scale function of Eq. \ref{Sx} 
\begin{eqnarray}
s(x)  =  \int_0^x dy  e^{U_I(y) }    =  \int_0^x dy \left( 1+y^2 \right)^{ \frac{\mu-1}{2}}
\label{SxStudent}
\end{eqnarray}
with the corresponding expressions of
the two functions of Eqs \ref{Rxplus}
and \ref{Rxmoins} 
\begin{eqnarray}
R^+_{\Gamma}(x) &&=  \frac{ s'(x) }{ s(x)-s(x-\Gamma) } 
= \frac{ \left( 1+x^2 \right)^{ \frac{\mu-1}{2}} }{ \int_{x-\Gamma}^x dy \left( 1+y^2 \right)^{ \frac{\mu-1}{2}} } 
\nonumber \\
R^-_{\Gamma}(x) && =  \frac{ s'(x) }{ s(x+\Gamma)-s(x) } 
=\frac{ \left( 1+x^2 \right)^{ \frac{\mu-1}{2}} }{ \int_x^{x+\Gamma} dy \left( 1+y^2 \right)^{ \frac{\mu-1}{2}} } 
  \label{RxStudent} 
\end{eqnarray}

Since $s(-\infty)=-\infty$ and $s(+\infty)=+\infty$
the densities of maxima and minima are given by Eqs \ref{ngammackappa0}
\begin{eqnarray}
n_{\Gamma}^{+}(x) &&  
=c   \frac{ s'(x) }{[s(x)-s(x-\Gamma)]^2}
= c   \frac{ \left( 1+x^2 \right)^{ \frac{\mu-1}{2}} }{\left[\int_{x-\Gamma}^x dy \left( 1+y^2 \right)^{ \frac{\mu-1}{2}}\right]^2}
\nonumber \\
n_{\Gamma}^{-}(x)&& 
= c  \frac{s'(x) }{[s(x+\Gamma)-s(x)]^2}
=  c  \frac{\left( 1+x^2 \right)^{ \frac{\mu-1}{2}} }{\left[\int_x^{x+\Gamma} dy \left( 1+y^2 \right)^{ \frac{\mu-1}{2}}\right]^2}
  \label{npmsolfinalstudent}
\end{eqnarray}
with the asymptotic behavior of Eq. \ref{ngammackappa0symp}
\begin{eqnarray}
n_{\Gamma}^{\pm}(x) && \opsimeq_{x \to \pm \infty}
   \frac{ c }{\Gamma^2 s'(x) } =  \frac{ c }{\Gamma^2  \left( 1+x^2 \right)^{ \frac{\mu-1}{2}} }
   \opsimeq_{x \to \pm \infty} \frac{ c }{\Gamma^2  \vert x\vert^{ \mu-1} }
  \label{npmsolfinalasympstudent}
\end{eqnarray}
that corresponds to a fatter tail than the power-law of the steady state of Eq. \ref{studentpower}
and that is not normalizable in the region $0<\mu\leq 2$ while the steady state is normalizable.

For $\mu=1$ corresponding to the Cauchy steady state in Eq. \ref{student}
\begin{eqnarray}
\text{ Case $\mu=1$} : \ \  
P_{eq}(x) = \frac{1 }{ \pi ( 1+x^2 )}  \ \ \ {\rm } \ \ {\rm for } \ \ x \in ]-\infty,+\infty[
\label{cauchy}
\end{eqnarray}
the scale function of Eq. \ref{SxStudent} 
\begin{eqnarray}
s(x)  =   x
\label{SxCauchy}
\end{eqnarray}
coincides with the scale function of the pure Brownian motion of Eq. \ref{spxlinear}
with the corresponding properties described in subsection
\ref{subsec_universalityBrown}.

%%%%%%%%%%%%%%%%%%%%%%%%%%%%%%%%%%%%%%%%%%%

%%%%%%%%%%%%%%%%%%%%%%%%%%%%%%%%%%%%

\subsection{ Brownian motion with positive drift $F_I(x)=F>0$ } 

\label{subsec_BrownF}

For the Brownian motion with positive drift $F_I(x)=F>0$
\begin{eqnarray}
D(x) && =1
\nonumber \\
F_I(x) && = F
  \label{FDBrown}
\end{eqnarray}
the two potentials $U(x)$ and $U_I(x)$ of Eq. \ref{U'} coincide and are linear
\begin{eqnarray}
 U_I(x)=U(x)  = - F x
\label{Uxbrownian}
\end{eqnarray}
The scale function of Eq. \ref{Sx} 
\begin{eqnarray}
s'(x) && = e^{U_I(x) }   =   e^{- F x } 
\nonumber \\
s(x)  && =  \int_0^x dy  e^{- F y }  = \frac{1-e^{- F x } }{F}  
\label{Sxbrownian}
\end{eqnarray}
 yields that the two functions of Eqs \ref{Rxplus}
and \ref{Rxmoins} 
read 
\begin{eqnarray}
R^+_{\Gamma}(x) &&=  \frac{ s'(x) }{ s(x)-s(x-\Gamma) } 
=  \frac{  e^{- F x }   }{ \frac{e^{- F (x-\Gamma) }-e^{- F x } }{F}    } 
=\frac{  F  }{ e^{ F \Gamma } -1 }  \equiv r^+_{\Gamma}
\nonumber \\
R^-_{\Gamma}(x) && =  \frac{ s'(x) }{ s(x+\Gamma)-s(x) } 
=  \frac{ e^{- F x }   }{ \frac{e^{- F x }-e^{- F (x+\Gamma) } }{F}    }
 =\frac{  F  }{ 1- e^{- F \Gamma }  }  \equiv r^-_{\Gamma}
  \label{RxBrowndrift} 
\end{eqnarray}

Since they are independent of $x$, the probability distributions of the ascending and descending renormalized bonds of Eqs
\ref{Pplus}
and \ref{Pmoins}
are simply exponential
\begin{eqnarray}
P_{\Gamma}^+(x \vert x_0)  
= r_{\Gamma}^+ e^{ - (x-x_0-\Gamma) r_{\Gamma}^+ } \ \ \ \ \text{ for } \ \ x \in [x_0+\Gamma,+\infty[
\nonumber \\
P_{\Gamma}^-(x \vert x_0)  
= r_{\Gamma}^- e^{ - (x_0-\Gamma-x) r_{\Gamma}^- } \ \ \ \ \text{ for } \ \  x \in ]-\infty,x_0-\Gamma]
  \label{PplusmoinsBorown}
\end{eqnarray}

Here the asymptotic behaviors of the scale function $s(x)$ of Eq. \ref{Sxbrownian}
for $F>0$ are
\begin{eqnarray}
s(x)  && =  \frac{1-e^{- F x } }{F}  
\begin{cases}
\displaystyle \opsimeq_{x \to - \infty} -\infty  \\
\displaystyle \opsimeq_{x \to + \infty} \frac{1}{F}  <+\infty
\end{cases}
\label{Sxbrownianasymp}
\end{eqnarray}
while the general expression of Eq. \ref{ngammackappa}
for the possible extrema densities 
\begin{eqnarray}
n_{\Gamma}^{+}(x) &&
=  \frac{ s'(x) \left[ c - \kappa s(x-\Gamma)\right] }{[s(x)-s(x-\Gamma)]^2} 
=\frac{ e^{- F x } \left[ c - \kappa \frac{1-e^{- F x +F \Gamma} }{F} \right] }{\left[ \frac{e^{- F (x-\Gamma) }-e^{- F x } }{F}\right]^2} 
=\frac{ e^{+ F x } \left[ c -  \frac{\kappa }{F} \right]
+\kappa \frac{e^{F \Gamma} }{F}  }{\left[ \frac{e^{F\Gamma} -1 }{F}\right]^2} 
\nonumber \\
n_{\Gamma}^{-}(x)&& 
=   \frac{ s'(x) [c - \kappa s(x+\Gamma)] }{[s(x+\Gamma)-s(x)]^2} 
=\frac{ e^{- F x } \left[c - \kappa \frac{1-e^{- F x-F \Gamma } }{F} \right] }{\left[ \frac{e^{- F x }-e^{- F (x+\Gamma) } }{F} \right]^2} 
= \frac{ e^{+ F x } \left[c - \frac{\kappa }{F} \right] + \kappa \frac{e^{-F \Gamma } }{F}  }{\left[ \frac{1-e^{- F \Gamma } }{F} \right]^2} 
  \label{ngammackappaBrownDrift}
\end{eqnarray}
yields that the expected translation-invariant physical solutions correspond to 
$\left[c - \frac{\kappa }{F} \right]=0$  leading to 
\begin{eqnarray}
n_{\Gamma}^{\pm}(x) &&
=  \kappa\frac{ F   }{\left[ e^{\frac{F\Gamma}{2}} - e^{-\frac{F\Gamma}{2}} \right]^2} 
= \kappa\frac{ F   }{ 4\sinh^2 \left(\frac{F\Gamma}{2}\right)} 
  \label{ngammackappaBrownDriftinv}
\end{eqnarray}
with a non-vanishing current parameter $\kappa \propto F >0$
associated to the constant drift towards infinity.

%%%%%%%%%%%%%%%%%%%%%%%%%%%%%%%%%%%%%%%%%%

\section{ Analysis of renormalized trajectories for diffusions on the half-line $x \in ]0,+\infty[$}

\label{sec_HalfLine}

When the diffusion takes place on $x \in ]0,+\infty[$, the analysis of the ascending bonds of Eq. \ref{Pplus}
that are defined for $x \in [x_0+\Gamma,+\infty[ $
does not change, apart from the facts that the starting point belongs to $x_0 \in ]0,+\infty[$
and the end-point belongs to $x \in ]\Gamma,+\infty[$
\begin{eqnarray}
P_{\Gamma}^+(x \vert x_0)  
= R_{\Gamma}^+(x) e^{ - \int_{x_0+\Gamma}^x dz R_{\Gamma}^+(z)  } \ \ \ \ \text{ for } \ \ x \in ]x_0+\Gamma,+\infty[ 
\ \ \ \ \text{ with } \ \ x_0 \geq 0
  \label{Pplushalf}
\end{eqnarray}
but one needs to reconsider the descending bonds 
as discussed in the next subsection.

\subsection{ Analysis of the descending bonds in the presence of the boundary at $x=0$ }

The conditional probability of Eq. \ref{Pmoins} for the descending bonds
 was defined for $x \in ]-\infty,x_0-\Gamma]$
while now the same expression is valid only for $0<x<x_0-\Gamma$ 
\begin{eqnarray}
P_{\Gamma}^-(x \vert x_0) =  \partial_x C_{\Gamma}^-(x\vert x_0)
= R_{\Gamma}^-(x) e^{ - \int_x^{x_0-\Gamma} dz R_{\Gamma}^-(z)  } \ \ \ \ \text{ for } \ \  x \in ]0,x_0-\Gamma]
  \label{Pmoins0}
\end{eqnarray}
The corresponding contribution of this domain $x \in ]0,x_0-\Gamma] $ to the normalization
\begin{eqnarray}
\int_0^{x_0-\Gamma}  P_{\Gamma}^-(x \vert x_0) 
=   \left[ e^{ - \int_x^{x_0-\Gamma} dz R_{\Gamma}^-(z)  } \right]_0^{x_0-\Gamma}
= 1 - e^{ - \int_0^{x_0-\Gamma} dz R_{\Gamma}^-(z)  } \equiv 1 - W_{\Gamma}(x_0)
  \label{Pmoins0norma}
\end{eqnarray}
where the possible missing weight $W_{\Gamma}(x_0) $ with respect to the normalization 
depends on the scale function $s(.)$ that determines $R_{\Gamma}^-(z)$ of Eq. \ref{Rxmoins}
\begin{eqnarray}
W_{\Gamma}(x_0) \equiv  e^{ - \int_0^{x_0-\Gamma} dz R_{\Gamma}^-(z)  } 
= e^{ - \int_0^{x_0-\Gamma} dz \frac{ s'(z) }{ s(z+\Gamma)-s(z) }   }
  \label{Wmissing}
\end{eqnarray}

So the discussion depends on the three possibilities discussed around Eqs \ref{ExitRx0} and \ref{MFTPCriterion}

(N) When the boundary $x=0$ is never reached $s(0^+)=-\infty$ :
then the asymptotic behaviors
\begin{eqnarray}
 R_{\Gamma}^-(z)  && \opsimeq_{z \to 0^+}  - \frac{ s'(z) }{ s(z) }    = - \frac{d}{dz} \ln \vert s(z) \vert 
 \nonumber \\
 \int_{\epsilon}^{x_0-\Gamma} dz R_{\Gamma}^-(z) && \opsimeq_{\epsilon  \to 0^+} \ln \vert s(\epsilon) \vert = +\infty
  \label{WmissingN}
\end{eqnarray}
yields that the possible missing weight $W_{\Gamma}(x_0) $ of Eq. \ref{Wmissing}
\begin{eqnarray}
W_{\Gamma}(x_0) \equiv  e^{ - \int_0^{x_0-\Gamma} dz R_{\Gamma}^-(z)  } 
= 0
  \label{Wmissing0}
\end{eqnarray}
actually vanishes, i.e. the normalization of Eq. \ref{Pmoins0norma} is satisfied.

When $s(0^+)$ is finite, then the missing weight $W_{\Gamma}(x_0) $ of Eq. \ref{Wmissing} is finite,
but has a different interpretation in the two cases of Eq. \ref{MFTPCriterion} :

(R) If $s(0^+)$ is  finite and $\tau^-_{[0,x]}(x) $ finite, then the finite weight $W_{\Gamma}(x_0) $
represents the probability that the minimum is exactly at the boundary $x=0$ where the particle is reflected,
so that Eq. \ref{Pmoins0}
becomes
\begin{eqnarray}
P_{\Gamma}^-(x \vert x_0) 
= R_{\Gamma}^-(x) e^{ - \int_x^{x_0-\Gamma} dz R_{\Gamma}^-(z)  } \theta(0<x<x_0-\Gamma)
+ W_{\Gamma}(x_0) \delta(x)
  \label{Pmoins0R}
\end{eqnarray}

(I) If $s(0^+)$ is finite but $\tau^-_{[0,x]}(x) $ diverges, then the finite weight $W_{\Gamma}(x_0) $
represents the probability that the particle flows towards zero without any further drawup $\Gamma$.

%%%%%%%%%%%%%%%%%%%%%%%%%%%%%%%%%%%%%%%%%%%%%%%%%%

\subsection{ Densities $n_{\Gamma}^{\pm}$ of extrema in the presence of the boundary at $x=0$ }

The extrema densities of Eqs \ref{ngammackappa} are now defined on the domains
\begin{eqnarray}
n_{\Gamma}^{-}(x)&& 
=   \frac{ s'(x) [c - \kappa s(x+\Gamma)] }{[s(x+\Gamma)-s(x)]^2} \geq 0 \ \ \text{ for} \ \ x  \in ]0,+\infty[
\nonumber \\
n_{\Gamma}^{+}(x) &&
=  \frac{ s'(x) \left[ c - \kappa s(x-\Gamma)\right] }{[s(x)-s(x-\Gamma)]^2} \geq 0 \ \ \text{ for} \ \ x  \in ]\Gamma,+\infty[
  \label{ngammackappaHalf}
\end{eqnarray}
with the following properties :

(i) The density of maxima at near lowest position $x \to \Gamma^+$ 
is determined by the convergence or divergence of $s(0^+)$ of Eq. \ref{ExitRx0}
\begin{eqnarray}
n_{\Gamma}^{+}(x=\Gamma^+) &&
\opsimeq  \frac{ s'(\Gamma) \left[ c - \kappa s(0^+)\right] }{[s(\Gamma)-s(0^+)]^2}
= \begin{cases} 
0 \ \ & \text{if } \ s(0^+)=-\infty \ \ \ \text{ The boundary $x=0$ is never reached (Never) } \\
>0 & \text{if } \ s(0^+) > -\infty \ \ \ \text{Cases (I) and (R)}
\end{cases}
  \label{nplusmin}
\end{eqnarray}

(ii) The density of minima 
near the boundary $x=0^+$ behaves as
\begin{eqnarray}
n_{\Gamma}^{-}(x=0^+)&& 
=   \frac{ s'(0^+) [c - \kappa s(\Gamma)] }{[s(\Gamma)-s(0^+)]^2} 
\opsimeq_{x \to 0} 
\begin{cases} 
  \frac{s'(0^+)[c - \kappa s(\Gamma)] }{[s(0^+)]^2}   \ \ & \text{if } \ s(0^+)=-\infty \ \ \ \text{ The boundary $x=0$ is never reached (Never) } \\
 \frac{ s'(0^+) [c - \kappa s(\Gamma)] }{[s(\Gamma)-s(0^+)]^2}  & \text{if } \ s(0) > -\infty \ \ \ \text{Cases (I) and (R)}
\end{cases}
  \label{nmoinsmin}
\end{eqnarray}

(iii) For reflecting boundary conditions (R) at the origin, there is an additional 
finite density $n_{\Gamma}^{-}(x=0)  $ exactly at $x=0$ as discussed in the next subsection.

%%%%%%%%%%%%%%%%%%%%%%%%%%%%%%%%%%%%%%%%%%%%%%

\subsection{ Analysis of the steady Markov chain for reflecting boundary conditions (R) at the origin }

\label{subsec_Reflected}

For reflecting boundary conditions (R) at the origin with Eq. \ref{Pmoins0R},
a finite density $n_{\Gamma}^{-}(x=0)  $ exactly at $x=0$ appears 
and changes the Markov chain of Eq. \ref{MarkovChain}
as follows
\begin{eqnarray}
 n_{\Gamma}^{+}(x) && = \int_{0}^{x-\Gamma} dx_0 P_{\Gamma}^+(x \vert x_0) 
\left[ n_{\Gamma}^-(x_0)  + n_{\Gamma}^{-}(0) \delta(x_0) \right]
\nonumber \\
&& = \int_{0}^{x-\Gamma} dx_0 
R_{\Gamma}^+(x)   e^{ - \int_{x_0+\Gamma}^x dz R_{\Gamma}^+(z)  }
 n_{\Gamma}^-(x_0)  
+  R_{\Gamma}^+(x)   e^{ - \int_{\Gamma}^x dz R_{\Gamma}^+(z)  } n_{\Gamma}^{-}(0) 
\nonumber \\
 n_{\Gamma}^{-}(x)+ n_{\Gamma}^{-}(0) \delta(x) 
&&= \int_{x+\Gamma}^{+\infty} dx_0  P_{\Gamma}^-(x \vert x_0)  n_{\Gamma}^+(x_0)
\nonumber \\
&&= \int_{x+\Gamma}^{+\infty} dx_0  \left[R_{\Gamma}^-(x) e^{ - \int_x^{x_0-\Gamma} dz R_{\Gamma}^-(z)  }
+ W_{\Gamma}(x_0) \delta(x)
 \right]  n_{\Gamma}^+(x_0)
  \label{MarkovChainW}
\end{eqnarray}
The two terms in $\delta(x)$ of the second line determine the value 
of the finite density $n_{\Gamma}^{-}(x=0)  $ in terms of $W_{\Gamma}(x_0) $
of Eq. \ref{Wmissing}
and in terms of the maxima density $n_{\Gamma}^+(x_0) $ of Eq. \ref{ngammackappaHalf}
\begin{eqnarray}
n_{\Gamma}^{-}(x=0) && = \int_{\Gamma}^{+\infty} d x_0  W_{\Gamma}(x_0) n_{\Gamma}^+(x_0)
 = \int_{\Gamma}^{+\infty} d x_0  
e^{ - \int_0^{x_0-\Gamma} dz R_{\Gamma}^-(z)   }
n_{\Gamma}^+(x_0)
  \label{nm0Wmissingdef}
\end{eqnarray}

The changes in the Markov chain induce changes for the auxiliary functions 
$\Phi^{\pm}_{\Gamma}(.)$ that become
\begin{eqnarray}
\Phi_{\Gamma}^{+}(x) && \equiv  \frac{n_{\Gamma}^{+}(x)}{R_{\Gamma}^+(x)} 
= \int_{0}^{x-\Gamma} dx_0 
   e^{ - \int_{x_0+\Gamma}^x dz R_{\Gamma}^+(z)  }
 n_{\Gamma}^-(x_0)  
+    e^{ - \int_{\Gamma}^x dz R_{\Gamma}^+(z)  } n_{\Gamma}^{-}(0) \ \ \ \text{ for } \ \ x \in ]\Gamma,+\infty[
\nonumber \\
\Phi_{\Gamma}^{-}(x) && \equiv \frac{n_{\Gamma}^{-}(x)}{R_{\Gamma}^-(x)} 
= \int_{x+\Gamma}^{+\infty} dx_0   e^{ - \int_x^{x_0-\Gamma} dz R_{\Gamma}^-(z)  }n_{\Gamma}^+(x_0)
\ \ \ \text{ for } \ \ x \in ]0,+\infty[
  \label{PhiHalf}
\end{eqnarray}
with the following values at their lower positions
\begin{eqnarray}
\Phi_{\Gamma}^{+}(x=\Gamma) && =  n_{\Gamma}^{-}(0) 
\nonumber \\
\Phi_{\Gamma}^{-}(x=0) && 
= \int_{\Gamma}^{+\infty} dx_0   e^{ - \int_0^{x_0-\Gamma} dz R_{\Gamma}^-(z)  }n_{\Gamma}^+(x_0)
=n_{\Gamma}^{-}(0) 
  \label{PhiHalfmin}
\end{eqnarray}
The partial derivatives of Eqs \ref{PhiHalf} 
 with respect to $x$ are given by the same expression as in Eq. \ref{Phideri}
\begin{eqnarray}
\partial_x\Phi_{\Gamma}^{+}(x) && 
= n_{\Gamma}^-(x-\Gamma) - R_{\Gamma}^+(x) \Phi_{\Gamma}^{+}(x)  
=n_{\Gamma}^-(x-\Gamma) - n_{\Gamma}^+(x) 
\nonumber \\
\partial_x \Phi_{\Gamma}^{-}(x) && 
=  - n_{\Gamma}^+(x+\Gamma)+ R_{\Gamma}^-(x) \Phi_{\Gamma}^{-}(x)
= - n_{\Gamma}^+(x+\Gamma)+ n_{\Gamma}^-(x)
  \label{PhiHalfderi}
\end{eqnarray}
so that their constant difference of Eq. \ref{vanishinteg}
\begin{eqnarray}
\kappa =\Phi_{\Gamma}^{+}(x+\Gamma) - \Phi_{\Gamma}^{-}(x)   
  \label{vanishinteghalf}
\end{eqnarray}
can be computed at $x=0$ with Eq. \ref{PhiHalfmin}
to obtain that it vanishes
\begin{eqnarray}
\kappa =\Phi_{\Gamma}^{+}(\Gamma) - \Phi_{\Gamma}^{-}(0)   =0
  \label{vanishinteghalf0}
\end{eqnarray}
so that 
the extrema densities of Eqs \ref{ngammackappaHalf} reduce to
\begin{eqnarray}
n_{\Gamma}^{-}(x)&& 
= c  \frac{ s'(x)  }{[s(x+\Gamma)-s(x)]^2}  \ \ \text{ for} \ \ x  \in ]0,+\infty[
\nonumber \\
n_{\Gamma}^{+}(x) &&
= c \frac{ s'(x)  }{[s(x)-s(x-\Gamma)]^2}  \ \ \text{ for} \ \ x  \in ]\Gamma,+\infty[
  \label{ngammackappaHalfkappazero}
\end{eqnarray}

The finite density $n_{\Gamma}^{-}(x=0)  $ 
of Eq. \ref{nm0Wmissingdef}
can be then computed in terms of the maxima density $n_{\Gamma}^+(x_0) $ of Eq. \ref{ngammackappaHalfkappazero}
\begin{eqnarray}
n_{\Gamma}^{-}(x=0) &&  = \int_{\Gamma}^{+\infty} d x_0  
e^{ - \int_0^{x_0-\Gamma} dz R_{\Gamma}^-(z)   }
n_{\Gamma}^+(x_0)
= \int_{\Gamma}^{+\infty} d x_0  
e^{ - \int_0^{x_0-\Gamma} dz \frac{ s'(z) }{ s(z+\Gamma)-s(z) }   }
 c   \frac{ s'(x_0) }{[s(x_0)-s(x_0-\Gamma)]^2}
 \nonumber \\
&& =  \int_{\Gamma}^{+\infty} d x_0  
e^{ - \int_{\Gamma}^{x_0} dy \frac{ s'(y-\Gamma) }{ s(y)-s(y-\Gamma) }   }
 c   \frac{ s'(x_0) }{[s(x_0)-s(x_0-\Gamma)]^2}
  \nonumber \\
&& =  \int_{\Gamma}^{+\infty} d x_0  
e^{ - \int_{\Gamma}^{x_0} dy \left( \frac{ s'(y)}{ s(y)-s(y-\Gamma) } - \frac{ s'(y) - s'(y-\Gamma) }{ s(y)-s(y-\Gamma) }  \right) }
 c   \frac{ s'(x_0) }{[s(x_0)-s(x_0-\Gamma)]^2}
  \nonumber \\
&& =  \int_{\Gamma}^{+\infty} d x_0  
e^{ - \int_{\Gamma}^{x_0} dy  \frac{ s'(y)}{ s(y)-s(y-\Gamma) }  }
e^{   \ln \left(  s(x_0)-s(x_0-\Gamma)   \right) -  \ln \left(  s(\Gamma)-s(0)   \right)}
 c   \frac{ s'(x_0) }{[s(x_0)-s(x_0-\Gamma)]^2}
  \nonumber \\
&& = \frac{c}{s(\Gamma)-s(0)}  \int_{\Gamma}^{+\infty} d x_0  
e^{ - \int_{\Gamma}^{x_0} dy  \frac{ s'(y)}{ s(y)-s(y-\Gamma) }  }
    \frac{ s'(x_0) }{s(x_0)-s(x_0-\Gamma)}
    =\frac{c}{s(\Gamma)-s(0)}
  \label{nm0Wmissing}
\end{eqnarray}
which has thus a very simple expression in terms of the scale function.

When they are finite, the total density $n_{\Gamma}^{+} $ of maxima is given by the integral 
of Eq. \ref{ngammackappaHalfkappazero} on $x \in ]\Gamma,+\infty[$ 
\begin{eqnarray}
n_{\Gamma}^{+}=\int_{\Gamma}^{+\infty} dx n_{\Gamma}^{+}(x)   
=c  \int_{\Gamma}^{+\infty} dx \frac{ s'(x) }{[s(x)-s(x-\Gamma)]^2}
= c  \int_{0}^{+\infty} dy \frac{ s'(y+\Gamma) }{[s(y+\Gamma)-s(y)]^2}
  \label{npmsolfinalWall+}
\end{eqnarray}
while the total density $n_{\Gamma}^{-} $ of minima is given by the sum of the integral 
of the continuous part of Eq. \ref{ngammackappaHalfkappazero} on $x \in ]0,+\infty[$
and of the finite density  of Eq. \ref{nm0Wmissing}
\begin{eqnarray}
n_{\Gamma}^{-}=\int_{0}^{+\infty} dx n_{\Gamma}^{-}(x) + n_{\Gamma}^{-}(x=0)
= c \int_{0}^{+\infty} dx \frac{s'(x) }{[s(x+\Gamma)-s(x)]^2} + \frac{c}{s(\Gamma)-s(0)}
  \label{npmsolfinalWall-}
\end{eqnarray}
and they coincide as it should
\begin{eqnarray}
n_{\Gamma}^{+} -n_{\Gamma}^{-}
= c \int_{0}^{+\infty} dx \frac{s'(x+\Gamma) -s'(x) }{[s(x+\Gamma)-s(x)]^2} - \frac{c}{s(\Gamma)-s(0)}
= c \left[- \frac{1 }{s(x+\Gamma)-s(x)} \right]_0^{+\infty} - \frac{c}{s(\Gamma)-s(0)} =0
  \label{npmsolfinalWall+-}
\end{eqnarray}

%%%%%%%%%%%%%%%%%%%%%%%%%%%%%%%%%%%%%%%

\section{ Universality classes parametrized by $s'(x) =  x^{\gamma-1} $ on the half-line $x \in ]0,+\infty[$ } 

\label{sec_HalfLineExamples}

In this section we focus on the family of diffusion processes whose scale function derivative $s'(x)$ of Eq. \ref{Sx}
is a power-law
\begin{eqnarray}
s'(x) \equiv e^{U_I(x)}=  x^{\gamma-1} \ \ \ \text{ with } \ \ 
\begin{cases}
s(x) = \frac{x^{\gamma}}{\gamma} \ \ & \text{ for} \ \gamma \ne 0 \\
s(x) = \ln x \ \ & \text{ for} \ \gamma = 0
\end{cases}
  \label{spxpower}
\end{eqnarray}
The corresponding potential $U_I(x)$ is logarithmic
\begin{eqnarray}
 U_I(x)=  (\gamma-1) \ln x
  \label{UIlog}
\end{eqnarray}
while its derivative of Eq. \ref{U'}
\begin{eqnarray}
 U_I'(x) && = - \frac{F_I(x)}{D(x) } = \frac{ \gamma-1}{x}
\label{U'scale}
\end{eqnarray}
means that the corresponding models can be parametrized by their diffusion coefficient $D(x)$
and then the Ito force is determined by Eq. \ref{U'scale}
\begin{eqnarray}
 F_I(x) = (1-\gamma) \frac{ D(x)}{x} 
\label{FIscale}
\end{eqnarray}

%%%%%%%%%%%%%%%%%%%%%%%%%%%%%%%%%%%%%%%%%%%%%%

\subsection{ Analysis of the possible behaviors at the boundary $x=0$}

The possible behaviors at the boundary $x=0$ 
are determined as follows :

(i) Eq. \ref{spxpower}
yields that the behavior of $s(x)$ near the origin depends on the sign of $\gamma$
\begin{eqnarray}
s(x) \opsimeq_{x \to 0^+}
\begin{cases}
-\infty \ \ & \text{ for} \ \gamma \le 0 \ \ \ \text{ (Never reached) } \\
0 \ \ & \text{ for} \ \gamma > 0
\end{cases}
  \label{sxpower0}
\end{eqnarray}

(ii) When $\gamma>0$, one needs to consider how the diffusion coefficient $D(x)$ behaves near the origin
\begin{eqnarray}
D(x) \opsimeq_{x \to 0}  x^{\nu} \ \ \ \text{ with } \ \ \nu \geq 0 
  \label{Dxpower}
\end{eqnarray}
in order to analyze 
whether the Mean-First-Passage-Time $\tau^-_{[0,x]}(x) $ of Eq. \ref{MFTP} is finite or infinite
\begin{eqnarray}
\tau^-_{[0,x]}(x) 
&& = \int_0^{x}d y \frac{ s(y)-s(0) }{ D(y) s'(y) } \simeq \int_0^x d y \frac{ \frac{y^{\gamma}}{\gamma} }{ y^{\nu} y^{\gamma-1} }
\propto \int_0^x d y y^{1-\nu } = \left[ \frac{y^{2-\nu }}{2-\nu} \right]_0^x 
% \nonumber \\ && 
= \begin{cases} 
+ \infty   \text{ if $\nu \geq 2$ (Infinite-time) } \\
<+ \infty   \text{ if $ 0\leq \nu<2$  (Reflected)} 
\end{cases}
  \label{MFTPpower}
\end{eqnarray}

%%%%%%%%%%%%%%%%%%%%%%%%%%%%%

\subsection{ Examples with various diffusion coefficient $D(x)$ }

\subsubsection{ Geometric Brownian motion with the quadratic diffusion coefficient $D(x)=x^2$ and the linear Ito force $F_I(x)=(1-\gamma) x $} 

\label{subsec_GBM}

Let us consider the Geometric Brownian motion \cite{Peters,reviewGBM,c_carleman}
defined by the Ito SDE  
involving the quadratic diffusion coefficient $D(x)=x^2$ and the linear Ito force $F_I(x)=(1-\gamma) x $
\begin{eqnarray}
dx(t)  && =   (1-\gamma) x(t) dt + \sqrt{ 2} x(t) dB(t) \ \ \ \ \ \ \ \ \ \ \ \ \ \ 
\label{ItoGBM}
\end{eqnarray}
with the discussion of Eq. \ref{sxpower0} and \ref{MFTPpower} with $\nu=2$
\begin{eqnarray}
&& \text{ (Never reached ) for $\gamma \le 0$  }
\nonumber  \\
&& \text{  (Infinite-time) for $\gamma > 0$  }
  \label{GBM0}
\end{eqnarray}
To understand the physical meaning of these two behaviors, it is useful to recall 
that the logarithmic variable
\begin{eqnarray}
y(t) \equiv \ln x(t) 
\label{YlogGBM}
\end{eqnarray}
satisfies the SDE
\begin{eqnarray}
dy(t)  = - \gamma dt  +  \sqrt{ 2  } \ dB(t) 
\label{ItoSDEGBMY}
\end{eqnarray}
involving the constant drift $(- \gamma)$ and the constant diffusion coefficient unity.
The solution $y(t)$ is simply a Brownian motion with drift
\begin{eqnarray}
y(t)=  y(0)-\gamma  t+  \sqrt{ 2  } \ B(t)
\label{ItoSDEGBMYinteg}
\end{eqnarray}
and its exponential gives the following representation of the initial process $x(t)$ 
in terms of the noise $B(t)$
\begin{eqnarray}
x(t) \equiv e^{ y(t)} = x(0) e^{- \gamma t+  \sqrt{ 2  } \ B(t)}
\label{YlogGBMinverse}
\end{eqnarray}

At large time $t \to + \infty$, 
there is of course no steady state 
for the drifted Brownian motion $y(t)$ 
and for its exponential $x(t) \equiv e^{ y(t)} $,
but it is useful to discuss the various possibilities in terms of the sign of $\gamma$ :

$\bullet$ For $\gamma < 0$, then the drifted Brownian motion $y(t)$ with positive drift $(-\gamma)>0 $
 flows towards $(+\infty)$, 
and the Geometric-Brownian $x(t)=e^{y(t)}$ also flows towards 
$(+\infty)$, leading to the case (N) where the boundary at $x=0$ cannot be reached
 (as it corresponds to $y=-\infty$ for the drifted Brownian motion)

$\bullet$ For $\gamma = 0$, $y(t)= \ln x(t)$ is a Brownian motion without drift, still leading to the case (N)

$\bullet$ For $\gamma>0$, 
then the drifted Brownian motion $y(t)$ with negative drift $(-\gamma)<0 $ flows towards $(-\infty)$, 
and the Geometric Brownian motion $x(t)=e^{y(t)}$ flows towards $0$ but cannot reach it in finite time 
(as it corresponds to $y=-\infty$ for the drifted Brownian motion), corresponding to the case (I) of Eq. \ref{MFTPpower}

%%%%%%%%%%%%%%%%%%%%%%%%%%%%%%%%%%%%%%%%%%%

\subsubsection{  Bessel process with the constant diffusion coefficient $D(x)=1$ 
and the Ito force $F_I(x)=\frac{1-\gamma }{ x }=\frac{d-1}{ x }$} 

The Bessel process \cite{Bessel73,BesselQM,BesselYor} representing the norm of a d-dimensional Brownian motion 
satisfies
\begin{eqnarray}
dx(t)  && =  \frac{d-1}{x(t)}  dt + \sqrt{ 2}  dB(t) \ \ \ \ \ \ \ \ \ \ \ \ \ \ 
\text{with} \ \  \gamma=2-d
\label{Bessel}
\end{eqnarray}
with the discussion of Eq. \ref{sxpower0} and \ref{MFTPpower} with $\nu=0$
\begin{eqnarray}
&& \text{ (Never reached ) for $\gamma=2-d \le 0$ i.e. $d \geq 2$ }
\nonumber  \\
&& \text{  (Reflected) for $\gamma=2-d > 0$ i.e. $0< d < 2$ }
  \label{BesselEn0}
\end{eqnarray}
that corresponds to the well-known 
properties of recurrence to the origin in dimension $d<2$ and transience for $d > 2$.

%%%%%%%%%%%%%%%%%%%%%%%%%%%%%%%%%%%%%%%%%%%

\subsubsection{ Squared Bessel processes with the linear diffusion coefficient $D(x)=x$ and the constant force $F_I(x)=1-\gamma$} 

The Squared Bessel process \cite{BesselYor}
with the linear diffusion coefficient $D(x)=x$ and the constant force $F_I(x)=1-\gamma$
satisfies
\begin{eqnarray}
dx(t)  && =  \frac{d}{2} dt + \sqrt{ 2 x(t)}  dB(t) \ \ \ \ \ \ \ \ \ \ \ \ \ \ 
\text{with} \ \  \gamma=1-\frac{d}{2}
\label{SqBessel}
\end{eqnarray}
with the discussion of Eq. \ref{sxpower0} and \ref{MFTPpower} with $\nu=1$
\begin{eqnarray}
&& \text{ (Never reached ) for $\gamma=1-\frac{d}{2} \le 0$ i.e. $d \geq 2$ }
\nonumber  \\
&& \text{  (Reflected) for $\gamma=1-\frac{d}{2} > 0$ i.e. $0< d < 2$}
  \label{SqBesselEn0}
\end{eqnarray}
that corresponds again to the well-known 
properties of recurrence to the origin in dimension $d<2$ and transience for $d > 2$.

%%%%%%%%%%%%%%%%%%%%%%%%%%%%%%%%%%%%%%%%%%%

\subsubsection{ Fisher-Snedecor process with the quadratic diffusion coefficient $D(x)=x (x+1)$
and the linear Ito force $F_I(x)=(1-\gamma) (x+1)$}

The Fisher-Snedecor process \cite{pearson_fisher,pearson_fisherSnedecor}
that belongs to the Pearson family (see  \cite{pearson1895,pearson_wong,diaconis,autocorrelation,pearson_class,pearson2012,PearsonHeavyTailed,pearson2018,c_pearson}
and references therein)
with its quadratic diffusion coefficient $D(x)=x (x+1)$
and its linear Ito force $F_I(x)=(1-\gamma) (x+1)$
\begin{eqnarray}
dx(t)  && =  (1-\gamma) (x(t)+1) dt + \sqrt{ 2 x(t) [x(t)+1]}  dB(t) \ \ \ \ \ \ \ \ \ \ \ \ \ \ 
\label{FisherSn}
\end{eqnarray}
has a normalizable steady state for $\gamma \in ]0,1[$  (in contrast to the three previous examples)
\begin{eqnarray}
P_*(x) =  \frac{ 1 }{Z x^{\gamma} (1+x) }
 \ \ \ {\rm for } \ x \in ]0,+\infty[
 \ \ \ {\rm with } \ \ 0<\gamma<1\ \ 
\label{fisher}
\end{eqnarray}
(with the normalization $Z=\frac{\pi}{\sin(\pi \gamma)}$), 
with the special case for $\gamma=\frac{1}{2}$
\begin{eqnarray}
\gamma=\frac{1}{2} : \ \ \ P_*(x) =  \frac{ 1 }{ \pi \sqrt{x} (1+x) }
 \ \ \ {\rm for } \ x \in ]0,+\infty[
\label{fisherdemi}
\end{eqnarray}
 The discussion of Eq. \ref{sxpower0} and \ref{MFTPpower} with $ 0<\gamma<1$ and $\nu=1$ yields 
 the behavior at boundary  $x=0$ is always 
\begin{eqnarray}
 \text{  (Reflected) }
  \label{fisher-reflected}
\end{eqnarray}

%%%%%%%%%%%%%%%%%%%%%%%%%%%%%%%%%%%

\subsection{ Properties of the renormalized trajectories at scale $\Gamma$}

\subsubsection{  Bond probabilities $P_{\Gamma}^{\pm}(x \vert x_0) $  at scale $\Gamma$}

With the scale function of Eq. \ref{spxpower},
the functions $ R^{\pm}_{\Gamma}(x)$ of Eqs \ref{Rxplus}
and \ref{Rxmoins} governing the bond probabilities of Eqs \ref{Pplus} and \ref{Pmoins} read 
for $\gamma \ne 0$
\begin{eqnarray}
R^+_{\Gamma}(x) && =  \frac{ s'(x) }{ s(x)-s(x-\Gamma) } 
=   \frac{ \gamma x^{\gamma-1} }{ x^{\gamma}-(x-\Gamma)^{\gamma} } 
\nonumber \\
R^-_{\Gamma}(x) && =  \frac{ s'(x) }{ s(x+\Gamma)-s(x) } 
=  \frac{ \gamma x^{\gamma-1} }{(x+\Gamma)^{\gamma}-x^{\gamma} } 
  \label{RxScalepower}
\end{eqnarray}
and can be thus rewritten in scaling forms
\begin{eqnarray}
R^{\pm}_{\Gamma}(x)   
=   \frac{ 1 }{ \Gamma  } r^{\pm} \left( \frac{x}{\Gamma} \right)
  \label{RxScalepowerrescal}
\end{eqnarray}
As a consequence, the bond probabilities of Eqs \ref{Pplus} and \ref{Pmoins} 
can be rewritten in the scale-invariant forms
\begin{eqnarray}
P_{\Gamma}^{+}(x \vert x_0)  
&& = R_{\Gamma}^{+}(x) e^{ - \int_{x_0+\Gamma}^x dz R_{\Gamma}^+(z)  } 
= \frac{ 1 }{ \Gamma  } r^{+} \left( \frac{x}{\Gamma} \right) 
e^{ - \int_{x_0+\Gamma}^x \frac{ dz }{ \Gamma  } r^{+} \left( \frac{z}{\Gamma} \right)   } 
\nonumber \\
&& =  \frac{ 1 }{ \Gamma  } r^{+} \left( \frac{x}{\Gamma} \right) 
e^{ - \int_{\frac{x_0}{\Gamma}+1}^{\frac{x}{\Gamma}} dw r^{+} (w)   } 
\equiv  \frac{ 1 }{ \Gamma  } 
p^{+}\left( \frac{x}{\Gamma} \bigg\vert \frac{x_0}{\Gamma} \right)
  \label{Pplusrescal}
\end{eqnarray}
and
\begin{eqnarray}
P_{\Gamma}^-(x \vert x_0)  
&&= R_{\Gamma}^-(x) e^{ - \int_x^{x_0-\Gamma} dz R_{\Gamma}^-(z)  } 
=  \frac{ 1 }{ \Gamma  } r^{-} \left( \frac{x}{\Gamma} \right) e^{ - \int_x^{x_0-\Gamma} dz  \frac{ 1 }{ \Gamma  } r^{-} \left( \frac{z}{\Gamma} \right)  } 
\nonumber \\
&& = \frac{ 1 }{ \Gamma  } r^{-} \left( \frac{x}{\Gamma} \right) e^{ - \int_{\frac{x}{\Gamma}}^{\frac{x_0}{\Gamma}-1} dw r^{-} (w)  } \equiv  \frac{ 1 }{ \Gamma  } 
p^{-}\left( \frac{x}{\Gamma} \bigg\vert \frac{x_0}{\Gamma} \right)
  \label{Pmoinsrescal}
\end{eqnarray}
as expected since there is no intrinsic scale besides the RG scale $\Gamma$.

%%%%%%%%%%%%%%%%%%%%%%%%%%%%%%%%%%%

\subsubsection{ Densities $n_{\Gamma}^{\pm}(x) $ of extrema at scale $\Gamma$}

The densities of extrema of Eq. \ref{ngammackappaHalf}
read
\begin{eqnarray}
n_{\Gamma}^{-}(x)&& 
=   \frac{ s'(x) [c - \kappa s(x+\Gamma)] }{[s(x+\Gamma)-s(x)]^2} 
=  \frac{ \gamma^2 x^{\gamma-1}[c-\frac{\kappa}{\gamma} (x+\Gamma)^{\gamma} ]}{[(x+\Gamma)^{\gamma}-x^{\gamma}]^2} 
\geq 0 \ \ \text{ for} \ \ x  \in ]0,+\infty[
\nonumber \\
n_{\Gamma}^{+}(x) &&
=  \frac{ s'(x) \left[ c - \kappa s(x-\Gamma)\right] }{[s(x)-s(x-\Gamma)]^2} 
=   \frac{ \gamma^2 x^{\gamma-1}[c-\frac{\kappa}{\gamma} (x-\Gamma)^{\gamma}] }{[x^{\gamma}-(x-\Gamma)^{\gamma}]^2}
\geq 0 \ \ \text{ for} \ \ x  \in ]\Gamma,+\infty[
  \label{ngammackappaHalfpower}
\end{eqnarray}

The density of minima
near the boundary $x \to 0^+$ 
depends on the sign of $\gamma$ and the sign of $(\vert \gamma \vert -1) $
\begin{eqnarray}
n_{\Gamma}^{-}(x)&& \opsimeq_{x \to 0^+} 
 \frac{ \gamma^2 x^{\gamma-1}[c-\frac{\kappa}{\gamma} \Gamma^{\gamma} ]}{[\Gamma^{\gamma}-x^{\gamma}]^2}
 \opsimeq_{x \to 0} 
 \begin{cases}
c  \frac{ \gamma^2  [c-\frac{\kappa}{\gamma} \Gamma^{\gamma} ]}{\Gamma^{2\gamma}} x^{\gamma-1}
& \text { for  $\gamma>0$ } \\
     \gamma^2 x^{-\gamma-1} [c-\frac{\kappa}{\gamma} \Gamma^{\gamma} ]
     =   \gamma^2 [c-\frac{\kappa}{\gamma} \Gamma^{\gamma} ] x^{\vert \gamma \vert -1}   & \text { for  $\gamma<0$ } 
\end{cases}
  \label{npmscalepowernear0}
\end{eqnarray}

The density of maxima at its minimal position $x=\Gamma$ depends on the sign of $\gamma$
\begin{eqnarray}
n_{\Gamma}^{+}(x) &&  
\opsimeq_{x \to \Gamma^+}  \frac{ \gamma^2 \Gamma^{\gamma-1}[c-\frac{\kappa}{\gamma} (x-\Gamma)^{\gamma}] }{[\Gamma^{\gamma}-(x-\Gamma)^{\gamma}]^2}
\opsimeq_{x \to \Gamma^+}  \begin{cases} 
   \frac{ \gamma^2  }{\Gamma^{1+\gamma}} [c-\frac{\kappa}{\gamma} (x-\Gamma)^{\gamma}]
& \text {  for  $\gamma>0$ } \\
     \gamma^2 \Gamma^{\gamma-1}[c-\frac{\kappa}{\gamma} (x-\Gamma)^{\gamma}] (x-\Gamma)^{-2\gamma}  
  & \text { for  $\gamma<0$ } 
\end{cases}
  \label{npmscalepowermin}
\end{eqnarray}
and where the remaining discussion for the factor $[c-\frac{\kappa}{\gamma} (x-\Gamma)^{\gamma}] $ depends whether one of the two parameters $(c,\kappa)$ vanishes or not for the process one is interested in.

%%%%%%%%%%%%%%%%%%%%%%%%%%%%%%%%%%%

\subsection{ Simple special values of $\gamma$  as representative examples of the various boundary behaviors}

%%%%%%%%%%%%%%%%%%%%%%%%%%%%%%%%%%%%%%%%

\subsubsection{ Case $\gamma=1$ : Brownian motion reflected at $0$ (Bessel $d=1$) }

\label{subsec_BrownianReflected}

For the Brownian motion reflected at $0$, the 
probability distribution of ascending bonds is the same as for the unconstrained Brownian motion  
\begin{eqnarray}
P_{\Gamma}^+(x \vert x_0)  
= \frac{ 1}{ \Gamma } e^{ -  \frac{x-(x_0+\Gamma)}{\Gamma}  } 
\ \ \ \ \text{ for } \ \ x \in [x_0+\Gamma,+\infty[
  \label{Pplus+1}
\end{eqnarray}
while the probability distribution of descending bonds is of the form of Eq. \ref{Pmoins0R}
\begin{eqnarray}
P_{\Gamma}^-(x \vert x_0) 
=\frac{ 1}{ \Gamma } e^{ -  \frac{x_0-\Gamma-x}{\Gamma}  } 
\theta(0<x<x_0-\Gamma)
+ W_{\Gamma}(x_0) \delta(x)
  \label{Pmoins+1}
\end{eqnarray}
where the weight $W_{\Gamma}(x_0) $ discussed
 in Eqs \ref{Pmoins0norma} and \ref{Wmissing}
\begin{eqnarray}
W_{\Gamma}(x_0) 
=  e^{ - \int_0^{x_0-\Gamma} dz R_{\Gamma}^-(z)  }  =
e^{ -  \frac{x_0-\Gamma}{\Gamma}  } 
  \label{Wmissing+1}
\end{eqnarray}
represents the finite probability to find a minimum on the boundary $x=0$ at scale $\Gamma$.

As discussed in subsection \ref{subsec_Reflected},
the densities of extrema are given by Eq \ref{ngammackappaHalfkappazero} 
\begin{eqnarray}
n_{\Gamma}^{+}(x) &&  
=   \frac{ c }{\Gamma^2} \ \ \text{ for $x \in ]\Gamma,+\infty[$ }
\nonumber \\
n_{\Gamma}^{-}(x)&& 
=    \frac{ c }{\Gamma^2} \ \ \text{ for $x \in ]0,+\infty[$ }
  \label{npmscalepower+1}
\end{eqnarray}
and are independent of $x$ on their domains as expected,
while the additional finite density of minima at the boundary $x=0$ of Eq. \ref{nm0Wmissing} reads
\begin{eqnarray}
n_{\Gamma}^{-}(x=0) && = \frac{c}{s(\Gamma)-s(0)} = \frac{c  }{\Gamma} 
  \label{nm0Wmissingpower+1}
\end{eqnarray}

\subsubsection{ Case $\gamma=-1$ (Never Reached) : GBM, Bessel in $d=3$, Squared-Bessel in $d=4$ }

\label{subsec_gamma-1}

For $\gamma=-1$ with the scale function
\begin{eqnarray}
s(x) && = - \frac{1}{x}
\nonumber \\
s'(x) && = \frac{1}{x^2}
  \label{gamma-1}
\end{eqnarray}
the two functions of Eq. \ref{RxScalepower} reduce to
\begin{eqnarray}
R^+_{\Gamma}(x) && =  \frac{1}{\Gamma}-  \frac{1}{x}
\nonumber \\
R^-_{\Gamma}(x) && = \frac{1}{\Gamma} +  \frac{1}{x}
  \label{RxScalepower-1}
\end{eqnarray}
i.e. the Brownian result $ \frac{1}{\Gamma}$ is corrected by $\left( \mp \frac{1}{x} \right)$.
The bond distributions of Eq. \ref{Pplus}  
\begin{eqnarray}
P_{\Gamma}^+(x \vert x_0) 
= \left(\frac{1}{\Gamma}-  \frac{1}{x} \right) e^{ - \int_{x_0+\Gamma}^x dz \left(\frac{1}{\Gamma}-  \frac{1}{z} \right)  } 
= \frac{ x-\Gamma}{ \Gamma (x_0+\Gamma)} e^{ -  \frac{x-(x_0+\Gamma)}{\Gamma}  } 
\ \ \ \ \text{ for } \ \ x \in [x_0+\Gamma,+\infty[
  \label{Pplus-1}
\end{eqnarray}
and \ref{Pmoins} 
\begin{eqnarray}
P_{\Gamma}^-(x \vert x_0) 
= \left(\frac{1}{\Gamma}+  \frac{1}{x} \right) e^{ - \int_x^{x_0-\Gamma} dz \left(\frac{1}{\Gamma}+  \frac{1}{z} \right)  }
= \frac{ x+\Gamma}{ \Gamma (x_0-\Gamma)} e^{ -  \frac{x_0-\Gamma-x}{\Gamma}  } 
 \ \ \ \ \text{ for } \ \  x \in ]0,x_0-\Gamma]
  \label{Pmoins-1}
\end{eqnarray}
are tilted with respect to the Brownian distributions via the rational prefactors 
$\left( \frac{ x\mp\Gamma}{  x_0\pm\Gamma}\right)$.

For $(c=0,\kappa>0)$ that is appropriate for the Geometric Brownian motion
in relation with the case of the drifted Brownian motion, 
the densities of extrema of Eq. \ref{ngammackappaHalfpower}
read
\begin{eqnarray}
n_{\Gamma}^{-}(x)&& 
= \kappa \frac{  x+\Gamma }{\Gamma^2} 
 \ \ \text{ for} \ \ x  \in ]0,+\infty[
\nonumber \\
n_{\Gamma}^{+}(x) &&
= \kappa  \frac{  x-\Gamma }{\Gamma^2}
 \ \ \text{ for} \ \ x  \in ]\Gamma,+\infty[
  \label{ngammackappaHalfpower-1}
\end{eqnarray}

%%%%%%%%%%%%%%%%%%%%%%%%%%%%%%%%%%%%%%%%

\subsubsection{ Case $\gamma=2$ : Geometric Brownian motion in the case (Infinite time)}

\label{subsec_gamma2}

For $\gamma=2$ with the scale function
\begin{eqnarray}
s(x) = \frac{x^2}{2}
  \label{gamma+2}
\end{eqnarray}
the two functions of Eq. \ref{RxScalepower} reduce to
\begin{eqnarray}
R^+_{\Gamma}(x) && =  \frac{1}{\Gamma} \left( 1+ \frac{ \Gamma}{2x-\Gamma} \right)
\nonumber \\
R^-_{\Gamma}(x) && =   \frac{1}{\Gamma} \left( 1- \frac{ \Gamma}{2x+\Gamma} \right)
  \label{RxScalepower+2}
\end{eqnarray}
The bond distributions of Eqs \ref{Pplus}  \ref{Pmoins} 
\begin{eqnarray}
P_{\Gamma}^+(x \vert x_0) 
&& = \frac{1}{\Gamma} \left( 1+ \frac{ \Gamma}{2x-\Gamma} \right) 
e^{ - \int_{x_0+\Gamma}^x dz \frac{1}{\Gamma} \left( 1+ \frac{ \Gamma}{2z-\Gamma} \right)   } 
= \frac{1}{\Gamma} \left( 1+ \frac{ \Gamma}{2x-\Gamma} \right)   \sqrt{ \frac{2 x_0+\Gamma}{2 x -\Gamma} }
e^{ -  \frac{x-(x_0+\Gamma)}{\Gamma}  } 
 \ \text{ for }  x \in [x_0+\Gamma,+\infty[
\nonumber \\
P_{\Gamma}^-(x \vert x_0) 
&& =  \frac{1}{\Gamma} \left( 1- \frac{ \Gamma}{2x+\Gamma} \right) 
e^{ - \int_x^{x_0-\Gamma} dz  \frac{1}{\Gamma} \left( 1- \frac{ \Gamma}{2z+\Gamma} \right)   }
=\frac{1}{\Gamma} \left( 1- \frac{ \Gamma}{2x+\Gamma} \right) \sqrt{ \frac{2 x_0-\Gamma}{2 x +\Gamma} }
e^{ -  \frac{x_0-\Gamma-x}{\Gamma}  } 
  \text{ for }  x \in ]0,x_0-\Gamma]
  \label{Pmoins+2}
\end{eqnarray}
are again tilted with respect to the Brownian distributions via prefactors,
while the descending bonds also have the missing normalization discussed
 in Eqs \ref{Pmoins0norma} and \ref{Wmissing}
\begin{eqnarray}
W_{\Gamma}(x_0) = 1-\int_0^{x_0-\Gamma}  P_{\Gamma}^-(x \vert x_0) 
=  e^{ - \int_0^{x_0-\Gamma} dz R_{\Gamma}^-(z)  } =e^{ - \int_0^{x_0-\Gamma} dz \frac{1}{\Gamma} \left( 1- \frac{ \Gamma}{2z+\Gamma} \right)  }
 =\sqrt{ \frac{2 x_0-\Gamma}{\Gamma} }
e^{ -  \frac{x_0-\Gamma}{\Gamma}  } 
  \label{Wmissing+2}
\end{eqnarray}
but the interpretation is different from the previous case of Eq. \ref{Wmissing+1} with reflecting at the boundary :
for the Geometric Brownian motion with $D(x)=x^2$, the boundary is of the type (Infinite time),
 $W_{\Gamma}(x_0) $ of Eq. \ref{Wmissing+2} then represents the probability 
that $x_0$ is the last maximum of the renormalized trajectory at scale $\Gamma$,
i.e. after the maximum $x_0$, the particle flows to zero without any drawup $\Gamma$.

For $(c=0,\kappa<0)$ that is appropriate for the Geometric Brownian motion
in relation with the case of the drifted Brownian motion, 
the densities of extrema of Eq. \ref{ngammackappaHalfpower}
read
\begin{eqnarray}
n_{\Gamma}^{-}(x)&& 
= (-\kappa) \frac{ 2 x (x+\Gamma)^2 }{ \Gamma^2 (2x+\Gamma)^2} 
\ \ \text{ for} \ \ x  \in ]0,+\infty[
\nonumber \\
n_{\Gamma}^{+}(x) &&
=  (- \kappa) \frac{ 2 x (x-\Gamma)^2 }{ \Gamma^2 (2x-\Gamma)^2} 
\ \ \text{ for} \ \ x  \in ]\Gamma,+\infty[
  \label{ngammackappaHalfpower+2}
\end{eqnarray}

%%%%%%%%%%%%%%%%%%%%%%%%%%%%%%%%%%%%%%%%

\subsubsection{ Case $\gamma=0$ : Bessel and Squared-Bessel in $d=2$ (Never reached)}

For $\gamma=0$, the logarithmic scale function $s(x)=\ln x$ of Eq. \ref{spxpower}
produces logarithms in the two functions
\begin{eqnarray}
R^+_{\Gamma}(x) &&  
=   \frac{ 1 }{ x \ln \left( \frac{x}{x-\Gamma} \right) } 
\nonumber \\
R^-_{\Gamma}(x) && 
=  \frac{ 1 }{ x \ln \left( \frac{x+\Gamma}{x} \right) } 
  \label{RxScalepower0}
\end{eqnarray}
as well as in the densities of extrema of Eq. \ref{ngammackappa0}
\begin{eqnarray}
n_{\Gamma}^{+}(x) &&  
=c   \frac{ 1 }{ x \ln^2 \left( \frac{x}{x-\Gamma} \right) } 
 \ \ \text{ for $x \in ]\Gamma,+\infty[$ }
\nonumber \\
n_{\Gamma}^{-}(x)&& 
= c   \frac{ 1 }{x \ln^2 \left( \frac{x+\Gamma}{x} \right) }
 \ \ \text{ for $x \in ]0,+\infty[$ }
  \label{npmscalepower0}
\end{eqnarray}
The density of maxima vanishes at its left position $x \to \Gamma^+$ as
\begin{eqnarray}
n_{\Gamma}^{+}(x) &&  
\opsimeq_{ x \to \Gamma^+ } c   \frac{ 1 }{ \Gamma \ln^2 \left( \frac{\Gamma}{x-\Gamma} \right) } 
  \label{npmscalepower0+min}
\end{eqnarray}
while the density of minima diverges near the boundary $x \to 0^+$ while remaining integrable
\begin{eqnarray}
n_{\Gamma}^{-}(x)&& 
\opsimeq_{ x \to 0^+ } c   \frac{ 1 }{x \ln^2 \left( \frac{\Gamma}{x} \right) }
  \label{npmscalepower0-min}
\end{eqnarray}
Far from the boundary $x \to + \infty$, the two densities behave as
\begin{eqnarray}
n_{\Gamma}^{\pm}(x) &&  
\opsimeq_{ x \to +\infty} 
c   \frac{ x }{\Gamma^2}
  \label{npmscalepowerxlarge0}
\end{eqnarray}

%%%%%%%%%%%%%%%%%%%%%%%%%%%%%%%%%%%%%%%%

\section{ Conclusions}

\label{sec_conclusion}

For one-dimensional diffusion processes with arbitrary diffusion coefficient $D(x)$ and Ito force $F_I(x)$, we have analyzed  the statistical properties of the renormalized trajectories at scale $\Gamma$, where one keeps only the alternating sequence of local minima and local maxima that are separated by drawups and drawdowns of amplitude bigger than $\Gamma$, both via direct probabilistic methods and via real-space-RG flows.
The main result is that the universality classes are determined by the scale function $s(x)$ satisfying $D(x) s''(x)+F_I(x) s'(x)=0$ that governs the exit probabilities from an interval : diffusions sharing the same scale function $s(x)$ have the same statistics of renormalized trajectories.  The simplest universality classes for diffusions on the whole line $x \in ]-\infty,+\infty[$ and on the half-line $x \in ]0,+\infty[$ have been illustrated with several examples of processes, with or without steady states: Cauchy, Student, Bessel, Fisher-Snedecor, Geometric Brownian. This framework concerning diffusion processes is then adapted to discrete-time and to continuous-time random walks on the one-dimensional lattice with arbitrary nearest-neighbor jump probabilities or rates in the last two appendices.

%%%%%%%%%%%%%%%%%%%%%%%%%%%%%%%%%

\appendix

\section{ Distribution $\pi^+_{[x_L,x_R]} (t \vert x) $ of the boundary-exit-time and its Laplace transform ${\hat \pi}^{[p]+}_{[x_L,x_R]} ( x) $}

\label{app_ExitTime}

\subsection{ Equations satisfied by the probability distribution $\pi^+_{[x_L,x_R]} (t \vert x)  $ and by 
its Laplace transform ${\hat \pi}^{[p]+}_{[x_L,x_R]} ( x) $ }

The exit probability $E^+_{[x_L,x_R]}$ discussed in Eq. \ref{ExitRx}
can be decomposed
\begin{eqnarray}
E^+_{[x_L,x_R]}(x) = \int_0^{+\infty} dt \pi^+_{[x_L,x_R]} (t \vert x)
  \label{ExitRxt}
\end{eqnarray}
in terms of the probability $\pi^+_{[x_L,x_R]} (t \vert x)$ of the first-passage at the right boundary $x_R$ at time $t$ without having visited the other boundary $x_L$ when starting at $x$,
whose dynamics is governed by the adjoint operator ${\cal F}^{\dagger} $ \cite{gardiner,vankampen,risken}
 \begin{eqnarray}
\partial_t \pi^+_{[x_L,x_R]} (t \vert x)= {\cal F}^{\dagger} \pi^+_{[x_L,x_R]} (t \vert x)
\label{adjointzeroExitt}
\end{eqnarray}
while the two boundary conditions at $x=x_R$ and $x=x_L$ read
 \begin{eqnarray}
 \pi^+_{[x_L,x_R]} (t \vert x=x_R) && =\delta(t)
 \nonumber \\
 \pi^+_{[x_L,x_R]} (t \vert x=x_L) && =0
\label{BCExitt}
\end{eqnarray}

The Laplace transform
\begin{eqnarray}
{\hat \pi}^{[p]+}_{[x_L,x_R]} ( x) \equiv \int_0^{+\infty} dt e^{-p t}  \pi^+_{[x_L,x_R]} (t \vert x) 
  \label{Laplace}
\end{eqnarray}
 then satisfies
 \begin{eqnarray}
{\cal F}^{\dagger}  {\hat \pi}^{[p]+}_{[x_L,x_R]} ( x) && 
=   \int_0^{+\infty} dt e^{-p t} {\cal F}^{\dagger} \pi^+_{[x_L,x_R]} (t \vert x)
=   \int_0^{+\infty} dt e^{-p t} \partial_t  \pi^+_{[x_L,x_R]} (t \vert x)  
\nonumber \\
&& = -  \int_0^{+\infty} dt   \pi^+_{[x_L,x_R]} (t \vert x)  \partial_t  e^{-p t} = p \int_0^{+\infty} dt   \pi^+_{[x_L,x_R]} (t \vert x)    e^{-p t} = p {\hat \pi}^{[p]+}_{[x_L,x_R]} ( x)
  \label{LaplaceEqadjoint}
\end{eqnarray}
with the boundary conditions obtained from Eq. \ref{BCExitt}
 \begin{eqnarray}
{\hat \pi}^{[p]+}_{[x_L,x_R]} ( x=x_R) && = 1
 \nonumber \\
 {\hat \pi}^{[p]+}_{[x_L,x_R]} ( x=x_L) && =0
\label{BCLaplace}
\end{eqnarray}

%%%%%%%%%%%%%%%%%%%%%%%%%%%%%%%%%%%%%%%

\subsection{ Solution for the Laplace transform  ${\hat \pi}^{[p]+}_{[x_L,x_R]} ( x) $ }

 The second-order differential equation of Eq. \ref{LaplaceEqadjoint} for a function $g(x)$ 
 involves the adjoint
operator that can be written either with the expanded form of Eq. \ref{adjointIto}
\begin{eqnarray}
p g(x) && ={\cal F}^{\dagger} g(x)   =    F_I(x) g'(x)  + D(x) g''(x)
\label{LaplaceEqadjointExpanded}
\end{eqnarray}
or with the fully-factorized form of Eqs \ref{FPgeneratorprodadjointfactorized} and \ref{FPgeneratorprodadjointmath}
 \begin{eqnarray}
p g(x) && ={\cal F}^{\dagger} g(x) = e^{ U(x) }   \frac{d}{dx} D(x) e^{- U(x) }   \frac{d}{dx} g(x)
=  \frac{ 1}{m'(x) }  \frac{d}{dx}  \frac{ 1}{s'(x) }  \frac{d}{dx} g(x)
\nonumber \\
i.e. \ \ \ p m'(x) g(x) && = \frac{d}{dx} \left( \frac{ g'(x)}{s'(x) } \right)= \left( \frac{ g'(x)}{s'(x) } \right)'
  \label{LaplaceEqadjointfactorized}
\end{eqnarray}

Let us introduce two independent solutions $g^{[p]}_{1,2}(x)$ of this equation,
that shows that the appropriate wronskian that is conserved is
 \begin{eqnarray}
W^{[p]}(x) &&\equiv  \frac{ g^{[p]}_{1} (x) g^{[p]'}_2 (x) -  g^{[p]'}_1 (x) g^{[p]}_{2}(x)}{s'(x) } 
\nonumber \\
W^{[p]'}(x) && = g^{[p]'}_1 (x)\left( \frac{ g^{[p]'}_{2}}{s'(x) } \right) +  g^{[p]}_1 (x)\left( \frac{ g^{[p]'}_{2}}{s'(x) } \right)'
-\left( \frac{ g^{[p]'}_{1}}{s'(x) } \right)'g^{[p]}_2 (x) - \left( \frac{ g^{[p]'}_{1}}{s'(x) } \right)g^{[p]'}_2 (x)
\nonumber \\
&& =  g^{[p]}_1 (x)\left( \frac{ g^{[p]'}_{2}}{s'(x) } \right)'
-\left( \frac{ g^{[p]'}_{1}}{s'(x) } \right)'g^{[p]}_2 (x)
=  g^{[p]}_1 (x)  p m'(x) g^{[p]}_{2}(x)
-  p m'(x) g^{[p]}_{1}(x) g^{[p]}_2 (x) =0
  \label{Wronskian}
\end{eqnarray}

The Laplace transform $ {\hat \pi}^{[p]+}_{[x_L,x_R]} ( x) $  can be written as a linear combination of the two independent solutions $g^{[p]}_{1,2}(x) $
 \begin{eqnarray}
 {\hat \pi}^{[p]+}_{[x_L,x_R]} ( x) = c_1 g^{[p]}_1(x) + c_2 g^{[p]}_2(x)
  \label{CLLaplace}
\end{eqnarray}
where the constants $c_{1,2}$ are determined by the boundary conditions of Eq. \ref{BCLaplace}
 \begin{eqnarray}
1&& ={\hat \pi}^{[p]+}_{[x_L,x_R]} ( x=x_R) =c_1 g^{[p]}_1(x_R) + c_2 g^{[p]}_2(x_R)
 \nonumber \\
0 && = {\hat \pi}^{[p]+}_{[x_L,x_R]} ( x=x_L)  =c_1 g^{[p]}_1(x_L) + c_2 g^{[p]}_2(x_L)
\label{BCLaplacec}
\end{eqnarray}

As a consequence, it is useful to introduce the kernel $K^{[p]}(a,b) $ of two variables $ (a,b)$
 that involves the two functions $g^{[p]}_{1,2} $ and their constant Wronskian 
 $W^{[p]}(x)=W^{[p]} $ of Eq. \ref{Wronskian}
 \begin{eqnarray}
K^{[p]}(a,b) \equiv \frac{ g^{[p]}_1(a)g^{[p]}_2(b)- g^{[p]}_2(a)g^{[p]}_1(b) }{ W^{[p]} }
\label{kab}
\end{eqnarray}
in order to write the final solution of Eq. \ref{CLLaplace} as the ratio
 \begin{eqnarray}
 {\hat \pi}^{[p]+}_{[x_L,x_R]} ( x) = c_1 g^{[p]}_1(x) + c_2 g^{[p]}_2(x)
 = \frac{ K^{[p]}(x_L,x)  }{ K^{[p]}(x_L,x_R) } 
  \label{CLLaplace+}
\end{eqnarray}

%%%%%%%%%%%%%%%%%%%%%%%%%%%%%%%%%%%%%%%

\subsection{ Some useful properties of the kernel $K^{[p]}(a,b) $ of Eq. \ref{kab} }

The constant wronskian $W^{[p]} $ disappears in the ratio giving the solution of Eq. \ref{CLLaplace+},
but this normalization of Eq. \ref{kab}
is nevertheless useful when one considers
the partial derivatives of Eq. \ref{kab}
with respect to $a$ and with respect to $b$
 \begin{eqnarray}
 \partial_a K^{[p]}(a,b) && = \frac{ g^{[p]'}_1(a)g^{[p]}_2(b)- g^{[p]'}_2(a)g^{[p]}_1(b) }{W^{[p]}}
 \nonumber \\
\partial_b K^{[p]}(a,b) && = \frac{ g^{[p]}_1(a)g^{[p]'}_2(b)- g^{[p]}_2(a)g^{[p]'}_1(b) }{W^{[p]}}
\label{kabderib}
\end{eqnarray}
and then evaluate it at coinciding points using the constant wronskian $W^{[p]}(x)=W^{[p]} $ of Eq. \ref{Wronskian}
 \begin{eqnarray}
 \partial_a K^{[p]}(a,b) \vert_{b=a}&& = \frac{ g^{[p]'}_1(a)g^{[p]}_2(a)- g^{[p]'}_2(a)g^{[p]}_1(a) }{W^{[p]}}=-s'(a)
 \nonumber \\
\partial_b K^{[p]}(a,b) \vert_{a=b} && = \frac{g^{[p]}_1(b)g^{[p]'}_2(b)- g^{[p]}_2(b)g^{[p]'}_1(b) }{W^{[p]}}= s'(b) 
\label{kabderibcoinciding}
\end{eqnarray}
to obtain that they do not depend on $p$, so that the limit $p \to 0$ goes smoothly to
\begin{eqnarray}
 K^{[p=0]}(a,b) && = s(b)-s(a)
\label{kabpzero}
\end{eqnarray}
that appears in the exit probabilities of Eq. \ref{ExitRx}
of the main text.

For later purposes, it is also useful to compute the cross-derivative
 \begin{eqnarray}
 \partial_a \partial_b K^{[p]}(a,b) && = \frac{ g^{[p]'}_1(a)g^{[p]'}_2(b)- g^{[p]'}_2(a)g^{[p]'}_1(b) }{W^{[p]}}
\label{kabderibcross}
\end{eqnarray}
as well as 
 \begin{eqnarray}
&& [W^{[p]}]^2  \bigg(K^{[p]}(a,b)  \partial_a \partial_b K^{[p]}(a,b)
-  [\partial_a K^{[p]}(a,b) ] [ \partial_b K^{[p]}(a,b) ]
    \bigg)
    \nonumber \\
 && = \bigg(g^{[p]}_1(a)g^{[p]}_2(b)- g^{[p]}_2(a)g^{[p]}_1(b) \bigg) 
 \bigg(g^{[p]'}_1(a)g^{[p]'}_2(b)- g^{[p]'}_2(a)g^{[p]'}_1(b)  \bigg)
\nonumber \\
 && - \bigg(  g^{[p]'}_1(a)g^{[p]}_2(b)- g^{[p]'}_2(a)g^{[p]}_1(b)  \bigg)
 \bigg( g^{[p]}_1(a)g^{[p]'}_2(b)- g^{[p]}_2(a)g^{[p]'}_1(b)  \bigg)
 \nonumber \\
 && = 
 \left[ g^{[p]}_1(a)g^{[p]'}_2(a) -g^{[p]'}_1(a)g^{[p]}_2(a) \right] \left[g^{[p]}_1(b) g^{[p]'}_2(b)- g^{[p]'}_1(b)g^{[p]}_2(b) \right]
 =  [W^{[p]}]^2 s'(a) s'(b)
\label{kabcrossdiffcalcul}
\end{eqnarray}
where one recognizes the Wronskian of Eq. \ref{Wronskian}
 at position $a$ and at position $b$ to simplify the last line,
 so that the final result that will be useful in the next Appendix is
  \begin{eqnarray}
  K^{[p]}(a,b)  \partial_a \partial_b K^{[p]}(a,b)
-  [\partial_a K^{[p]}(a,b) ] [ \partial_b K^{[p]}(a,b) ]
  && =  s'(a) s'(b)
\label{kabcrossdiff}
\end{eqnarray}

%%%%%%%%%%%%%%%%%%%%%%%%%%%%%%%%%%%%%%%

\subsection{ Perturbative expansion of the kernel $K^{[p]}(a,x) =K_0(a,x) +p K_1(a,x)+p^2 K_2(a,x) +...$  }

For $b=x$, the kernel of Eq. \ref{kab} 
 \begin{eqnarray}
K^{[p]}(a,x) \equiv \frac{ g^{[p]}_1(a)g^{[p]}_2(x)- g^{[p]}_2(a)g^{[p]}_1(x) }{ W^{[p]} }
\label{kax}
\end{eqnarray}
is a solution of the differential equation of Eq. \ref{LaplaceEqadjointfactorized}
 \begin{eqnarray}
 p m'(x) K^{[p]}(a,x) && =  \left( \frac{ \partial_x K^{[p]}(a,x)}{s'(x) } \right)'
  \label{LaplaceEqadjointfactorizedx}
\end{eqnarray}
where one can plug the perturbative expansion in $p$
\begin{eqnarray}
 K^{[p]}(a,x) =K_0(a,x) +p K_1(a,x)+p^2 K_2(a,x) +...+p^n K_n(a,x)
  \label{Kper}
\end{eqnarray}
around the known solution of Eq. \ref{kabpzero}
for $p=0$
\begin{eqnarray}
K_0(a,x)= K^{[p=0]}(a,x) && = s(x)-s(a)
\label{kabpzero0}
\end{eqnarray}
to obtain the simple recurrence between the two consecutive orders $(n-1)$ and $n$
 \begin{eqnarray}
\left( \frac{ \partial_x K_n(a,x)}{s'(x) } \right)' =  m'(x) K_{n-1}(a,x)   
  \label{Reckn}
\end{eqnarray}
For any $p$, the vanishing at coinciding point $K^{[p]}(a,a) =0$ 
and the partial derivative at coinciding point $\partial_a K^{[p]}(a,b) \vert_{b=a}=-s'(a)$ of Eq. \ref{kabderibcoinciding} 
yields the following conditions at coinciding points for any $n\geq1$
\begin{eqnarray}
K_n(a,a) && =0
\nonumber \\
 \partial_x K_n(a,x) \vert_{x=a} && =0
  \label{coinciding}
\end{eqnarray}
that are useful to integrate Eq. \ref{Reckn} 
 \begin{eqnarray}
 \frac{ \partial_x K_n(a,x)}{s'(x) }  && = \int_a^x dy  m'(y) K_{n-1}(a,y)   
 \nonumber \\
 K_n(a,b)  && = \int_a^b dx s'(x)\int_a^x dy  m'(y) K_{n-1}(a,y) 
  \label{Reckninteg}
\end{eqnarray}
In particular, the first correction $n=1$ computed from
$K_0(a,x) $ of Eq. \ref{kabpzero0}
reads in terms of the scale function $s(.)$ and speed-measure $m'(.)$
\begin{eqnarray}
 K_1(a,b)  && = \int_a^b dx s'(x)\int_a^x dy  m'(y) \left[ s(y)-s(a)\right]
 = \int_a^b dy  m'(y) \left[ s(y)-s(a)\right] \int_y^b dx s'(x)
 \nonumber \\
 && = \int_a^b dy  m'(y) \left[ s(y)-s(a)\right] \left[ s(b)-s(y)\right] 
  \label{K1ab}
\end{eqnarray}
with its partial derivative  
\begin{eqnarray}
 \partial_b K_1(a,b)  
  = 0  +  s'(b) \int_a^b dy  m'(y) \left[ s(y)-s(a)\right] 
  \label{K1abderib}
\end{eqnarray}
that will be useful later.

%%%%%%%%%%%%%%%%%%%%%%%%%%%%%%%%%%%%%%%

\section{ Direct computation of the probabilities $P_{\Gamma}^{\pm}(x,t \vert x_0) $ that a renormalized bond 
at scale $\Gamma$ starting at $x_0$ ends at the position $x$ after the duration $t$ }

\subsection{ Decomposition of the duration $t=t_1+t_2$ of an ascending bond into two times }

In this Appendix, one wishes to decompose
the probability $P_{\Gamma}^+(x \vert x_0) $ of an ascending bond computed in Eq. \ref{Pplus}
of the main text according to its duration $t$  
\begin{eqnarray}
P_{\Gamma}^+(x \vert x_0) = \int_0^{+\infty} dt P_{\Gamma}^+(x , t\vert x_0)
  \label{Pplustdef}
\end{eqnarray}
This duration $t$ of an ascending bond starting at $x_0$ is the sum of two contributions
\begin{eqnarray}
 && t=t_1+t_2
\nonumber \\
&& \text{$t_1$ is the time to go from the minimum $x_0$ to $(x_0+\Gamma)$ }
\nonumber \\
&& \text{$t_2$ is the time to go from $(x_0+\Gamma)$ up to the top $x$ without any drawdown $\Gamma$}
  \label{t1t2}
\end{eqnarray}
so that $P_{\Gamma}^+(x , t\vert x_0) $ is given by the time-convolution 
\begin{eqnarray}
P_{\Gamma}^+(x , t\vert x_0) = \int_0^{+\infty} dt_1  \Pi^+_{\Gamma}(t_1\vert x_0)
 \int_0^{+\infty} dt_2  Q^+_{\Gamma}(x , t_2\vert x_0+\Gamma) \delta(t-(t_1+t_2))
  \label{Pplustconvolt1t2}
\end{eqnarray}
where the two probabilities $\Pi^+_{\Gamma}(t_1\vert x_0) $ and $Q^+_{\Gamma}(x , t_2\vert x_0+\Gamma) $
are analyzed in the two next subsections.

\subsection{ Probability distribution $\Pi^+_{\Gamma}(t_1\vert x_0)$ of the time $t_1$ }

The probability distribution $\Pi^+_{\Gamma}(t_1\vert x_0)$ of the time $t_1$
can be evaluated from the probability distribution $\pi^+_{[x_0,x_0+\Gamma]} (t_1 \vert x_0+\epsilon) $
of the exit-time at the right boundary
of the interval $[x_L=x_0,x_R=x_0+\Gamma ]$ when starting at $x=x_0+\epsilon$
that was analyzed in Appendix \ref{app_ExitTime}
\begin{eqnarray}
  \Pi^+_{\Gamma}(t_1\vert x_0) = \lim_{\epsilon \to 0} \left( \frac{\pi^+_{[x_0,x_0+\Gamma]} (t_1 \vert x_0+\epsilon)}{ \int_0^{+\infty} dt_1' \pi^+_{[x_0,x_0+\Gamma]} (t_1' \vert x_0+\epsilon) }\right)
  \label{Q1}
\end{eqnarray}
The Laplace transform ${\hat  \Pi}^{[p]+}_{\Gamma}( x_0) $
\begin{eqnarray}
{\hat  \Pi}^{[p]+}_{\Gamma}( x_0) \equiv \int_0^{+\infty} dt_1 e^{-pt_1} \Pi^+_{\Gamma}(t_1\vert x_0)
 = \lim_{\epsilon \to 0} \left( \frac{ {\hat \pi}^{[p]+}_{[x_0,x_0+\Gamma]} ( x_0+\epsilon)}
 { {\hat \pi}^{[p=0]+}_{[x_0,x_0+\Gamma]} ( x_0+\epsilon) }\right)
 =  \lim_{\epsilon \to 0} \left( \frac{ {\hat \pi}^{[p]+}_{[x_0,x_0+\Gamma]} ( x_0+\epsilon)}
 { E^{+}_{[x_0,x_0+\Gamma]} ( x_0+\epsilon) }\right)
  \label{Q1Laplacedef}
\end{eqnarray}
thus involves the Laplace-transform ${\hat \pi}^{[p]+}_{[x_0,x_0+\Gamma]} ( x_0+\epsilon) $ 
computed in Eq. \ref{CLLaplace+} with the property of Eq. \ref {kabderibcoinciding}
\begin{eqnarray}
{\hat \pi}^{[p]+}_{[x_L=x_0,x_R=x_0+\Gamma]} (x= x_0+\epsilon)
 && = \frac{ K^{[p]}(x_0,x_0+\epsilon)  }{ K^{[p]}(x_0,x_0+\Gamma) } 
 = \epsilon  \frac{ \partial_b K^{[p]}(x_0,x_0)  }{ K^{[p]}(x_0,x_0+\Gamma) } 
 +O(\epsilon^2)
 \nonumber \\
 && = \epsilon  \frac{  s'(x_0)  }{ K^{[p]}(x_0,x_0+\Gamma) } 
 +O(\epsilon^2)
  \label{pit1}
\end{eqnarray}
and the exit probability $E^{+}_{[x_0,x_0+\Gamma]} ( x_0+\epsilon) ={\hat \pi}^{[p=0]+}_{[x_0,x_0+\Gamma]} ( x_0+\epsilon)$ of Eq. \ref{ExitRxt}
written in Eq. \ref{ExitRx} of the main text in terms of the scale function
\begin{eqnarray}
E^+_{[x_L=x_0,x_R=x_0+\Gamma]}(x) && = \frac{ s(x_0+\epsilon)-s(x_0) }{ s(x_0+\Gamma)-s(x_0) }
= \epsilon   \frac{ s'(x_0) }{ s(x_0+\Gamma)-s(x_0) } + O(\epsilon^2)
  \label{ExitRxeps}
\end{eqnarray}
so that Eq. \ref{Q1Laplacedef} reads
\begin{eqnarray}
{\hat  \Pi}^{[p]+}_{\Gamma}( x_0) 
 =  \lim_{\epsilon \to 0} \left( \frac{ {\hat \pi}^{[p]+}_{[x_0,x_0+\Gamma]} ( x_0+\epsilon)}
 { E^{+}_{[x_0,x_0+\Gamma]} ( x_0+\epsilon) }\right)
 = \frac{ s(x_0+\Gamma)-s(x_0) }{ K^{[p]}(x_0,x_0+\Gamma) } 
  \label{Pi1Laplace}
\end{eqnarray}
The series expansion of the Laplace transform of Eq. \ref{Q1Laplacedef}
\begin{eqnarray}
{\hat  \Pi}^{[p]+}_{\Gamma}( x_0) =\int_0^{+\infty} dt_1 (1-pt_1+O(p^2)) \Pi^+_{\Gamma}(t_1\vert x_0)
 = 1 -p \int_0^{+\infty} dt_1 t_1  \Pi^+_{\Gamma}(t_1\vert x_0)+ O(p^2)
  \label{PiLaplaceseries}
\end{eqnarray}
is useful to compute the moments of the duration $t_1$.
In particular the first moment requires only the first-order expansion of the result of Eq. \ref{Pi1Laplace}
using $K_0(a,b)= s(b)-s(a) $ of Eq. \ref{kabpzero0}
and $K_1(a,b) $ of Eq. \ref{K1ab} 
\begin{eqnarray}
{\hat  \Pi}^{[p]+}_{\Gamma}( x_0) 
 = \frac{ s(x_0+\Gamma)-s(x_0) }{ K_0(x_0,x_0+\Gamma) +p K_1(x_0,x_0+\Gamma) +.. } 
 = 1- p \frac{K_1(x_0,x_0+\Gamma)}{s(x_0+\Gamma)-s(x_0)} + O(p^2)
  \label{PiLaplaceseriesbis}
\end{eqnarray}
so that the first moment $\tau_1^+(x_0,\Gamma) $ of $t_1$ for an ascending bond at scale $\Gamma$ starting at $x_0$ reads in terms of the scale function $s(.)$ and speed-measure $m'(.)$
\begin{eqnarray}
\tau_1^+(x_0,\Gamma) && \equiv \int_0^{+\infty} dt_1 t_1  \Pi^+_{\Gamma}(t_1\vert x_0)  =\frac{K_1(x_0,x_0+\Gamma)}{s(x_0+\Gamma)-s(x_0)}
\nonumber \\
&&  = \frac{\int_{x_0}^{x_0+\Gamma} dy  m'(y) \left[ s(y)-s(x_0)\right] \left[ s(x_0+\Gamma)-s(y)\right] }{s(x_0+\Gamma)-s(x_0)}
  \label{Q1LaplaceMoment}
\end{eqnarray}

%%%%%%%%%%%%%%%%%%%%%%%%%%%%%%%%%%%%%%%%%%%%

\subsection{ Probability $Q^+_{\Gamma}(x , t_2\vert x_0+\Gamma)$ 
to go from $(x_0+\Gamma)$ up to the top $x$ in time $t_2$ without any drawdown $\Gamma$ }

 The probability $Q^+_{\Gamma}(x , t_2\vert x_0+\Gamma)$ 
to go from $(x_0+\Gamma)$ up to the top $x$ in time $t_2$ without any drawdown $\Gamma$
has for Laplace transform
\begin{eqnarray}
{\hat Q}^{[p]+} _{\Gamma}(x \vert x_0+\Gamma) \equiv \int_0^{+\infty} dt_2 e^{- p t_2}
Q^+_{\Gamma}(x , t_2\vert x_0+\Gamma)
= R_{\Gamma}^+(x) e^{ - \int_{x_0+\Gamma}^x dz R_{\Gamma}^{[p]+}(z)  } 
  \label{QLplace}
\end{eqnarray}
that generalizes Eq. \ref{Pplus} of the main text concerning $p=0$ as follows :
the prefactor $R_{\Gamma}^+(x)$ is still the rate of Eq. \ref{Rxplus}
at which the running maximum at $x$ is stopped by a drawdown $\Gamma$,
while the function $R_{\Gamma}^{[p]+}(x) $ in the exponential 
appears when one evaluates the Laplace-transform 
${\hat \pi}^{[p]+}_{[x_L=x-\Gamma,x_R=x]} ( x-dx) $ of the boundary-exit-time on the interval $ [x_L=x-\Gamma,x_R=x]$ when starting at $(x-dx)$ computed in Eq. \ref{CLLaplace+}
 \begin{eqnarray}
{\hat \pi}^{[p]+}_{[x_L=x-\Gamma,x_R=x]} ( x-dx) = \frac{ K^{[p]}(x-\Gamma,x-dx)  }{ K^{[p]}(x-\Gamma,x) } 
=1 - dx  \frac{ \partial_b K^{[p]}(x-\Gamma,x)  }{ K^{[p]}(x-\Gamma,x) } 
\equiv 1-dx R_{\Gamma}^{[p]+}(x)
  \label{CLLaplace+t2}
\end{eqnarray}
and reads in terms of Eqs \ref{kab} and \ref{kabderib}
 \begin{eqnarray}
R_{\Gamma}^{[p]+}(x) \equiv  \frac{ \partial_b K^{[p]}(x-\Gamma,x)  }{ K^{[p]}(x-\Gamma,x) } 
%= \partial_b \ln K^{[p]}(x-\Gamma,x)
  \label{Rp+}
\end{eqnarray}
The first-order series expansion in $p$ around the value for $p=0$ that corresponds to 
$R_{\Gamma}^{+}(x)=R_{\Gamma}^{[p=0]+}(x) $ of Eq. \ref{Rxplus}
 \begin{eqnarray}
R_{\Gamma}^{[p=0]+}(x) = R_{\Gamma}^{+}(x) =  \frac{ s'(x) }{ s(x)-s(x-\Gamma) } 
  \label{Rp+exp0}
\end{eqnarray}
can be computed using Eq. \ref{kabpzero}
and $K_1$ of Eq. \ref{K1ab} 
 \begin{eqnarray}
&&R_{\Gamma}^{[p]+}(x) =  \frac{ s'(x) + p \partial_b K_1(x-\Gamma,x) }
{ s(x)-s(x-\Gamma) + p K_1(x-\Gamma,x)} +O(p^2)
\nonumber \\
&&= \frac{s'(x)}{s(x)-s(x-\Gamma)} \left(1 + p \frac{\partial_b K_1(x-\Gamma,x)}{s'(x) } - p \frac{K_1(x-\Gamma,x)}{s(x)-s(x-\Gamma) }\right) +O(p^2)
\nonumber \\
&&= R_{\Gamma}^{+}(x) + p {\cal R}_{\Gamma}^+(x) +O(p^2)
 \nonumber \\
  \label{Rp+exp}
\end{eqnarray}
where the first correction reads using Eqs \ref{K1ab} and \ref{K1abderib}
\begin{eqnarray}
{\cal R}_{\Gamma}^+(x) && \equiv \frac{s'(x)}{s(x)-s(x-\Gamma)} \left( \frac{\partial_b K_1(x-\Gamma,x)}{s'(x) } 
-  \frac{K_1(x-\Gamma,x)}{s(x)-s(x-\Gamma) }\right) 
\nonumber \\
&&  = \frac{s'(x)}{s(x)-s(x-\Gamma)} 
\left(  \int_{x-\Gamma}^x dy  m'(y) \left[ s(y)-s(x-\Gamma)\right]
 -  \frac{\int_{x-\Gamma}^x dy  m'(y) \left[ s(y)-s(x-\Gamma)\right] \left[ s(x)-s(y)\right] 
 }{s(x)-s(x-\Gamma) }\right) 
 \nonumber \\
&&  = \frac{s'(x)}{[s(x)-s(x-\Gamma)]^2} 
 \int_{x-\Gamma}^x dy  m'(y) \left[ s(y)-s(x-\Gamma)\right]^2
  \label{CalR}
\end{eqnarray}

As a consequence, the corresponding expansion for the Laplace transform of Eq. \ref{QLplace}
is given by
\begin{eqnarray}
{\hat Q}^{[p]+} _{\Gamma}(x \vert x_0+\Gamma) 
&& = R_{\Gamma}^+(x) e^{ - \int_{x_0+\Gamma}^x dz \left[ R_{\Gamma}^{+}(z) + p {\cal R}_{\Gamma}^+(z) +O(p^2)\right]  } 
\nonumber \\
&& = R_{\Gamma}^+(x) e^{ - \int_{x_0+\Gamma}^x dz  R_{\Gamma}^{+}(z) }
\left[ 1  - p \int_{x_0+\Gamma}^x dz  {\cal R}_{\Gamma}^+(z) +O(p^2)\right]   
  \label{QLplaceSeriesp}
\end{eqnarray}
with the corresponding integration over the end-point $x \in ]x_0+\Gamma,+\infty[$ of the ascending bond
\begin{eqnarray}
\int_{x_0+\Gamma}^{+\infty} dx {\hat Q}^{[p]+} _{\Gamma}(x \vert x_0+\Gamma) 
&& = \int_{x_0+\Gamma}^{+\infty} dx \left[ 1  - p \int_{x_0+\Gamma}^x dz  {\cal R}_{\Gamma}^+(z) +O(p^2)\right]   
\left[ - \frac{d}{dx} e^{ - \int_{x_0+\Gamma}^x dz  R_{\Gamma}^{+}(z) } \right]
\nonumber \\
&& = 
1-p \int_{x_0+\Gamma}^{+\infty} dx \left[  \int_{x_0+\Gamma}^x dz  {\cal R}_{\Gamma}^+(z) \right]   
\left[ - \frac{d}{dx} e^{ - \int_{x_0+\Gamma}^x dz  R_{\Gamma}^{+}(z) } \right]+O(p^2)
  \label{QLplaceSeriespx}
\end{eqnarray}
so the first moment $\tau_2^+(x_0,\Gamma) $ of $t_2$ for an ascending bond at scale $\Gamma$ starting at $x_0$ 
reads after an integration by parts
\begin{eqnarray}
\tau_2^+(x_0,\Gamma)= \int_{x_0+\Gamma}^{+\infty} dx   {\cal R}_{\Gamma}^+(x) 
 e^{ - \int_{x_0+\Gamma}^x dz  R_{\Gamma}^{+}(z) } 
  \label{t2av}
\end{eqnarray}
where one can plug the explicit expressions of ${\cal R}_{\Gamma}^+(x) $ of Eq. \ref{CalR}
and use $R_{\Gamma}^{+}(x) = \frac{ s'(x) }{ s(x)-s(x-\Gamma) } $ of Eq. \ref{Rxplus}
\begin{eqnarray}
\tau_2^+(x_0,\Gamma) && = \int_{x_0+\Gamma}^{+\infty} dx 
 \frac{1}{[s(x)-s(x-\Gamma)]} 
 \int_{x-\Gamma}^x dy  m'(y) \left[ s(y)-s(x-\Gamma)\right]^2
 R_{\Gamma}^{+}(x)
 e^{ - \int_{x_0+\Gamma}^x dz  R_{\Gamma}^{+}(z) } 
\nonumber \\
&& = \int_{x_0+\Gamma}^{+\infty} dx 
 \frac{\int_{x-\Gamma}^x dy  m'(y) \left[ s(y)-s(x-\Gamma)\right]^2}{s(x)-s(x-\Gamma)} 
  P_{\Gamma}^+(x \vert x_0) 
  \label{t2avtop}
\end{eqnarray}
to identify the contributions associated to the end-point $x$ appearing with the probability $P_{\Gamma}^+(x \vert x_0)  $ of Eq. \ref{Pplus}.

%%%%%%%%%%%%%%%%%%%%%%%%%%%%%%%%%%%%%%%%%

\subsection{ Probability $P_{\Gamma}^{+}(x,t \vert x_0)  $ of ascending bonds and its Laplace transform ${\hat P}_{\Gamma}^{[p]+}(x \vert x_0) $ }

The two previous subsections yield that the Laplace transform $ {\hat P}_{\Gamma}^{[p]+}(x \vert x_0)$ of the ascending-bond probability 
$P_{\Gamma}^{+}(x,t \vert x_0) $ of Eq. \ref{Pplustconvolt1t2}
is the product of the Laplace transforms of Eqs \ref{Pi1Laplace}
and \ref{QLplace}
\begin{eqnarray}
{\hat P}_{\Gamma}^{[p]+}(x \vert x_0)
&& \equiv \int_0^{+\infty} dt e^{- p t} P_{\Gamma}^+(x , t\vert x_0) 
 = {\hat  \Pi}^{[p]+}_{\Gamma}( x_0) {\hat Q}^{[p]+} _{\Gamma}(x \vert x_0+\Gamma)
 \nonumber \\
 && = \frac{ s(x_0+\Gamma)-s(x_0) }{ K^{[p]}(x_0,x_0+\Gamma) } 
  R_{\Gamma}^+(x) e^{ - \int_{x_0+\Gamma}^x dz R_{\Gamma}^{[p]+}(z)  } 
  \ \ \ \ \text{ for } \ \ x \in [x_0+\Gamma,+\infty[
  \label{Pplusp}
\end{eqnarray}
that generalizes $P_{\Gamma}^+(x \vert x_0)={\hat P}_{\Gamma}^{[p=0]+}(x \vert x_0) $ of Eq. \ref{Pplus}
of the main text.

In particular, the probability $P_{\Gamma}^+(x=x_0+\Gamma , t\vert x_0) $ of ascending-bond about to be decimated have for Laplace transform using the explicit expression of $R_{\Gamma}^+(.)$ of Eq. \ref{Rxplus}
\begin{eqnarray}
{\hat P}_{\Gamma}^{[p]+}(x=x_0+\Gamma \vert x_0)
&&  = \frac{ s(x_0+\Gamma)-s(x_0) }{ K^{[p]}(x_0,x_0+\Gamma) } 
  R_{\Gamma}^+(x_0+\Gamma) 
  = \frac{ s'(x_0+\Gamma) }{ K^{[p]}(x_0,x_0+\Gamma) } 
  \label{Ppluspdeci}
\end{eqnarray}

%%%%%%%%%%%%%%%%%

\subsection{ Probability $P_{\Gamma}^{-}(x,t \vert x_0)  $ of descending bonds and its Laplace transform ${\hat P}_{\Gamma}^{[p]-}(x \vert x_0) $ }

For a descending bond starting at $x_0$, 
the analog of ${\hat  \Pi}^{[p]+}_{\Gamma}( x_0) $ of Eq. \ref{Pi1Laplace}
is the Laplace transform of the time $t_1$ needed to go from $x_0$ to $(x_0-\Gamma)$
\begin{eqnarray}
  {\hat  \Pi}^{[p]-}_{\Gamma}( x_0)  = \frac{ s(x_0)-s(x_0-\Gamma) }{ K^{[p]}(x_0-\Gamma,x_0) } 
   \label{Pi1Laplace-}
\end{eqnarray}
while the analog of ${\hat Q}^{[p]+} _{\Gamma}(x \vert x_0+\Gamma) $ of Eq. \ref{QLplace}
is the Laplace transform of 
the time $t_2$ needed to go from $(x_0-\Gamma)$ to $x$
\begin{eqnarray}
 {\hat Q}^{[p]-} _{\Gamma}(x \vert x_0-\Gamma)
 =  R_{\Gamma}^-(x) e^{ - \int_x^{x_0-\Gamma} dz R_{\Gamma}^{[p]-}(z)  } 
 \ \ \ \ \text{ for } \ \  x \in ]-\infty,x_0-\Gamma]
   \label{QLplace-}
\end{eqnarray}
where
 \begin{eqnarray}
R_{\Gamma}^{[p]-}(x) \equiv  - \frac{ \partial_a K^{[p]}(x,x+\Gamma)  }{ K^{[p]}(x,x+\Gamma) } 
  \label{Rp-}
\end{eqnarray}
generalizes $R_{\Gamma}^{-}(x)=R_{\Gamma}^{[p=0]-}(x) $ of Eq. \ref{Rxmoins} using Eq. \ref{kabpzero}.

So the Laplace transform ${\hat P}_{\Gamma}^{[p]-}(x \vert x_0)$
of the descending-bond probability $ P_{\Gamma}^{-}(x,t \vert x_0)$
that generalizes $P_{\Gamma}^-(x \vert x_0)={\hat P}_{\Gamma}^{[p=0]-}(x \vert x_0) $ of Eq. \ref{Pmoins}
of the main text
reads
\begin{eqnarray}
{\hat P}_{\Gamma}^{[p]-}(x \vert x_0)
&& \equiv \int_0^{+\infty} dt e^{- p t} P_{\Gamma}^-(x , t\vert x_0) =  {\hat  \Pi}^{[p]-}_{\Gamma}( x_0) {\hat Q}^{[p]-} _{\Gamma}(x \vert x_0-\Gamma)
 \nonumber \\
 && = \frac{ s(x_0)-s(x_0-\Gamma) }{ K^{[p]}(x_0-\Gamma,x_0) } 
  R_{\Gamma}^-(x) e^{ - \int_x^{x_0-\Gamma} dz R_{\Gamma}^{[p]-}(z)  } 
 \ \ \ \ \text{ for } \ \  x \in ]-\infty,x_0-\Gamma]
   \label{Pmoinsp}
\end{eqnarray}

In particular, the probability $P_{\Gamma}^-(x=x_0-\Gamma , t\vert x_0) $ of descending-bond about to be decimated have for Laplace transform using the explicit expression of $R_{\Gamma}^-(.)$ of Eq. \ref{Rxmoins}
\begin{eqnarray}
{\hat P}_{\Gamma}^{[p]-}(x=x_0-\Gamma \vert x_0)
 = \frac{ s(x_0)-s(x_0-\Gamma) }{ K^{[p]}(x_0-\Gamma,x_0) } 
  R_{\Gamma}^-(x_0-\Gamma) 
  =  \frac{ s'(x_0-\Gamma) }{ K^{[p]}(x_0-\Gamma,x_0) } 
   \label{Pmoinspdeci}
\end{eqnarray}

For a descending bond at scale $\Gamma$ starting at $x_0$,
the first moment $\tau_1^-(x_0,\Gamma) $ is given by the formula analog to Eq. \ref{Q1LaplaceMoment}
\begin{eqnarray}
\tau_1^-(x_0,\Gamma) 
 = \frac{\int_{x_0-\Gamma}^{x_0} dy  m'(y) \left[ s(y)-s(x_0-\Gamma)\right] \left[ s(x_0)-s(y)\right] }
 {s(x_0)-s(x_0-\Gamma)}
  \label{Q1LaplaceMoment-}
\end{eqnarray}
and the first moment $\tau_2^-(x_0,\Gamma) $ is given by the formula analog to Eq. \ref{t2avtop}
\begin{eqnarray}
\tau_2^-(x_0,\Gamma) &&  = \int_{-\infty}^{x_0-\Gamma} dx 
 \frac{\int_x^{x+\Gamma} dy  m'(y) \left[ s(x+\Gamma)- s(y)\right]^2}{s(x+\Gamma)-s(x)} 
  P_{\Gamma}^-(x \vert x_0) 
  \label{t2avtop-}
\end{eqnarray}
that shows the contributions associated to the end-point $x$ appearing with the probability $P_{\Gamma}^-(x \vert x_0)  $ of Eq. \ref{Pmoins}.

%%%%%%%%%%%%%%%%%%%%%%%%%%%%%%%%%%%%%%%%%

\section{ RG-flows for the probabilities $P_{\Gamma}^{\pm}(x,t \vert x_0) $ of ascending and descending bonds  }

In this Appendix, one includes the duration $t$ in the bonds densities $N_{\Gamma}(x_0\asc x ; t ) $ and $N_{\Gamma}(x_0\des x ; t)$
and in the bonds probabilities 
computed as the ratios that still involves the densities $n_{\Gamma}^{\pm}(x_0)$ of Eq. \ref{Pgamma}
\begin{eqnarray}
P_{\Gamma}(x_0\nearrow x;t) && = \frac{N_{\Gamma}(x_0\asc x;t)}{n^-_{\Gamma}(x_0)} 
\nonumber \\
P_{\Gamma}(x_0\searrow x;t) && = \frac{ N_{\Gamma}(x_0\des x ;t)}{n^+_{\Gamma}(x_0)} 
  \label{Pgammat}
\end{eqnarray}
whose Laplace transforms
should coincide with ${\hat P}_{\Gamma}^{[p]\pm}(x \vert x_0) $ of Eqs \ref{Pplusp}
and \ref{Pmoinsp}.

The RG-flow for the bond densities of Eq. \ref{RGNasc}
is generalized into
\begin{eqnarray}
&& \partial_{\Gamma} N_{\Gamma}(x_0\asc x ; t ) =- P_{\Gamma}(x_0+\Gamma\nwarrow x_0)N_{\Gamma}(x_0\asc x ; t)
 -  N_{\Gamma}(x_0\asc x ; t) P_{\Gamma}(x\searrow x-\Gamma)
\nonumber \\
&&+\int_{x_0+\Gamma}^{x}d z \int_0^{+\infty} dt_1 \int_0^{+\infty} dt_2 \int_0^{+\infty} dt_3
N_{\Gamma}(x_0\asc z; t_1) P_{\Gamma}(z\searrow z-\Gamma ; t_2 ) P_{\Gamma}(z-\Gamma\nearrow x ; t_3)
\delta(t- (t_1+t_2+t_3))
  \label{RGNasct}
\end{eqnarray}
while the RG-flow for the ascending-bond probabilities of Eq. \ref{RGPasc}
is generalized into
\begin{eqnarray}
&& \partial_{\Gamma} P_{\Gamma}(x_0\nearrow x ; t)  = 
   P_{\Gamma}(x_0\nearrow x ; t) \left[ - P_{\Gamma}(x\searrow x-\Gamma)
 +   P_{\Gamma}(x_0\nearrow x_0+\Gamma) \right]
\nonumber \\
&&+\int_{x_0+\Gamma}^{x}d z \int_0^{+\infty} dt_1 \int_0^{+\infty} dt_2 \int_0^{+\infty} dt_3
P_{\Gamma}(x_0\nearrow z ; t_1)
P_{\Gamma}(z\searrow z-\Gamma ; t_2) P_{\Gamma}(z-\Gamma\nearrow x ; t_3)
\delta(t- (t_1+t_2+t_3))
  \label{RGPasct}
\end{eqnarray}
that translates for the Laplace transforms into
\begin{eqnarray}
 \partial_{\Gamma} {\hat P}^{[p]}_{\Gamma}(x_0\nearrow x )  
&&= 
   {\hat P}^{[p]}_{\Gamma}(x_0\nearrow x ) \left[ - P_{\Gamma}(x\searrow x-\Gamma)
 +   P_{\Gamma}(x_0\nearrow x_0+\Gamma) \right]
\nonumber \\
&&+\int_{x_0+\Gamma}^{x}d z 
{\hat P}^{[p]}_{\Gamma}(x_0\nearrow z )
{\hat P}^{[p]}_{\Gamma}(z\searrow z-\Gamma ) 
{\hat P}^{[p]}_{\Gamma}(z-\Gamma\nearrow x )
  \label{RGPascp}
\end{eqnarray}
Similarly, the RG-flow for the descending-bond probabilities of Eq. \ref{RGPdes}
is generalized into the following flow for the Laplace transform ${\hat P}^{[p]}_{\Gamma}(x_0\searrow x) $
\begin{eqnarray}
\partial_{\Gamma} {\hat P}^{[p]}_{\Gamma}(x_0\searrow x)
&& = {\hat P}^{[p]}_{\Gamma}(x_0\searrow x) \left[-P_{\Gamma}(x\nearrow x+\Gamma)+ P_{\Gamma}(x_0\searrow x_0-\Gamma)\right]
\nonumber \\
&&+\int_{x}^{x_0-\Gamma} d z {\hat P}^{[p]}_{\Gamma}(x_0\searrow z) 
{\hat P}^{[p]}_{\Gamma}(z\nearrow z+\Gamma) 
{\hat P}^{[p]}_{\Gamma}(z+\Gamma\searrow x)
  \label{RGPdesp}
\end{eqnarray}

Since these Laplace transforms should coincide with the Laplace transforms computed 
 Eqs \ref{Pplusp}
and \ref{Pmoinsp}

\begin{eqnarray}
{\hat P}^{[p]}_{\Gamma}(x_0\nearrow x ) = {\hat P}_{\Gamma}^{[p]+}(x \vert x_0)
&& = \frac{ s(x_0+\Gamma)-s(x_0) }{ K^{[p]}(x_0,x_0+\Gamma) } 
  R_{\Gamma}^+(x) e^{ - \int_{x_0+\Gamma}^x dz R_{\Gamma}^{[p]+}(z)  } 
  \ \ \ \ \text{ for } \ \ x \in [x_0+\Gamma,+\infty[
\nonumber \\
{\hat P}^{[p]}_{\Gamma}(x_0\searrow x) ={\hat P}_{\Gamma}^{[p]-}(x \vert x_0)
&& = \frac{ s(x_0)-s(x_0-\Gamma) }{ K^{[p]}(x_0-\Gamma,x_0) } 
  R_{\Gamma}^-(x) e^{ - \int_x^{x_0-\Gamma} dz R_{\Gamma}^{[p]-}(z)  } 
 \ \ \ \ \text{ for } \ \  x \in ]-\infty,x_0-\Gamma]
   \label{Pplusmoinsp}
\end{eqnarray}
it is interesting to plug these Ansatz
into the RG-flows of Eqs \ref{RGPascp} and \ref{RGPdesp}.

Let us write the log-derivatives of Eq. \ref{Pplusmoinsp}
that can be simplified using Eqs \ref{Rp+} and \ref{Rp-}
as well as Eq \ref{RGforRpm}
of the main text
\begin{eqnarray}
\partial_{\Gamma} \ln {\hat P}^{[p]}_{\Gamma}(x_0\nearrow x ) 
&& = \frac{ s'(x_0+\Gamma) }{ s(x_0+\Gamma)-s(x_0) } 
 - \frac{ \partial_b K^{[p]}(x_0,x_0+\Gamma) }{ K^{[p]}(x_0,x_0+\Gamma) } 
+  \partial_{\Gamma} \ln R_{\Gamma}^+(x) 
 +   R_{\Gamma}^{[p]+}(x_0+\Gamma)   
  -  \int_{x_0+\Gamma}^x dz \partial_{\Gamma} R_{\Gamma}^{[p]+}(z)    
\nonumber \\
  && =
  R_{\Gamma}^+(x_0+\Gamma)
 -  R_{\Gamma}^-( x-\Gamma) 
  -  \int_{x_0+\Gamma}^x dz \partial_{\Gamma} R_{\Gamma}^{[p]+}(z)  
\nonumber \\
\partial_{\Gamma} \ln {\hat P}^{[p]}_{\Gamma}(x_0\searrow x) 
  &&= \frac{ s'(x_0-\Gamma) }{ s(x_0)-s(x_0-\Gamma)  } 
 + \frac{ \partial_a K^{[p]}(x_0-\Gamma,x_0) }{ K^{[p]}(x_0-\Gamma,x_0) } 
 + \partial_{\Gamma} \ln R_{\Gamma}^-(x) 
+    R_{\Gamma}^{[p]-}(x_0-\Gamma)   
-   \int_x^{x_0-\Gamma} dz \partial_{\Gamma} R_{\Gamma}^{[p]-}(z)   
\nonumber \\
 &&= 
 R_{\Gamma}^-(x_0-\Gamma)  
- R_{\Gamma}^+( x+\Gamma)
-   \int_x^{x_0-\Gamma} dz \partial_{\Gamma} R_{\Gamma}^{[p]-}(z) 
   \label{Pplusmoinspderi}
\end{eqnarray}
and evaluate the convolution terms of Eq \ref{RGPascp} using Eq. \ref{Pmoinspdeci}
\begin{eqnarray}
&& \int_{x_0+\Gamma}^{x}d z 
{\hat P}^{[p]}_{\Gamma}(x_0\nearrow z )
{\hat P}^{[p]}_{\Gamma}(z\searrow z-\Gamma ) 
{\hat P}^{[p]}_{\Gamma}(z-\Gamma\nearrow x )
\nonumber \\
&& = \int_{x_0+\Gamma}^{x}d z 
 \frac{ s(x_0+\Gamma)-s(x_0) }{ K^{[p]}(x_0,x_0+\Gamma) } 
  R_{\Gamma}^+(z) e^{ - \int_{x_0+\Gamma}^z dy R_{\Gamma}^{[p]+}(y)  } \times
  \frac{ s'(z-\Gamma) }{ K^{[p]}(z-\Gamma,z) } \times
 \frac{ s(z)-s(z-\Gamma) }{ K^{[p]}(z-\Gamma,z) } 
  R_{\Gamma}^+(x) e^{ - \int_{z}^x dy R_{\Gamma}^{[p]+}(y)  } 
  \nonumber \\
&& =\frac{ s(x_0+\Gamma)-s(x_0) }{ K^{[p]}(x_0,x_0+\Gamma) } 
 R_{\Gamma}^+(x)e^{ - \int_{x_0+\Gamma}^x dy R_{\Gamma}^{[p]+}(y)  } 
\int_{x_0+\Gamma}^{x}d z 
  R_{\Gamma}^+(z) 
  \frac{ s'(z-\Gamma) [ s(z)-s(z-\Gamma) ]}{ [K^{[p]}(z-\Gamma,z)]^2 } 
   \nonumber \\
&& ={\hat P}^{[p]}_{\Gamma}(x_0\nearrow x )
\int_{x_0+\Gamma}^{x}d z 
   \frac{ s'(z) s'(z-\Gamma) }{ [K^{[p]}(z-\Gamma,z)]^2 } 
  \label{RGPascpconvol}
\end{eqnarray}
where we have used Eq. \ref{Rxplus} to simplify the integrand.

and Eq \ref{RGPdesp} using Eq. \ref{Ppluspdeci}
\begin{eqnarray}
&&\int_{x}^{x_0-\Gamma} d z {\hat P}^{[p]}_{\Gamma}(x_0\searrow z) 
{\hat P}^{[p]}_{\Gamma}(z\nearrow z+\Gamma) 
{\hat P}^{[p]}_{\Gamma}(z+\Gamma\searrow x)
\nonumber \\
&& = \int_{x}^{x_0-\Gamma} d z 
 \frac{ s(x_0)-s(x_0-\Gamma) }{ K^{[p]}(x_0-\Gamma,x_0) } 
  R_{\Gamma}^-(z) e^{ - \int_z^{x_0-\Gamma} dy R_{\Gamma}^{[p]-}(y)  } \times
\frac{ s'(z+\Gamma) }{ K^{[p]}(z,z+\Gamma) } \times
 \frac{ s(z+\Gamma)-s(z) }{ K^{[p]}(z,z+\Gamma) } 
  R_{\Gamma}^-(x) e^{ - \int_x^{z} dy R_{\Gamma}^{[p]-}(y)  } 
\nonumber \\
&& =  \frac{ s(x_0)-s(x_0-\Gamma) }{ K^{[p]}(x_0-\Gamma,x_0) } 
 R_{\Gamma}^-(x)  e^{ - \int_x^{x_0-\Gamma} dy R_{\Gamma}^{[p]-}(y)  } 
\int_{x}^{x_0-\Gamma} d z 
  R_{\Gamma}^-(z) 
\frac{ s'(z+\Gamma)[ s(z+\Gamma)-s(z)] }{ [K^{[p]}(z,z+\Gamma)]^2 }
\nonumber \\
&& =  {\hat P}^{[p]}_{\Gamma}(x_0\searrow x)
\int_{x}^{x_0-\Gamma} d z 
  \frac{ s'(z)  s'(z+\Gamma) }{ [K^{[p]}(z,z+\Gamma)]^2 }   
  \label{RGPdespconvol}
\end{eqnarray}
where we have used Eq. \ref{Rxmoins} to simplify the integrand.

Let us plug Eqs \ref{Pplusmoinspderi}
and \ref{RGPascpconvol}
into the RG-flow of Eq. \ref{RGPascp} using also Eqs \ref{Pplusdecim} and \ref{Pmoinsdecim}
of the main text
\begin{eqnarray}
0 && = - \partial_{\Gamma} \ln {\hat P}^{[p]}_{\Gamma}(x_0\nearrow x )  
 - P_{\Gamma}(x\searrow x-\Gamma)
 +   P_{\Gamma}(x_0\nearrow x_0+\Gamma) 
\nonumber \\
&&+ \frac{1}{{\hat P}^{[p]}_{\Gamma}(x_0\nearrow x )} \int_{x_0+\Gamma}^{x}d z 
{\hat P}^{[p]}_{\Gamma}(x_0\nearrow z )
{\hat P}^{[p]}_{\Gamma}(z\searrow z-\Gamma ) 
{\hat P}^{[p]}_{\Gamma}(z-\Gamma\nearrow x )
\nonumber \\
&& =
- R_{\Gamma}^+(x_0+\Gamma) 
 +  R_{\Gamma}^-( x-\Gamma) 
  +  \int_{x_0+\Gamma}^x dz \partial_{\Gamma} R_{\Gamma}^{[p]+}(z)  
\nonumber \\
&&- R_{\Gamma}^-(x-\Gamma)
 +   R_{\Gamma}^+( x_0+\Gamma) + \int_{x_0+\Gamma}^{x}d z 
  R_{\Gamma}^+(z) 
  \frac{ s'(z-\Gamma) [ s(z)-s(z-\Gamma) ]}{ [K^{[p]}(z-\Gamma,z)]^2 } 
  \nonumber \\
&& =
  \int_{x_0+\Gamma}^x dz   
  \left[ \partial_{\Gamma}  R_{\Gamma}^{[p]+}(z)  
  +    \frac{ s'(z) s'(z-\Gamma) }{ [K^{[p]}(z-\Gamma,z)]^2 }  \right]
  \label{RGPascpexpli}
\end{eqnarray}

and plug Eqs \ref{Pplusmoinspderi}
and \ref{RGPdespconvol}
into the RG-flow of Eq. \ref{RGPdesp} using also Eqs \ref{Pplusdecim} and \ref{Pmoinsdecim}
of the main text
\begin{eqnarray}
0 && = - \partial_{\Gamma} \ln {\hat P}^{[p]}_{\Gamma}(x_0\searrow x)
-P_{\Gamma}(x\nearrow x+\Gamma)+ P_{\Gamma}(x_0\searrow x_0-\Gamma)
\nonumber \\
&&+  \frac{1}{ {\hat P}^{[p]}_{\Gamma}(x_0\searrow x)}
\int_{x}^{x_0-\Gamma} d z {\hat P}^{[p]}_{\Gamma}(x_0\searrow z) 
{\hat P}^{[p]}_{\Gamma}(z\nearrow z+\Gamma) 
{\hat P}^{[p]}_{\Gamma}(z+\Gamma\searrow x)
\nonumber \\
&& =
- R_{\Gamma}^-(x_0-\Gamma)  
+ R_{\Gamma}^+( x+\Gamma)
+   \int_x^{x_0-\Gamma} dz \partial_{\Gamma} R_{\Gamma}^{[p]-}(z) 
\nonumber \\
&&-R_{\Gamma}^+( x+\Gamma)+ R_{\Gamma}^-( x_0-\Gamma)+  \int_{x}^{x_0-\Gamma} d z 
  R_{\Gamma}^-(z) 
\frac{ s'(z+\Gamma)[ s(z+\Gamma)-s(z)] }{ [K^{[p]}(z,z+\Gamma)]^2 }  
\nonumber \\
&& =
   \int_x^{x_0-\Gamma} dz \left[ \partial_{\Gamma} R_{\Gamma}^{[p]-}(z) 
+       \frac{ s'(z)  s'(z+\Gamma) }{ [K^{[p]}(z,z+\Gamma)]^2 }      \right]
  \label{RGPdespexpli}
\end{eqnarray}

In summary, the RG-flows of Eq. \ref{RGPascpexpli}
and \ref{RGPdespexpli}
require
\begin{eqnarray}
\partial_{\Gamma}  R_{\Gamma}^{[p]+}(z)  
  && = -    \frac{s'(z-\Gamma) s'(z)  }{ [K^{[p]}(z-\Gamma,z)]^2 }
\nonumber \\
 \partial_{\Gamma} R_{\Gamma}^{[p]-}(z) 
&& =  -    \frac{ s'(z)  s'(z+\Gamma) }{ [K^{[p]}(z,z+\Gamma)]^2 }      
  \label{rgRp}
\end{eqnarray}

The derivatives with respect to $\Gamma$ of $R_{\Gamma}^{[p]\pm}(x) $ given in Eq. \ref{Rp+} 
and \ref{Rp-} can be simplified using the identity of Eq. \ref{kabcrossdiff}
 \begin{eqnarray}
\partial_{\Gamma} R_{\Gamma}^{[p]+}(x) && = -  \frac{\partial_a \partial_b K^{[p]}(x-\Gamma,x)  }{ K^{[p]}(x-\Gamma,x) } 
+ \frac{ [\partial_b K^{[p]}(x-\Gamma,x) ]  [\partial_a K^{[p]}(x-\Gamma,x) ]}{ [K^{[p]}(x-\Gamma,x)]^2 }
= -  \frac{s'(x-\Gamma) s'(x)  }{ [K^{[p]}(x-\Gamma,x)]^2 }
\nonumber \\
\partial_{\Gamma}R_{\Gamma}^{[p]-}(x) && =  - \frac{ \partial_b \partial_a K^{[p]}(x,x+\Gamma)  }{ K^{[p]}(x,x+\Gamma) } 
+ \frac{ [\partial_a K^{[p]}(x,x+\Gamma) ] [\partial_b K^{[p]}(x,x+\Gamma) ]}{ [K^{[p]}(x,x+\Gamma)]^2 } 
= -  \frac{ s'(x)  s'(x+\Gamma) }{ [K^{[p]}(x,x+\Gamma)]^2 }    
  \label{Rp+-deri}
\end{eqnarray}
and thus satisfy the RG-flow of Eq. \ref{rgRp}
as it should for consistency between the two perspectives.

%%%%%%%%%%%%%%%%%%%%%%%%%%%%%%%%%%%%%%%%%%

\section{ Detailed-Balance properties at scale $\Gamma$ and similarity transformation towards symmetric blocks }

The properties discussed in subsection \ref{subsec_similarity} of the main text concerning $p=0$
can be generalized to $p>0$ as follows:

(i) Using the extrema densities $n_{\Gamma}^{\pm}(.) $ of Eq \ref{ngammackappa0}
and $R_{\Gamma}^{\pm}(.)$ of Eqs \ref{Rxplus}
\ref{Rxmoins}, the blocks obtained via similarity from Eq. \ref{Pplusp}
\begin{eqnarray}
{\hat B}_{\Gamma}^{[p]+}(x , x_0) && \equiv 
\frac{1}{\sqrt{ n_{\Gamma}^+(x)}}
{\hat P}_{\Gamma}^{[p]+}(x \vert x_0) 
\sqrt{n_{\Gamma}^-(x_0) }
\nonumber \\
&&  = 
\frac{ \sqrt{ s'(x) s'(x_0)} }{ K^{[p]}(x_0,x_0+\Gamma) } 
   e^{ - \int_{x_0+\Gamma}^x dz R_{\Gamma}^{[p]+}(z)  } 
  \ \ \ \ \text{ for } \ \ x \in [x_0+\Gamma,+\infty[
  \label{Bplusp}
\end{eqnarray}
and from Eq. \ref{Pmoinsp}
\begin{eqnarray}
{\hat B}_{\Gamma}^{[p]-}(x , x_0) && \equiv 
\frac{1}{\sqrt{ n_{\Gamma}^-(x)}}
{\hat P}_{\Gamma}^{[p]-}(x \vert x_0) 
\sqrt{n_{\Gamma}^+(x_0) }
\nonumber \\
&&  =
\frac{ \sqrt{ s'(x) s'(x_0) } }{ K^{[p]}(x_0-\Gamma,x_0) } 
   e^{ - \int_x^{x_0-\Gamma} dz R_{\Gamma}^{[p]-}(z)  } 
 \ \ \ \ \text{ for } \ \  x \in ]-\infty,x_0-\Gamma]
   \label{Bmoinsp}
\end{eqnarray}

(ii) The two functions of Eqs \ref{Rp+} and \ref{Rp-}
 \begin{eqnarray}
R_{\Gamma}^{[p]+}(x) && 
= \partial_b \ln K^{[p]}(x-\Gamma,x)
\nonumber \\
R_{\Gamma}^{[p]-}(x) && 
= - \partial_a \ln K^{[p]}(x,x+\Gamma)
  \label{Rp+-}
\end{eqnarray}
yields that the integral that appear in the block ${\hat B}_{\Gamma}^{[p]-}(x , x_0) $ of Eq. \ref{Bmoinsp}
\begin{eqnarray}
 \int_x^{x_0-\Gamma} dz R_{\Gamma}^{[p]-}(z)  
&& = -  \int_x^{x_0-\Gamma} dz  \partial_a \ln K^{[p]}(z,z+\Gamma)
= - \int_{x+\Gamma}^{x_0} dy \partial_a \ln K^{[p]}(y-\Gamma,y)
\nonumber \\
&& = - \int_{x+\Gamma}^{x_0} dy \left( \partial_y \ln K^{[p]}(y-\Gamma,y) - \partial_b \ln K^{[p]}(y-\Gamma,y)\right) 
\nonumber \\
&& = - \left[  \ln K^{[p]}(y-\Gamma,y) \right]_{x+\Gamma}^{x_0}
+ \int_{x+\Gamma}^{x_0} dy  \partial_b \ln K^{[p]}(y-\Gamma,y)
\nonumber \\
&& =- \ln K^{[p]}(x_0-\Gamma,x_0) + \ln K^{[p]}(x,x+\Gamma)
+ \int_{x+\Gamma}^{x_0} dy R_{\Gamma}^{[p]+}(y)
   \label{integR-p}
\end{eqnarray}
is related up to boundary terms to the integral that appears in the block ${\hat B}_{\Gamma}^{[p]+}(x_0 , x) $
of Eq. \ref{Bplusp}, and one obtains the symmetry property of the blocks for ascending and descending points
via the exchange of their end-points
\begin{eqnarray}
{\hat B}_{\Gamma}^{[p]-}(x , x_0) 
&&  =
\frac{ \sqrt{ s'(x) s'(x_0) } }{ K^{[p]}(x_0-\Gamma,x_0) } 
   e^{ - \int_x^{x_0-\Gamma} dz R_{\Gamma}^{[p]-}(z)  } 
=\frac{ \sqrt{ s'(x) s'(x_0) } }{ K^{[p]}(x_0-\Gamma,x_0) } 
   e^{ \ln K^{[p]}(x_0-\Gamma,x_0) - \ln K^{[p]}(x,x+\Gamma)
- \int_{x+\Gamma}^{x_0} dy R_{\Gamma}^{[p]+}(y) }
\nonumber \\
&& = \frac{ \sqrt{ s'(x) s'(x_0) } }{ K^{[p]}(x,x+\Gamma) } 
   e^{- \int_{x+\Gamma}^{x_0} dy R_{\Gamma}^{[p]+}(y) } = {\hat B}_{\Gamma}^{[p]+}(x_0 , x) 
   \label{SymBlockBexp}
\end{eqnarray}
that generalizes Eq. \ref{SymBlockBex}
of the main text concerning $p=0$.

(iii) Taking the logarithms of the explicit forms of Eqs \ref{Bplusp} and \ref{Bmoinsp}
\begin{eqnarray}
\ln {\hat B}_{\Gamma}^{[p]+}(x , x_0) 
&&  = 
\ln( \sqrt{ s'(x) s'(x_0)}) - \ln ( K^{[p]}(x_0,x_0+\Gamma)  )
    - \int_{x_0+\Gamma}^x dz R_{\Gamma}^{[p]+}(z)  
    \nonumber \\
\ln {\hat B}_{\Gamma}^{[p]-}(x , x_0) &&  =
\ln( \sqrt{ s'(x) s'(x_0) }) - \ln(  K^{[p]}(x_0-\Gamma,x_0)  )
    - \int_x^{x_0-\Gamma} dz R_{\Gamma}^{[p]-}(z)  
   \label{ExpliciBlockBp}
\end{eqnarray}
one obtains that their RG-flows reduce to
\begin{eqnarray}
\partial_{\Gamma} \ln {\hat B}_{\Gamma}^{[p]+}(x , x_0) 
&&  = 
 - \partial_b \ln ( K^{[p]}(x_0,x_0+\Gamma)  )
    + R_{\Gamma}^{[p]+}(x_0+\Gamma)  
    - \int_{x_0+\Gamma}^x dz \partial_{\Gamma}R_{\Gamma}^{[p]+}(z)  
    = - \int_{x_0+\Gamma}^x dz \partial_{\Gamma}R_{\Gamma}^{[p]+}(z)  
    \nonumber \\
\partial_{\Gamma}\ln {\hat B}_{\Gamma}^{[p]-}(x , x_0) &&  =
  \partial_a  \ln(  K^{[p]}(x_0-\Gamma,x_0)  )
    + R_{\Gamma}^{[p]-}(x_0-\Gamma)  
    - \int_x^{x_0-\Gamma} dz \partial_{\Gamma} R_{\Gamma}^{[p]-}(z)  
    = - \int_x^{x_0-\Gamma} dz \partial_{\Gamma} R_{\Gamma}^{[p]-}(z) 
   \label{ExpliciBlockBpderi}
\end{eqnarray}
where one can plug Eq. \ref{rgRp}
\begin{eqnarray}
\partial_{\Gamma} \ln {\hat B}_{\Gamma}^{[p]+}(x , x_0) 
&&  = 
   \int_{x_0+\Gamma}^x dz \frac{s'(z-\Gamma) s'(z)  }{ [K^{[p]}(z-\Gamma,z)]^2 }
    \nonumber \\
\partial_{\Gamma}\ln {\hat B}_{\Gamma}^{[p]-}(x , x_0) &&  =
    \int_x^{x_0-\Gamma} dz   \frac{ s'(z)  s'(z+\Gamma) }{ [K^{[p]}(z,z+\Gamma)]^2 }  
   \label{RGBlnp}
\end{eqnarray}
that generalizes Eq. \ref{RGBln}
of the main text concerning $p=0$.

For the blocks ${\hat B}_{\Gamma}^{[p]\pm}$ themselves, these RG-flows 
are similar to Eqs \ref{RGPascp} and \ref{RGPdesp}, except that only the 
merging term involving three blocks is present on the right handside,
i.e. 
\begin{eqnarray}
 \partial_{\Gamma} {\hat B}^{[p]}_{\Gamma}(x_0\nearrow x )  
&&= 
  \int_{x_0+\Gamma}^{x}d z 
{\hat B}^{[p]}_{\Gamma}(x_0\nearrow z )
{\hat B}^{[p]}_{\Gamma}(z\searrow z-\Gamma ) 
{\hat B}^{[p]}_{\Gamma}(z-\Gamma\nearrow x )
\nonumber \\
\partial_{\Gamma} {\hat B}^{[p]}_{\Gamma}(x_0\searrow x)
&& = \int_{x}^{x_0-\Gamma} d z {\hat B}^{[p]}_{\Gamma}(x_0\searrow z) 
{\hat B}^{[p]}_{\Gamma}(z\nearrow z+\Gamma) 
{\hat B}^{[p]}_{\Gamma}(z+\Gamma\searrow x)
   \label{RGBpmerging}
\end{eqnarray}
in agreement with \cite{rgtoy}.

%%%%%%%%%%%%%%%%%%%%%%%%%%%%%%%%%%%%%%%%%

\section{ Laplace properties for diffusions on $x \in ]0,+\infty[$ }

As in section \ref{sec_HalfLine} of the main text, the properties of ascending bonds of Eq. \ref{Pplusp}
that are defined for $x \in [x_0+\Gamma,+\infty[ $
does not change, apart from the facts that the starting point belongs to $x_0 \in ]0,+\infty[$
and the end-point belongs to $x \in ]\Gamma,+\infty[$,
but one needs to reconsider the descending bonds.

For descending bonds, the Laplace transform of Eq. \ref{Pmoinsp} is now valid for $x \in ]0,x_0-\Gamma] $
\begin{eqnarray}
{\hat P}_{\Gamma}^{[p]-}(x \vert x_0)
 = \frac{ s(x_0)-s(x_0-\Gamma) }{ K^{[p]}(x_0-\Gamma,x_0) } 
  R_{\Gamma}^-(x) e^{ - \int_x^{x_0-\Gamma} dz R_{\Gamma}^{[p]-}(z)  } 
 \ \ \ \ \text{ for } \ \  x \in ]0,x_0-\Gamma]
   \label{Pmoinsphalf}
\end{eqnarray}
and one needs to discuss the various possibilities for the boundary :

(N) When the boundary $x=0$ is never reached $s(0^+)=-\infty$, there is no missing weight as explained
in Eq. \ref{Wmissing0} of the main text, so Eq. \ref{Pmoinsphalf} is the full result for descending bonds.

(I) If $s(0^+)$ is finite but $\tau^-_{[0,x]}(x) $ diverges, then the finite weight $W_{\Gamma}(x_0) $ of Eq. \ref{Wmissing}
represents the probability that the particle flows towards zero without any further drawup $\Gamma$.
However, these events have an infinite duration, so they are present only for $p=0$ considered in the main text,
but they are killed for any $p>0$ by the factor $e^{-p (t=\infty) }=0$, so that 
Eq. \ref{Pmoinsphalf} is still the full result for descending bonds for any $p>0$.

(R) If $s(0^+)$ is  finite and $\tau^-_{[0,x]}(x) $ finite, then the finite weight $W_{\Gamma}(x_0) $
represents the probability that the descending bond starting at $x_0$
ends at the boundary $x=0$, where the particle will be reflected until the last-passage at $x=0$
that will be the start of the next ascending bond able to reach at least $\Gamma$. 
As a consequence, the duration associated to this descending bond
is the sum of three contributions instead of only two as in Eq. \ref{t1t2}
\begin{eqnarray}
 && t=t_1+t_2+t_3
\nonumber \\
&& \text{$t_1$ is the time to go from the maximum $x_0$ to $(x_0-\Gamma)$ }
\nonumber \\
&& \text{$t_2$ is the time to go from $(x_0-\Gamma)$ down to the boundary $x=0$ without any drawup $\Gamma$}
\nonumber \\
&& \text{$t_3$ is the time between the first passage at $x=0$ and the last-passage at $x=0$ representing the starting point }
\nonumber \\
&& \ \ \ \ \ \ \ \ \ \ \text{ of  the next ascending bond  }
  \label{t1t2t3}
\end{eqnarray}
In the presence of the Laplace parameter $p>0$, Eq. \ref{Pmoins0R}
will be thus generalized into
\begin{eqnarray}
{\hat P}_{\Gamma}^{[p]-}(x \vert x_0)
 = \frac{ s(x_0)-s(x_0-\Gamma) }{ K^{[p]}(x_0-\Gamma,x_0) } 
  R_{\Gamma}^-(x) e^{ - \int_x^{x_0-\Gamma} dz R_{\Gamma}^{[p]-}(z)  } 
   \theta(0<x<x_0-\Gamma)
+ {\hat W}^{[p]}_{\Gamma}(x_0) \delta(x)
   \label{Pmoinsphalf0}
\end{eqnarray}
where the amplitude ${\hat W}^{[p]}_{\Gamma}(x_0) $ of the delta function $\delta(x)$
is the product of the three Laplace transforms associated to the three times of Eq. \ref{t1t2t3}
\begin{eqnarray}
 {\hat W}^{[p]}_{\Gamma}(x_0)  
 = \frac{ s(x_0)-s(x_0-\Gamma) }{ K^{[p]}(x_0-\Gamma,x_0) }
 \times  e^{ - \int_0^{x_0-\Gamma} dz R_{\Gamma}^{[p]-}(z)  } 
 \times {\hat w}^{[p]}_{\Gamma}
   \label{WpProuct3}
\end{eqnarray}
with the following physical meaning :

$\bullet$ The first factor $ \frac{ s(x_0)-s(x_0-\Gamma) }{ K^{[p]}(x_0-\Gamma,x_0) }$
represents the Laplace transform ${\hat  \Pi}^{[p]-}_{\Gamma}( x_0)$ of the time $t_1$ needed to go from the maximum $x_0$ towards $(x_0-\Gamma)$ that is exactly the same as the first term in the bulk expression of Eq. \ref{Pmoinsphalf}

$\bullet $ The second factor $ e^{ - \int_0^{x_0-\Gamma} dz R_{\Gamma}^{[p]-}(z)  } $ represents the Laplace transform 
 of the time $t_2$, the only difference with respect to the bulk contribution is the absence of factor $R_{\Gamma}^-(x) $
 that imposes the end of the bond in the bulk, while here the end is imposed by the boundary $x=0$.
 
 $\bullet$ The third factor ${\hat w}^{[p]}_{\Gamma} $ represents the Laplace transform of the time $t_3$,
 that is a new type of contribution associated to the boundary, and that can be computed as follows.
 Let us consider the sum $(t_3+t_1^+)$ with the first contribution $t_1^+$ of the next ascending bond :
 the Laplace transform of $t_1^+$ alone is given by ${\hat  \Pi}^{[p]+}_{\Gamma}( x_0=0)  $ of Eq. \ref{Pi1Laplace}
 \begin{eqnarray}
{\hat  \Pi}^{[p]+}_{\Gamma}( 0) 
 = \frac{ s(\Gamma)-s(0) }{ K^{[p]}(0,\Gamma) } 
  \label{Pi1Laplace0}
\end{eqnarray}
 The sum $(t_3+t_1^+)$ correspond to the first-passage time at $\Gamma$ when starting at $x=0$ in the presence 
 of a reflecting boundary at $x=0$. For an arbitrary starting point $x \in ]0,\Gamma[$, the Laplace transform of this time is given by the ratio $\frac{ g_R^{[p]}(x)}{ g_R^{[p]}(\Gamma)} $, where $g_R^{[p]}$ is the solution
of the same equation as in Eq. \ref{LaplaceEqadjointExpanded}
\begin{eqnarray}
p g_R^{[p]}(x) && ={\cal F}^{\dagger} g_R^{[p]}(x)   =    F_I(x) \frac{dg_R^{[p]}(x) }{dx}  + D(x) \frac{d^2g_R^{[p]}(x) }{dx^2}
 =  \frac{ 1}{m'(x) }  \frac{d}{dx} \left( \frac{ \frac{dg_R^{[p]}(x) }{dx}}{s'(x) }  \right)
\label{LaplacegR}
\end{eqnarray}
that satisfies the reflecting boundary condition at $x=0$, i.e. the vanishing current boundary condition at $x=0$
that read for observable satisfying the backward Fokker-Planck equation
 \begin{eqnarray}
0&& = \lim_{x \to 0} \left( D(x) e^{-U(x)} \frac{dg_R^{[p]}(x) }{dx} \right) = \lim_{x \to 0} \left( \frac{ \frac{dg_R^{[p]}(x) }{dx}}{s'(x) } \right)
\label{ReflectingBC}
\end{eqnarray}
So the Laplace transform ${\hat w}^{[p]}_{\Gamma} $ of the time $t_3$ can be obtained from the ratio
of the Laplace transform $\frac{ g_R^{[p]}(x=0)}{ g_R^{[p]}(\Gamma)} $ of the sum $(t_3+t_1^+)$
divided by the Laplace transform ${\hat  \Pi}^{[p]+}_{\Gamma}( x_0=0)  $ of the time $t_1^+$
 \begin{eqnarray}
{\hat w}^{[p]}_{\Gamma}  = \frac{\frac{ g_R^{[p]}(0)}{ g_R^{[p]}(\Gamma)}}{{\hat  \Pi}^{[p]+}_{\Gamma}( 0) }
 = \frac{g_R^{[p]}(0) K^{[p]}(0,\Gamma) }{g_R^{[p]}(\Gamma) [s(\Gamma)-s(0)] } 
  \label{wpt3}
\end{eqnarray}

%%%%%%%%%%%%%%%%%%%%%%%%%%%%%%%%%

\section{ Examples of statistics of renormalized trajectories with the Laplace parameter $p$} 

In contrast to the main text concerning $p=0$ where the properties of the renormalized trajectories
can be written in terms of the scale function alone,
the analysis of renormalized trajectories where one keeps the information on the durations of the bonds
require to work in Laplace with the parameter $p$ and to be able to solve 
the second-order differential Equation of Eq. \ref{LaplaceEqadjointExpanded} for any $p$.
In many simple examples considered in the main text, 
the solutions involves hypergeometric functions and special functions (see the example of Bessel processes in \cite{schehr10}),
so in this Appendix, we have chosen to focus only on examples where 
the solutions involve only elementary functions in order to be more concrete.

\subsection{ Brownian motion with or without drift on $]-\infty,+\infty[$ }

Let us describe how the formula above reproduce the well-known results for 
the Brownian motion with or without drift.

(i) For the Brownian motion with $D=1$ and drift $F_I(x)=F$ already considered in subsection \ref{subsec_BrownF}
of the main text concerning $p=0$, 
the two independent solutions $g^{[p]}_{1,2}(x)$ of Eq. \ref{LaplaceEqadjointExpanded}
\begin{eqnarray}
p g(x) && ={\cal F}^{\dagger} g(x)   =    F g'(x)  + g''(x)
\label{LaplaceEqadjointExpandedBrown}
\end{eqnarray}
are two exponential functions
\begin{eqnarray}
g^{[p]}_1(x) && = e^{ \left( - \frac{F}{2}- \frac{\sqrt{F^2+4 p}}{2} \right) x} 
\nonumber \\
g^{[p]}_2(x) && =e^{ \left( - \frac{F}{2}+ \frac{\sqrt{F^2+4 p}}{2} \right) x} 
\label{LaplaceEqadjointExpandedBrownSol}
\end{eqnarray}
with their Wronskian $W^{[p]}(x) $ of Eq. \ref{Wronskian}
\begin{eqnarray}
W^{[p]}(x) &&\equiv  \frac{ g^{[p]}_{1} g^{[p]'}_2 (x) -  g^{[p]'}_1 (x) g^{[p]}_{2}}{s'(x) } 
= \sqrt{F^2+4 p}
  \label{WronskianBrown}
\end{eqnarray}
The kernel of Eq. \ref{kab} reads
\begin{eqnarray}
K^{[p]}(a,b) && = \frac{ g^{[p]}_1(a)g^{[p]}_2(b)- g^{[p]}_2(a)g^{[p]}_1(b) }{ W^{[p]} }
=e^{-\frac{F}{2}(a+b) }  \frac{ e^{   \frac{\sqrt{F^2+4 p}}{2}  (b-a) } 
-  e^{ - \frac{\sqrt{F^2+4 p}}{2}  (b-a)}  }
{ \sqrt{F^2+4 p} }
\nonumber \\
&& =e^{-\frac{F}{2}(a+b) }  \frac{  \sinh \left[ \sqrt{  \frac{F^2}{4}+p }  (b-a) \right] }
{ \sqrt{  \frac{F^2}{4}+p } }
\label{kabBrown}
\end{eqnarray}
with its logarithmic derivatives $ \partial_a \ln K^{[p]}(a,b)$ and $ \partial_b \ln K^{[p]}(a,b)$
 \begin{eqnarray}
\partial_b \ln K^{[p]}(a,b) && =-\frac{F}{2} + \sqrt{  \frac{F^2}{4}+p } \coth\left[ \sqrt{  \frac{F^2}{4}+p }  (b-a) \right]
\nonumber \\
 \partial_a \ln K^{[p]}(a,b) && =-\frac{F}{2}- \sqrt{  \frac{F^2}{4}+p } \coth\left[ \sqrt{  \frac{F^2}{4}+p }  (b-a) \right]
  \label{kabderiBrown}
\end{eqnarray}

So the two Laplace transforms of Eqs \ref{Pi1Laplace} and \ref{Pi1Laplace-} do not depend on $x_0$ and coincide
\begin{eqnarray}
{\hat  \Pi}^{[p]+}_{\Gamma}( x_0) 
 = \frac{ K^{[0]}(x_0,x_0+\Gamma) }{ K^{[p]}(x_0,x_0+\Gamma) } 
 =   \frac{\frac{  \sinh \left( \frac{F}{2} \Gamma \right) }{ \frac{F}{2} } }
 {   \frac{  \sinh \left( \sqrt{  \frac{F^2}{4}+p } \Gamma \right) }{ \sqrt{  \frac{F^2}{4}+p } } } \equiv {\hat  \Pi}^{[p]}_{\Gamma}
\nonumber \\
  {\hat  \Pi}^{[p]-}_{\Gamma}( x_0)  
  = \frac{ K^{[0]}(x_0-\Gamma,x_0) }{ K^{[p]}(x_0-\Gamma,x_0) } 
  = \frac{ \frac{  \sinh \left( \frac{F}{2} \Gamma \right) }{ \frac{F}{2} }   }
  {   \frac{  \sinh \left( \sqrt{  \frac{F^2}{4}+p } \Gamma \right) }{ \sqrt{  \frac{F^2}{4}+p } }  } \equiv {\hat  \Pi}^{[p]}_{\Gamma}
   \label{Pi1LaplaceBrown}
\end{eqnarray}
and the two functions of
Eqs \ref{Rp+} \ref{Rp-} do not depend on $x$
 \begin{eqnarray}
R_{\Gamma}^{[p]+}(x) 
= \partial_b \ln K^{[p]}(x-\Gamma,x)
=-\frac{F}{2} + \sqrt{  \frac{F^2}{4}+p } \coth\left[ \Gamma \sqrt{  \frac{F^2}{4}+p }  \right] \equiv r_{\Gamma}^{[p]+}
\nonumber \\
R_{\Gamma}^{[p]-}(x) 
= -\partial_a \ln K^{[p]}(x,x+\Gamma)
= +\frac{F}{2}+ \sqrt{  \frac{F^2}{4}+p } \coth\left[ \Gamma \sqrt{  \frac{F^2}{4}+p }   \right]\equiv r_{\Gamma}^{[p]-}
  \label{Rp+-Brown}
\end{eqnarray}

So one recovers that the Laplace transforms of the bond probabilities are given by
\begin{eqnarray}
{\hat P}_{\Gamma}^{[p]+}(x \vert x_0)
 && = {\hat  \Pi}^{[p]}_{\Gamma} 
  r_{\Gamma}^+ e^{ - (x-x_0-\Gamma) r_{\Gamma}^{[p]+}  } 
  \ \ \ \ \text{ for } \ \ x \in [x_0+\Gamma,+\infty[
\nonumber \\
{\hat P}_{\Gamma}^{[p]-}(x \vert x_0)
&& ={\hat  \Pi}^{[p]}_{\Gamma} 
  r_{\Gamma}^- e^{ -(x_0-\Gamma-x) r_{\Gamma}^{[p]-}  } 
 \ \ \ \ \text{ for } \ \  x \in ]-\infty,x_0-\Gamma]
   \label{PplusmoinspBrown}
\end{eqnarray}
where $ r_{\Gamma}^{\pm} $ were written in Eq \ref{RxBrowndrift}.

For the Brownian motion without drift when $F=0$, Eqs \ref{Pi1LaplaceBrown}
and Eqs \ref{Rp+-Brown}
simplify into
\begin{eqnarray}
{\hat  \Pi}^{[p]}_{\Gamma}
 && =   \frac{  \Gamma \sqrt{  p } }
 {     \sinh \left( \sqrt{  p } \Gamma \right)  } 
\nonumber \\
r_{\Gamma}^{[p]\pm}&&  = \sqrt{  p } \coth\left( \Gamma \sqrt{ p }   \right)
   \label{LaplaceBrown0}
\end{eqnarray}
leading to
\begin{eqnarray}
{\hat P}_{\Gamma}^{[p]+}(x \vert x_0)
 && =   \frac{   \sqrt{  p } }
 {     \sinh \left( \sqrt{  p } \Gamma \right)  }  e^{ - (x-x_0-\Gamma) \sqrt{  p } \coth\left( \Gamma \sqrt{ p }   \right)  } 
  \ \ \ \ \text{ for } \ \ x \in [x_0+\Gamma,+\infty[
\nonumber \\
{\hat P}_{\Gamma}^{[p]-}(x \vert x_0)
&& =  \frac{   \sqrt{  p } }
 {     \sinh \left( \sqrt{  p } \Gamma \right)  }  e^{ -(x_0-\Gamma-x) \sqrt{  p } \coth\left( \Gamma \sqrt{ p }   \right)  } 
 \ \ \ \ \text{ for } \ \  x \in ]-\infty,x_0-\Gamma]
   \label{PplusmoinspBrown0}
\end{eqnarray}

%%%%%%%%%%%%%%%%%%%%%%%%%%%%%%%%%%%%%%%%%%%%%%%%%%

\subsection{ Reflected Brownian motion on $x \in ]0,+\infty[$ (Bessel $d=1)$ }

The reflected Brownian motion on $x \in ]0,+\infty[$ is the simplest process on the half-line
and its renormalized trajectories have been already analyzed in \cite{schehr10}
together with other Brownian motions with various constraints.
Within the present perspective,
this is the simplest case to analyze the properties of the boundary behavior 'Reflected',
while the bulk properties are the same as the free Brownian case described above except
for the different domains of definitions :

The Laplace transform ${\hat P}_{\Gamma}^{[p]+}(x \vert x_0) $ for ascending bonds is given by Eq. \ref{PplusmoinspBrown0}, now defined for $x_0 \in ]0,+\infty[$.

The Laplace transform ${\hat P}_{\Gamma}^{[p]-}(x \vert x_0) $ for descending bonds is given by Eq. \ref{PplusmoinspBrown0} with  Eq. \ref{Pmoinsp} in the domain $x>0$ and an additional contribution in $\delta(x)$
as explained in Eq. \ref{Pmoinsphalf0}
\begin{eqnarray}
{\hat P}_{\Gamma}^{[p]-}(x \vert x_0)
 =  \frac{   \sqrt{  p } }
 {     \sinh \left( \sqrt{  p } \Gamma \right)  }  e^{ -(x_0-\Gamma-x) \sqrt{  p } \coth\left( \Gamma \sqrt{ p }   \right)  } 
   \theta(0<x<x_0-\Gamma)
+ {\hat W}^{[p]}_{\Gamma}(x_0) \delta(x)
   \label{Pmoinsphalf0Brown}
\end{eqnarray}
where the amplitude ${\hat W}^{[p]}_{\Gamma}(x_0) $ of the delta function $\delta(x)$
is given by Eq. \ref{WpProuct3} that reads in the present case
\begin{eqnarray}
 {\hat W}^{[p]}_{\Gamma}(x_0)  
 =  \frac{  \Gamma \sqrt{  p } }
 {     \sinh \left( \sqrt{  p } \Gamma \right)  }  e^{ -(x_0-\Gamma) \sqrt{  p } \coth\left( \Gamma \sqrt{ p }   \right)  }
 \times {\hat w}^{[p]}_{\Gamma}
   \label{WpProuct3Browndef}
\end{eqnarray}
where the Laplace transform $ {\hat w}^{[p]}_{\Gamma}$ of Eq. \ref{wpt3}
involves the solution $g_R(x)=\cosh({\sqrt p} x)$ satisfying $p g(x)  ={\cal F}^{\dagger} g(x)   =   g''(x) $ with reflecting boundary conditions at $x=0$
 \begin{eqnarray}
{\hat w}^{[p]}_{\Gamma}  = \frac{\frac{ g_R^{[p]}(0)}{ g_R^{[p]}(\Gamma)}}{{\hat  \Pi}^{[p]}_{\Gamma}( 0) }
 = \frac{\frac{1}{\cosh({\sqrt p} \Gamma)} }
 {  \frac{  \Gamma \sqrt{  p } } {     \sinh \left( \sqrt{  p } \Gamma \right)  } } = \frac{\tanh \left( \sqrt{  p } \Gamma \right) }{\Gamma \sqrt{  p }}
  \label{wpt3Brown}
\end{eqnarray}
while the total contribution of Eq. \ref{WpProuct3} reduces to
\begin{eqnarray}
 {\hat W}^{[p]}_{\Gamma}(x_0)  
 = \frac{1}{\cosh({\sqrt p} \Gamma)} e^{ -(x_0-\Gamma) \sqrt{  p } \coth\left( \Gamma \sqrt{ p }   \right)  }
   \label{WpProuct3Brown}
\end{eqnarray}

%%%%%%%%%%%%%%%%%%%%%%%%%%%%%%%%%%%%%%%%%%%%%%

\subsection{ Geometric Brownian motion with the quadratic diffusion coefficient $D(x)=x^2$ and the linear Ito force $F_I(x)=(1-\gamma) x $ }

For the Geometric Brownian motion of subsection \ref{subsec_GBM},
the two independent solutions $g^{[p]}_{1,2}(x)$ of Eq. \ref{LaplaceEqadjointExpanded}
\begin{eqnarray}
p g(x) && ={\cal F}^{\dagger} g(x)   =    (1-\gamma) x g'(x)  + x^2 g''(x)
\label{LaplaceEqadjointExpandedGBM}
\end{eqnarray}
are two power functions
\begin{eqnarray}
g^{[p]}_1(x) && = x^{ \frac{\gamma}{2}- \frac{\sqrt{\gamma^2+4 p}}{2}  } 
\nonumber \\
g^{[p]}_2(x) && =x^{   \frac{\gamma}{2}+ \frac{\sqrt{\gamma^2+4 p}}{2} } 
\label{LaplaceEqadjointExpandedGBMSol}
\end{eqnarray}
with their Wronskian $W^{[p]}(x) $ of Eq. \ref{Wronskian}
\begin{eqnarray}
W^{[p]}(x) &&\equiv  \frac{ g^{[p]}_{1} g^{[p]'}_2 (x) -  g^{[p]'}_1 (x) g^{[p]}_{2}}{s'(x) } 
= \sqrt{\gamma^2+4 p}
  \label{WronskianGBM}
\end{eqnarray}
The kernel of Eq. \ref{kab} reads
\begin{eqnarray}
K^{[p]}(a,b) && = \frac{ g^{[p]}_1(a)g^{[p]}_2(b)- g^{[p]}_2(a)g^{[p]}_1(b) }{ W^{[p]} }
 =(ab)^{\frac{\gamma}{2} }  \frac{  \sinh \left[ \sqrt{  \frac{\gamma^2}{4}+p } \ln \frac{b}{a}  \right] }
{ \sqrt{  \frac{\gamma^2}{4}+p } }
\label{kabGBM}
\end{eqnarray}
with its logarithmic derivatives $ \partial_a \ln K^{[p]}(a,b)$ and $ \partial_b \ln K^{[p]}(a,b)$
 \begin{eqnarray}
\partial_b \ln K^{[p]}(a,b) && =\frac{\gamma}{2b} + \frac{\sqrt{  \frac{\gamma^2}{4}+p } }{b}\coth\left[ \sqrt{  \frac{\gamma^2}{4}+p } \ln \frac{b}{a}\right]
\nonumber \\
 \partial_a \ln K^{[p]}(a,b) && =\frac{\gamma}{2a} - \frac{\sqrt{  \frac{\gamma^2}{4}+p } }{a}\coth\left[ \sqrt{  \frac{\gamma^2}{4}+p } \ln \frac{b}{a}\right]
  \label{kabderiGBM}
\end{eqnarray}

 So the two Laplace transforms of Eqs \ref{Pi1Laplace} and \ref{Pi1Laplace-} read
\begin{eqnarray}
{\hat  \Pi}^{[p]+}_{\Gamma}( x_0) 
 = \frac{ K^{[0]}(x_0,x_0+\Gamma) }{ K^{[p]}(x_0,x_0+\Gamma) } 
 =   \frac{ \frac{(x_0+\Gamma)^{\gamma}- x_0^{\gamma}}{\gamma}}
 {  x_0^{\frac{\gamma}{2} } (x_0+\Gamma)^{\frac{\gamma}{2} }  \frac{  \sinh \left[ \sqrt{  \frac{\gamma^2}{4}+p } \ln \frac{x_0+\Gamma}{x_0}  \right] }
{ \sqrt{  \frac{\gamma^2}{4}+p } }  }
\nonumber \\
  {\hat  \Pi}^{[p]-}_{\Gamma}( x_0)  
  = \frac{ s(x_0) - s(x_0-\Gamma) }{ K^{[p]}(x_0-\Gamma,x_0) } 
  = \frac{  \frac{x_0^{\gamma}-(x_0-\Gamma)^{\gamma}}{\gamma} }
  { (x_0-\Gamma)^{\frac{\gamma}{2} } x_0^{\frac{\gamma}{2} }  \frac{  \sinh \left[ \sqrt{  \frac{\gamma^2}{4}+p } \ln \frac{x_0}{x_0-\Gamma}  \right] }
{ \sqrt{  \frac{\gamma^2}{4}+p } }  }
   \label{Pi1LaplaceGBM}
\end{eqnarray}
and the two functions of
Eqs \ref{Rp+} \ref{Rp-} 
 \begin{eqnarray}
R_{\Gamma}^{[p]+}(x) 
= \partial_b \ln K^{[p]}(x-\Gamma,x)
=\frac{\gamma}{2x} + \frac{\sqrt{  \frac{\gamma^2}{4}+p } }{x}\coth\left[ \sqrt{  \frac{\gamma^2}{4}+p } \ln \frac{x}{x-\Gamma}\right]
\nonumber \\
R_{\Gamma}^{[p]-}(x) 
= -\partial_a \ln K^{[p]}(x,x+\Gamma)
= -\frac{\gamma}{2x} + \frac{\sqrt{  \frac{\gamma^2}{4}+p } }{x}\coth\left[ \sqrt{  \frac{\gamma^2}{4}+p } \ln \frac{x+\Gamma}{x}\right]
  \label{Rp+-GBM}
\end{eqnarray}

%%%%%%%%%%%%%%%%%%%%%%%%%%%%%%%%%%%%%%%%%%%

\section{ Adaptation to discrete-time random walks on the one-dimensional lattice }

\label{sec_DiscreteTimeRW}

In this section, we describe how the framework of the previous sections
can be adapted to discrete-time random walks on the one-dimensional lattice
with arbitrary nearest-neighbor jump probabilities
\begin{eqnarray}
P(x + 1,t+1 \vert x,t) && = q_x \equiv M(x+1,x)
\nonumber \\
P(x - 1,t+1 \vert x,t) && =1-q_x \equiv M(x-1,x) 
\label{discreteTimeRWdef}
\end{eqnarray}
i.e. the dynamics for the probability $P_t(x)$ to be on site $x$ at time $t$ reads
\begin{eqnarray}
 P_{t+1}(x)   && =  q_{x-1} P_t(x-1)  + (1-q_{x+1}) P_t(x+1) 
\label{discreteTimeRW}
\end{eqnarray}
Since all the results of the previous sections were based on the scale function $s(x)$ of Eq. \ref{Sx}
governing the exit probabilities of Eq. \ref{ExitRx}, one needs first to write their analogs on the lattice.

\subsection{ Scale function $s(x)$ governing the exit probabilities $E^{\pm}_{[x_L,x_R]}(x) $ on the lattice   } 

The exit probabilities $E^{\pm}_{[x_L,x_R]}(x) $ are given by the same formula as Eq. \ref{ExitRx}
on the lattice in terms of the discrete scale function $s(x)$ 
that satisfies the backward equation analogous to Eq. \ref{adjointzero}
\begin{eqnarray}
0= - s(x) + \sum_{y} M^{\dagger} (x,y) s(y) = - s(x) +\sum_{y}   s(y) M(y,x) 
=q_x  \left[ s(x+1) -s(x)\right] -   (1-q_x) \left[ s(x) -s(x-1) \right]
\label{adjointzerodiscrete}
\end{eqnarray}
besides the constant solution associated to the conservation of probabilities.
So the increments satisfy the recurrence
\begin{eqnarray}
\left[ s(x+1) -s(x)\right] =\frac{   (1-q_x)}{q_x}  \left[ s(x) -s(x-1) \right]
\label{adjointzerodiscreterec}
\end{eqnarray}
It is convenient to choose the condition
\begin{eqnarray}
s(1)-s(0)  =1
\label{Sxdiscretechoice}
\end{eqnarray}
that leads to
\begin{eqnarray}
s(x+1)  - s(x) && = \prod_{j=1}^x \frac{ 1-q_j}{q_j} \ \ \ \text{ for $x=1,2,..$}
\label{Sxdiscrete+}
\end{eqnarray}
and to 
\begin{eqnarray}
s(x+1)  - s(x) && = \prod_{i=1+x}^0 \frac{q_i}{ 1-q_i} \ \ \ \text{ for $x=...,-2,-1$ if defined on Z instead of N}
\label{Sxdiscrete-}
\end{eqnarray}

%%%%%%%%%%%%%%%%%%%%%%%%%%%%%%

\subsection{ Bond probabilities $P_{\Gamma}^{\pm}(x \vert x_0) $ on the lattice   } 

On the lattice, the function $R_{\Gamma}^{+}(x ) $ of Eq. \ref{Rxplus} 
 is replaced by the complementary exit probabilities
\begin{eqnarray}
r^+_{\Gamma}(x) && = E^-_{[x-\Gamma,x+1]}(x) = \frac{ s(x+1)-s(x) }{ s(x+1)-s(x-\Gamma) } 
\nonumber \\
 1-r^+_{\Gamma}(x) &&  = E^+_{[x-\Gamma,x+1]}(x)=  \frac{s(x)-s(x-\Gamma) }{ s(x+1)-s(x-\Gamma) } 
  \label{Rxplusdiscrete}
\end{eqnarray}
while the function $R_{\Gamma}^{-}(x ) $ of Eq. 
 \ref{Rxmoins}
 is replaced by the complementary exit probabilities
\begin{eqnarray}
r^-_{\Gamma}(x) &&  = E^+_{[x-1,x+\Gamma]}(x) = \frac{ s(x)-s(x-1) }{ s(x+\Gamma)-s(x-1) } 
\nonumber \\
1- r^-_{\Gamma}(x) &&  = E^-_{[x-1,x+\Gamma]}(x) = \frac{ s(x+\Gamma)-s(x) }{ s(x+\Gamma)-s(x-1) } 
  \label{Rxmoinsdiscrete}
\end{eqnarray}

The analog of the bond probabilities $P_{\Gamma}^{\pm}(x \vert x_0) $ of Eqs \ref{Pplus}
and \ref{Pmoins} then read for the lattice
\begin{eqnarray}
P_{\Gamma}^+(x \vert x_0)
= r_{\Gamma}^+(x) \prod_{y=x_0+\Gamma}^{x-1} [ 1-r^+_{\Gamma}(y)]
\ \ \ \  \text{ for  $ x > x_0+\Gamma$ 
with $P_{\Gamma}^+(x=x_0+\Gamma \vert x_0) =r_{\Gamma}^+(x_0+\Gamma) $ }
  \label{Pplusdiscrete}
\end{eqnarray}
and
\begin{eqnarray}
P_{\Gamma}^-(x \vert x_0) 
= r_{\Gamma}^-(x) \prod_{y=x+1}^{x_0-\Gamma} [ 1-r^-_{\Gamma}(y)] 
\ \ \ \ \text{ for  $ x < x_0-\Gamma$ 
with $P_{\Gamma}^-(x=x_0-\Gamma \vert x_0) =r_{\Gamma}^-(x_0-\Gamma) $}
  \label{Pmoinsdiscrete}
\end{eqnarray}

On the half-line $x=0,1,2..$ with reflecting boundary conditions $q_0=1$ at the origin,
 the analog of Eq. \ref{Pmoins0R} is that Eq. \ref{Pmoinsdiscrete} is valid for $1 \leq x \leq x_0-\Gamma$,
 while the weight at $x=0$ generalizing Eq. \ref{Wmissing}
 reads
 \begin{eqnarray}
W_{\Gamma}( x_0) \equiv  \prod_{y=1}^{x_0-\Gamma} [ 1-r^-_{\Gamma}(y)] 
  \label{Wmissingdiscrete}
\end{eqnarray}

\subsection{ Densities $n_{\Gamma}^{\pm}(x ) $ of extrema at scale $\Gamma$ on the lattice   } 

The analog of the densities $n_{\Gamma}^{\pm}(x ) $ of extrema at scale $\Gamma$ 
of Eq. \ref{ngammackappa}
read on the lattice
\begin{eqnarray}
n_{\Gamma}^{-}(x)&& 
=   \frac{[s(x) -s(x-1)][c- \kappa s(x+\Gamma) ] }{[s(x+\Gamma)-s(x)][s(x+\Gamma)-s(x-1)]}
\nonumber \\
n_{\Gamma}^{+}(x) &&  
=   \frac{ [s(x+1) -s(x) ] [c- \kappa s(x-\Gamma) ] }{[s(x)-s(x-\Gamma)][s(x+1)-s(x-\Gamma)]}
  \label{npmsolfinaldiscrete}
\end{eqnarray}

%%%%%%%%%%%%%%%%%%%%%%%%%%%%%%%%%%%%%%%

\subsection{ Example of the simple random walk with $q_y=\frac{1}{2}$ with the linear scale function $s(x)=x$  } 

Then the scale function $s(x)$ of Eq. \ref{adjointzerodiscrete}
is linear
\begin{eqnarray}
s(x)=x  
\label{sxlineardiscrete}
\end{eqnarray}
that is the analog Eq. \ref{spxlinear} for the Brownian universality class.
The exit probabilities of Eqs \ref{Rxplusdiscrete}
and \ref{Rxmoinsdiscrete}
\begin{eqnarray}
r^{\pm}_{\Gamma}(x) &&  = \frac{ 1 }{ \Gamma+1 } 
\nonumber \\
 1-r^{\pm}_{\Gamma}(x) &&  =  \frac{ \Gamma }{ \Gamma+1 } 
  \label{Rxdiscrete}
\end{eqnarray}
then lead to the geometric bond distribution of Eqs \ref{Pplusdiscrete} \ref{Pmoinsdiscrete}
\begin{eqnarray}
p_{\Gamma}^{+}(x \vert x_0)
= \frac{ 1 }{ \Gamma+1 }  \left(  \frac{ \Gamma }{ \Gamma+1 } \right)^{ x-x_0-\Gamma} 
\ \ \ \  \text{ for  $ x \geq x_0+\Gamma$ }
\nonumber \\
p_{\Gamma}^{-}(x \vert x_0)
= \frac{ 1 }{ \Gamma+1 }  \left(  \frac{ \Gamma }{ \Gamma+1 } \right)^{ x_0-\Gamma-x} 
\ \ \ \ \text{ for  $ x \leq x_0-\Gamma$ }
  \label{PRWdiscrete}
\end{eqnarray}

%%%%%%%%%%%%%%%%%%%%%%%%%%%%%%%%%%%%%%%

\subsection{ Example of the power-law scale function $s(x)=\frac{x^{\gamma} }{\gamma} $ on the half-lattice $x=0,1,2,..$}

The power-law scale function $s(x)=\frac{x^{\gamma} }{\gamma} $ analog to Eq. \ref{spxpower} corresponds on the half-lattice with reflecting B.C. $q_0=1$ at $0$
to the probabilities $q_x$ satisfying Eq. \ref{adjointzerodiscrete}
\begin{eqnarray}
\frac{   1-q_x}{q_x}= \frac{ s(x+1) -s(x) }{    s(x) -s(x-1) } 
= \frac{ (x+1)^{\gamma} - x^{\gamma} }{   x^{\gamma} -(x-1)^{\gamma} }
\label{qforspowerdiscrete}
\end{eqnarray}
leading to
\begin{eqnarray}
 q_x= \frac{   x^{\gamma} -(x-1)^{\gamma} }{ (x+1)^{\gamma}  -(x-1)^{\gamma} } \ \ \text{ for $ x \geq 1$ while $q_0=1$}
\label{qforspowerdiscretesol}
\end{eqnarray}
with the special cases :

(i) $\gamma=1$ (analog of subsection \ref{subsec_BrownianReflected}): the simple random walk $q_x=\frac{1}{2}$ reflected at the origin 

(ii) $\gamma=-1$ (analog of subsection \ref{subsec_gamma-1}) : 
\begin{eqnarray}
 q_x= \frac{  x+1 }{ 2x } \ \ \text{ for $ x \geq 1$ with $q_1=1$ (Reflection at $x=1$)}
\label{qforspowerdiscrete-1}
\end{eqnarray}

(iii) $\gamma=2$ (analog of subsection \ref{subsec_gamma2}) : 
\begin{eqnarray}
 q_x= \frac{  2x-1 }{ 4x } \ \ \text{ for $ x \geq 1$ while $q_0=1$}
\label{qforspowerdiscrete+2}
\end{eqnarray}

\subsection{ Discussion  }

In discrete-time random walks, each scale function $s(x)$ determines the probabilities $q_x$ of the model,
so it is interesting in the next Appendix to consider continuous-time random-walks
where each model is defined by two functions, as for diffusion processes discussed in the main text.

%%%%%%%%%%%%%%%%%%%%%%%%%%%%%%%%%%%%%%%%%%%

\section{ Adaptation to continuous-time random walks on the one-dimensional lattice }

\label{sec_ContinuousTimeRW}

In this section, we describe how the framework formulated in the previous sections
can be adapted to continuous-time random walks on the one-dimensional lattice
with arbitrary nearest-neighbor transition rates $w(y \pm 1,y)$ from $y$ towards its two neighbors $(y \pm 1)$.
The Master Equation for the probability $P_t(x)$ to be on site $x$ at time $t$
reads
\begin{eqnarray}
 \partial_t P_t(x)   && =  w(x,x-1) P_t(x-1)  + w(x,x+1) P_t(x+1) +w(x,x) P_t(x)
 \nonumber \\
 \ \ \text{ with the diagonal matrix element : } \ \ \ - w(x,x) && \equiv \bigg[ w(x-1,x) +w(x+1,x) \bigg] =  w^{out}(x) \text{ total rate out of $x$ }
\label{master}
\end{eqnarray}

\subsection{ Discrete scale function $s(x)$ and properties of the renormalized landscape that depend only on $s(x)$ }

The discrete scale function $s(x)$ 
satisfies the backward equation analogous to Eq. \ref{adjointzero}
\begin{eqnarray}
0 && = \sum_y s(y) w(y,x) = s(x+1) w(x+1,x) + s(x-1) w(x-1,x) +s(x) w(x,x)
\nonumber \\
&& = w(x+1,x)  \left[ s(x+1) -s(x)\right] - w(x-1,x) \left[ s(x) -s(x-1) \right]
\label{adjointmasterdiscrete}
\end{eqnarray}
that coincides with Eq. \ref{adjointzerodiscreterec} if one introduces the probabilities
to jump on the right or on the left when one jumps out of $x$ with rate $w^{out}(x) $
\begin{eqnarray}
q_x && \equiv \frac{w(x+1,x)}{ w^{out}(x)} =  \frac{w(x+1,x)}{ w(x-1,x) +w(x+1,x)}
\nonumber \\
1-q_x &&  \equiv \frac{w(x-1,x)}{ w^{out}(x)} =  \frac{w(x-1,x)}{ w(x-1,x) +w(x+1,x)} 
\label{qequivalent}
\end{eqnarray}
As a consequence, all the equations after Eq. \ref{adjointzerodiscreterec} written in the previous section,
 concerning the scale function $s(x)$ and the bond probabilities $p_{\Gamma}^{\pm}(.)$
 apply directly 
in terms of the effective probabilities $q_x$ computed from the rates in Eq. \ref{qequivalent}.

%%%%%%%%%%%%%%%%%%%%%%%%%%%%%%%%%

\subsection{ Discrete speed measure $m'(x)= e^{  -U(x) }$ and link with the notations of diffusion processes }

However the continuous-time RW are richer since once the scale function $s(x)$ is given, 
only the probabilities $q_x$ are determined, but one has still the freedom to choose 
the total rate $w^{out}(x) $ out of each site $x$, that plays a role similar to the diffusion coefficient $D(x)$ of diffusion processes. To make clearer the link with the notations of the main text concerning diffusion processes, it is useful to introduce the following parametrization of the transition rates \cite{c_susyFokkerPlanck,c_QES}
\begin{eqnarray}
 w(x+1,x)  =D \left(x+\frac{1}{2} \right) e^{ \frac{ U(x)-U(x+1) }{2} } = e^{- U_I\left(x+\frac{1}{2} \right) +U(x) }
 \nonumber \\
  w(x,x+1)  =D \left(x+\frac{1}{2} \right) e^{  \frac{ U(x+1)-U(x) }{2} }= e^{- U_I\left(x+\frac{1}{2} \right) +U(x+1) }
\label{rates}
\end{eqnarray}
where the diffusion coefficient $D \left(x+\frac{1}{2} \right)$ defined on the links 
and the potential $U(x)$ defined on the sites can be computed from the transition rates via
\begin{eqnarray}
D \left(x+\frac{1}{2} \right) && = \sqrt{ w(x+1,x)   w(x,x+1) }
\nonumber \\
U(x+1)-U(x) && = \ln \left( \frac{  w(x,x+1) }{  w(x+1,x) } \right)
\label{ratesInversion}
\end{eqnarray}
while the Ito potential $U_I(.)$ defined on the links satisfies
\begin{eqnarray}
e^{- U_I\left(x+\frac{1}{2} \right)  }  \equiv D \left(x+\frac{1}{2} \right) e^{ - \frac{ U(x)+U(x+1) }{2} }  
\label{UIdiscrete}
\end{eqnarray}
and has for increments using Eq. \ref{qequivalent}
\begin{eqnarray}
U_I\left(x+\frac{1}{2} \right) - U_I\left(x-\frac{1}{2} \right) 
= \ln \left(\frac{  w(x-1,x) }{w(x+1,x) } \right) =\ln \left(\frac{  1-q_x }{ q_x } \right)
\label{UIincrements}
\end{eqnarray}

Then the steady state is 
 the Boltzmann equilibrium in the potential $U(x)$
\begin{eqnarray}
  P_{eq}(x) \propto  e^{  -U(x) } = m'(x)
\label{steadyeqjump}
\end{eqnarray}
while the discrete scale function has for increments
\begin{eqnarray}
 s(y+1)-s(y) =e^{ U_I\left(y+\frac{1}{2} \right)  }
\label{Sxdiscrete}
\end{eqnarray}

%%%%%%%%%%%%%%%%%%%%%%%%%%%%%%%%%

\subsection{ Discussion }

As a final remark, let us mention that here we have considered arbitrary transition rates $w(y \pm 1,y)$ 
towards the two neighbors in the master equation of Eq. \ref{master}, while another
interesting case is when the walker performs a usual random walk $q_x=\frac{1}{2}$
but has to wait for a certain trapping time $\tau$ before each jump,
whose distribution displays a power-law tail in $\frac{1}{\tau^{1+\alpha}}$ with $0<\alpha<1$ 
leading to a diverging mean value and to a sub-diffusive behavior : its renormalized trajectories have been analyzed
 in \cite{schehr10} to obtain that the Laplace transform of the bond can be obtained from the Brownian case by the replacement of the Laplace variable $p \to p^{\alpha}$ (see \cite{schehr10} for more details).

%%%%%%%%%%%%%%%%%%%%%%%%%%%%%%%%%%%%%%%%%%%


\begin{thebibliography}{99}

%%%%%%%%%%%%%%%%%%%%%%%%%%%%%%%%%%%%%%%%%%%%%%%%%%%%%%
% References of Introduction

\bibitem{igloi05review}
% cond-mat/0502448
F. Igl\'oi and C. Monthus, 
%Strong disorder RG approach of random systems,
Phys. Rep. 412, 277 (2005).

\bibitem{igloiMiniReview}
%Strong Disorder RG approach - a short review of recent developments
F. Igl\'oi and C. Monthus, Eur. Phys. J. B (2018) 91: 290
%arXiv:1806.07684 [cond-mat.dis-nn]
 


\bibitem{ma79}       
       S.-K. Ma, C. Dasgupta, and C.-k. Hu,
%Random Antiferromagnetic Chain,
       Phys. Rev. Lett. 43, 1434 (1979) ; \\
      C. Dasgupta and S.-K. Ma
%Low-temperature properties of the random Heisenberg antiferromagnetic chain
       Phys. Rev. B 22, 1305 (1980).
       
  \bibitem{danielantiferro}
D. S. Fisher 
Phys. Rev. B 50, 3799 (1994) 
%``Random antiferromagnetic quantum spin chains".     

\bibitem{fisher92}
D. S. Fisher,
% Random transverse field Ising spin chains,
Phys. Rev. Lett. 69, 534 (1992) ; \\
D. S. Fisher,
% Critical behavior of random transverse-field Ising spin chains
Phys. Rev. B 51, 6411 (1995).

\bibitem{rgfinitesize}
D.S. Fisher and A. P. Young, 
Phys. Rev. B 58, 9131 (1998) 
%``Distributions of gaps and end-to-end correlations in random transverse-field Ising spin chains".

\bibitem{danielreview}
D.S. Fisher, Physica A 263 (1999) 222 
% ``Phase transitions and singularities in random quantum systems".


\bibitem{igloi98}
F. Igl\'oi and H. Rieger.
       Phys. Rev. B {\bf 57} (1998) 11404 
%``Random transverse Ising spin chain and random walks".

\bibitem{dharyoung}
A. Dhar and A.P. Young, 
Phys. Rev. B 68, 134441 (2003) 
%``Ensemble dependence in the random transverse-field Ising chain ".

\bibitem{microcano}
 C. Monthus, Phys. Rev. B 69, 054431 (2004)    
%``Finite-size scaling properties of random transverse-field Ising chains :
%comparison between canonical and microcanonical ensembles for the disorder". 


\bibitem{igloi98c} F. Igl\'oi and H. Rieger, Phys. Rev. E{\bf 58}, 4238 (1998)
%``Anomalous diffusion in disordered media and random quantum spin chains''

\bibitem{rgsinaishort}
 D. Fisher, P. Le Doussal and C. Monthus, Phys. Rev. Lett. 80  (1998) 3539
%  ``Random Walks, Reaction-Diffusion, and Nonequilibrium Dynamics 
% of Spin Chains in One-dimensional Random Environments". 


\bibitem{rgsinailong}
 D. S. Fisher, P. Le Doussal and C. Monthus,   Phys. Rev. E 59  (1999) 4795
%\\  ``Random Walkers in One-dimensional Random Environments : 
% Exact Renormalization Group Analysis".

\bibitem{rgrfim}
 D. S. Fisher, P. Le Doussal and C. Monthus,  Phys. Rev. E 64 (2001) 66107   
%``Nonequilibrium Dynamics of Random Field Ising Spin Chains :  exact results via real space RG". 

\bibitem{rgreadiff}
 P. Le Doussal and C. Monthus,  Phys. Rev. E 60 (1999) 1212   
%``Reaction Diffusion models in one dimension with disorder". 

\bibitem{locgolosov}
C. Monthus and  P. Le Doussal,   Phys. Rev. E 65 (2002) 66129 
% \\  ``Localization of thermal packets and metastable states in the Sinai model."


\bibitem{energysinai}
 C. Monthus and P. Le Doussal, Physica A 334 (2004) 78 
% \\  ``Energy dynamics in the Sinai model."



\bibitem{rgtoy}
P. Le Doussal and C. Monthus,  Physica A  317  (2003) 140   
%``Exact solutions for the statistics of extrema of some random 1D landscapes, Application to the equilibrium and the dynamics of the toy model."

\bibitem{schehr10}
% arxiv 0910.4913
G. Schehr and P. Le Doussal,
%Extreme value statistics from the real space renormalization group: Brownian motion, Bessel processes and continuous time random walks,
J. Stat. Mech.  P01009 (2010).


%%%%%%%%%%%%%%%%%%%%%%%%%%%%%%%%%%%%%%%%%%%
% Math 

\bibitem{Taylor75}
H.M. Taylor, Annals of Prob. 3, 234 (1975).

\bibitem{Lehoczky77}
J.P. Lehoczky, Annals of Prob. 5, 601 (1977).


\bibitem{2009}
L. Pospisil, J. Vecer and O. Hadjiliadis, Stochastic Processes and their Applications 119(8), 2563 (2009).
%“Formulas for stopped diffusion processes with stopping times based on drawdowns and drawups,” 

\bibitem{Z2012}
H. Zhang and O. Hadjiliadis, Methodology and Computing in Applied Probability 14(3), 739 (2012).
% “Drawdowns and the Speed of Market Crash,” 

\bibitem{Z2015}
H. Zhang, Advances in Applied Probability 47, 210 (2015)

\bibitem{proof}
%A Proof of Lehoczky's Theorem on Drawdowns
P.J. Fitzsimmons, arXiv:2209.04695


\bibitem{Salminen}
% arXiv:2411.18374
% Drawdowns of diffusions
P. Salminen and P. Vallois, 	
ESAIM: Probability and Statistics 29, 357 (2025)
% https://doi.org/10.1051/ps/2025010

\bibitem{BookZhang}
H. Zhang, "Stochastic drawdowns", World Scientific Publishing Singapore (2018)



%%%%%%%%%%%%%%%%%%%%%%%%%%%%%%%%%%%
% h-extrema in math

\bibitem{Neveu}
J. Neveu and J. Pitman, 
Séminaire de Probabilités XXIII, Lecture Notes in Mathematics 1372, 239 (1989).
%“Renewal property of the extrema and tree property of the excursion of a one-dimensional Brownian motion,” 

\bibitem{Foggionato}
A. Faggionato, Stochastic Processes and their Applications 119, 1765 (2009).
%“The alternating marked point process of hh-slopes of drifted Brownian motion,” 


%%%%%%%%%%%%%%%%%%%%%%%%%%%%%%%%
% Extremes
	
\bibitem{BookExtremes}
S. N Majumdar, G. Schehr
"Statistics of Extremes and Records in Random Sequences" 
Oxford University Press (2024)

\bibitem{SatyaReview}
%Extreme value statistics of correlated random variables: a pedagogical review
S. N. Majumdar, A. Pal, G. Schehr, Physics Reports 840, 1 (2020)
%arXiv:1910.10667

\bibitem{schehr_EVS}
% Extreme value statistics and some applications in statistical physics
M. P. Pruszczyk and G. Schehr, JSTAT 2026 (6), 064002
% arXiv:2603.18816 


%%%%%%%%%%%%%%%%%%%%%%%%%%%%
% Textbooks Math for scale function and speed measure


\bibitem{BookKarlin}
S. Karlin and H. Taylor, 
"A Second Course in Stochastic Processes", Academic Press, New York (1981).

\bibitem{BookYor}
D. Revuz and M. Yor, "Continuous martingales and Brownian motion",
Springer Verlag Berlin Heidelberg (1991)


\bibitem{HandBook}
A. N. Borodin and P. Salminen
"Handbook of Brownian Motion - Facts and Formulae"
Birkhäuser Verlag, Springer Basel (2002)


\bibitem{BookBorodin}
Andrei N. Borodin
"Stochastic Processes", Birkh\"auser
Springer International Publishing 
 Switzerland  (2017)
 
  
 %%%%%%%%%%%%%%%%%%%%%%%%%%%%%%%%%%%%%%%%%%
 % Textbook Exit

\bibitem{gardiner}
 C. W. Gardiner, “ Handbook of Stochastic Methods: for Physics, Chemistry and the Natural Sciences” (Springer Series
in Synergetics), Berlin (1985).

\bibitem{vankampen}
N.G. Van Kampen, “Stochastic processes in physics and chemistry”, Elsevier Amsterdam (1992).

\bibitem{risken}
 H. Risken, “The Fokker-Planck equation : methods of solutions and applications”, Springer Verlag Berlin (1989).
 
 %%%%%%%%%%%%%%%%%%%%%%%%%%%%%%%%%%%%%%%%%%%%
 % Detailed-Balance vers similarity towards symmetric
 
 \bibitem{glauber}
 R.J. Glauber, J. Math. Phys. 4, 294 (1963).
 
  \bibitem{Felderhof}
 B.U. Felderhof, Rep. Math. Phys. 1, 215 (1970); Rev. Math. Phys. 2, 151 (1971).
 
\bibitem{referee}
 C. F. Polnaszek, J. H. Freed,  J. Chem. Phys. 58, 3185 (1973).

\bibitem{siggia}
 E. D. Siggia, Phys. Rev. B 16, 2319 (1977).

\bibitem{kimball}
J. C. Kimball, J. Stat. Phys. 21, 289 (1979).

\bibitem{peschel}
 I. Peschel and V. J. Emery, Z. Phys. B 43, 241 (1981)
 
\bibitem{jpb_antoine}
 J.P. Bouchaud and A. Georges, Phys. Rep. 195, 127 (1990).
 
 
\bibitem{texier}
 C. Texier and C. Hagendorf, Europhys. Lett. 86 (2009) 37011.
 
\bibitem{us_eigenvaluemethod}
 C. Monthus and T. Garel, J. Stat. Mech. P12017 (2009).
 
\bibitem{Castelnovo}
 C. Castelnovo, C. Chamon and D. Sherrington, Phys. Rev. B 81, 184303 (2010).
 
 \bibitem{c_lyapunov}
% arXiv:2010.14994 
%Large deviations of the Lyapunov exponent in 2D matrix Langevin dynamics with applications to one-dimensional Anderson Localization models
C. Monthus, J. Stat. Mech. (2021) 033303.

\bibitem{us_gyrator}
%Nonequilibrium diffusion processes via non-Hermitian electromagnetic quantum mechanics with application to the statistics of entropy production in the Brownian gyrator
%arXiv:2210.05353
A. Mazzolo and C. Monthus, Phys. Rev. E 107, 014101 (2023).

\bibitem{us_kemeny}
% arXiv:2302.09965
% On the Kemeny time for continuous-time reversible and irreversible Markov processes with applications to stochastic resetting and to conditioning towards forever-survival
A. Mazzolo and C. Monthus, J. Stat. Mech. (2023) 063204.

 
\bibitem{c_pearson}
C. Monthus, J. Stat. Mech. (2023) 083204.
 
 \bibitem{c_boundarydriven}
C. Monthus,  J. Stat. Mech. (2023) 063206.



 \bibitem{c_largedevDBsusy}
 C. Monthus, J. Stat. Mech. (2024) 073203.
 %A supersymmetric quantum perspective on the explicit large deviations for reversible Markov jump processes, with applications to pure and random spin chains
%arXiv:2403.11525 


% my paper with m(.) et s(.) discrete
 \bibitem{c_susyFokkerPlanck}
C. Monthus, J. Phys. A: Math. Theor. 59 (2026) 105001
%2507.04941

%%%%%%%%%%%%%%%%%%%%%%%%%%%%%%%%%%%%%
% fully-factorized

\bibitem{c_QES}
%Using the slowest observable in one-dimensional Markov processes to construct quasi-exactly-solvable generators with two explicit levels
C. Monthus, J. Stat. Mech. (2026) 103204
%arXiv.2604.15908

%%%%%%%%%%%%%%%%%%%%%%%%%%%%%%%%%%%%%%%%%
\bibitem{pearson_student_processes}
%Student Processes
C. C. Heyde and N. N. Leonenko,
Advances in Applied Probability
Vol. 37, No. 2  pp. 342 (2005)

\bibitem{pearson_student}
% On exponential bounds for mixing and the rate of convergence for Student processes
N. Abourashchi and A. Yu. Veretennikov, Theor. Probability and Math. Statist. 81 (2010), 1



%%%%%%%%%%%%%%%%%%%%%%%%%%%%%%%%%%%
% REFS PEARSON


\bibitem{pearson1895}
K. Pearson,  Philos. Trans. R. Soc. Lond. Ser. A 186, 343–414 (1895)
% Contributions to the mathematical theory of evolution II. Skew variation in homogeneousmaterial.

\bibitem{pearson_wong}
E. Wong,  (1964) "The construction of a class of stationary Markoff processes",
in Stochastic processes, in mathematical physics and engineering (ed. R. Bellman), 264–276. American Mathematical Society, Rhode Island.

\bibitem{diaconis}
% Closed Form Summation for Classical Distributions: Variations on a Theme of De Moivre
P. Diaconis and S. Zabell, Statist. Sci. 6(3): 284-302 (1991)

\bibitem{autocorrelation}
% Diffusion-type models with given marginal distribution and autocorrelation function
B. M. Bibby, I. M. Skovgaard, M. Sorensen, Bernoulli 11(2), 2005, 191–220.

\bibitem{pearson_class}
% The Pearson Diffusions: A Class of Statistically Tractable Diffusion Processes
J. L. Forman and M. Sorensen,
Scandinavian Journal of Statistics, Vol. 35: 438–465, 2008

\bibitem{pearson2012}
G.M. Leonenko, T.N. Phillips,  Journal of Computational and Applied Mathematics 236 (2012) 2853–2868

\bibitem{PearsonHeavyTailed}
%On Spectral Analysis of Heavy-Tailed Kolmogorov-Pearson Diffusions
F. Avram, N.N. Leonenko, N. Suvak
Markov Processes and  Related Fields, v.19, Issue 2, 249-298 (2013)


\bibitem{pearson2018}
% "Optimal Diffusion Processes" 
S. Jafarizadeh, IEEE Control Systems Letters, vol. 2, no. 3, pp. 465-470,  2018

%%%%%%%%%%%%%%%%%%%%%%%%%%%%%%%%%%%%%%%%
% Geometric-BM

\bibitem{Peters}
O. Peters and W. Klein, Phys. Rev. Lett. 110, 100603 (2013).
%“Ergodicity breaking in geometric Brownian motion,” 
% 1209.4517

\bibitem{reviewGBM}
%Generalised geometric Brownian motion: Theory and applications to option pricing
V. Stojkoski, T. Sandev, L. Basnarkov, L. Kocarev, R. Metzler, Entropy  22, 1432 (2020)
% arXiv:2011.00312

\bibitem{c_carleman}
%Classification of diffusion processes in dimension d via the Carleman approach with applications to models involving additive, multiplicative or square-root noises
C. Monthus, J. Stat. Mech. (2026) 033204
% arXiv.2512.03857

%%%%%%%%%%%%%%%%%%%%%%%%%%%%%%%%%%%%%%%%%
% Bessel

\bibitem{Bessel73}
T. Shiga and S. Watanabe, Z. Wahrscheinlichkeitstheorie verw. Geb. 27, 37 (1973). 
%“Bessel diffusions as a one-parameter family of diffusion processes,” 


\bibitem{BesselQM}
%Bessel Process and Conformal Quantum Mechanics
M. A. Rajabpour, Journal of Statistical Physics 136 (2009) 785
%/arXiv.0906.1728

%%%%%%%%%%%%%%%%%%%%%%%%%%%%%%%%%
% Squared-Bessel

\bibitem{BesselYor}
A. Going-Jaeschke and M. Yor, Bernoulli 9, 249 (2003).
%“A survey and some generalizations of Bessel processes,” 



%%%%%%%%%%%%%%%%%%%%%%%%%%%%
% Fisher-Snedecor

\bibitem{pearson_fisher}
% On spectral properties and statistical analysis of Fisher-Snedecor diffusion arXiv:1007.4909
%Spectral representation of transition density of Fisher–Snedecor diffusion
F. Avram, N. N. Leonenko andN. Suvak,
International Journal of Probability and Stochastic Processes
Volume 85,  Pages 346-369 (2013)

\bibitem{pearson_fisherSnedecor}
% Ergodicity and mixing bounds for the Fisher—Snedecor diffusion
A.M. Kulik and N.N. Leonenko,
Bernoulli Vol. 19, No. 5B (2013)



\end{thebibliography}
\end{document}